\documentclass[prodmode,acmtkdd]{acmsmall} % Aptara syntax
\usepackage{url}
\usepackage{multirow}
\usepackage{array}
\usepackage{rotating}
\usepackage{amsmath}
\usepackage{graphicx}
\usepackage{caption}
\usepackage{amssymb}
\usepackage{rotating}

\usepackage[ruled]{algorithm2e}

\SetAlFnt{\small}
\SetAlCapFnt{\small}

\begin{document}

% Page heads
\markboth{T. Chakraborty et al.}{Permanence and Community Structure in Complex Networks}

% Title portion
\title{Permanence and Community Structure in Complex Networks\\{\color{blue} Accepted in ACM TKDD}}
\author{Tanmoy Chakraborty
\affil{University of Maryland, College Park, USA -- 20742}
Sriram Srinivasan
\affil{University of Nebraska at Omaha, USA -- 68106}
Niloy Ganguly
\affil{Indian Institute of Technology, Kharagpur, India -- 721302}
Animesh Mukherjee
\affil{Indian Institute of Technology, Kharagpur, India -- 721302}
Sanjukta Bhowmick
\affil{University of Nebraska at Omaha, USA -- 68106}
}

% NOTE! Affiliations placed here should be for the institution where the
%       BULK of the research was done. If the author has gone to a new
%       institution, before publication, the (above) affiliation should NOT be changed.
%       The authors 'current' address may be given in the "Author's addresses:" block (below).
%       So for example, Mr. Abdelzaher, the bulk of the research was done at UIUC, and he is
%       currently affiliated with NASA.

\begin{abstract}

The goal of community detection algorithms is to identify densely-connected units within large networks. 
An implicit assumption is that all the constituent nodes belong equally to their associated community.  However, some nodes are
more important in the community than others. To date, 
efforts have been primarily driven to identify communities as a whole, rather than understanding to what extent an individual node belongs to its community.  Therefore, most metrics for evaluating communities, for example modularity,  are global. These metrics produce a score for each community, not for each individual node. In this paper, we argue that the belongingness of nodes in a community is not uniform. %Some nodes are highly associated with the community and others are not. 
We quantify the degree of belongingness of a vertex within a community by a new vertex-based metric called
{\it permanence}. 

The central idea of permanence is based on the observation that the strength of membership of  a vertex to a community depends upon  two
factors: (i) the the extent of connections of the vertex within its community versus outside its community, and (ii) how
tightly the vertex is connected
internally. We present the formulation of permanence based on these two quantities. We demonstrate that compared to other  existing metrics
(such as modularity, conductance and cut-ratio), the change in permanence is more commensurate to the level of perturbation in  ground-truth
communities. We  discuss  how permanence can help us understand and utilize the structure and evolution of communities by demonstrating that
it can be used to -- (i)  measure the persistence of a vertex in a community, (ii)  design strategies to strengthen the community structure,
(iii) explore the core-periphery structure within a community, and  (iv) select suitable initiators for message spreading. 

We further show that permanence is an excellent metric for identifying communities.  We demonstrate that the process of maximizing permanence
(abbreviated as {\em MaxPerm}) produces meaningful communities that concur with the ground-truth community structure of the networks more
accurately than eight other popular community detection algorithms. Finally, we provide  mathematical proofs to demonstrate the correctness
of  finding communities by maximizing permanence. In particular, we show that the communities obtained by this method are  (i) less affected
by the changes in vertex-ordering, and (ii) more resilient to resolution limit, degeneracy of solutions and asymptotic growth of values.

\end{abstract}

\category{I.5.3}{Clustering}{Algorithms}

\terms{Design, Algorithms, Performance}

\keywords{Permanence, community discovery, community evaluation metric, modularity}

\maketitle

\section{Introduction}
Community detection is the process of finding closely related groups of entities in a network. Complex networks, such as those arising in biology, social sciences and epidemiology, represent  systems of  interacting entities. The entities are represented as vertices in the network and their pair-wise interactions are represented as edges. A community, then, is  a group of vertices that have more internal connections (i.e., connections to vertices within the group) than external connections (i.e., connections to vertices outside the group). 

Most community detection algorithms are based on  combinatorial optimization.  The goal is to find the community assignment that leads to the optimal value of a specified network parameter, such as modularity~\cite{NewGir04} or conductance~\cite{cond_09}. However, since many real-world communities are based on subjective measurements (as opposed to a formal mathematical definition), the validation of the results is done by comparing the obtained communities with known ``ground-truth" communities. 
Very rarely do the obtained communities exactly match with ground-truth communities. Moreover, due to
the phenomenon of resolution limit and degeneracy of solutions \cite{Barthelemy}, the optimum parameter value sometimes produces intuitively incorrect solutions. As a result, community detection is an active area of research with new optimization metrics  being regularly proposed \cite{Dongxiao,Yang:2012}, that either produce more accurate results on a certain subclass of networks and/or can address some of the above discussed issues.

Almost all community detection metrics and related algorithms contain the implicit notion that all the vertices in a community belong {\it
equally} to the community, i.e., the community membership is {\em homogeneous}. Therefore, the optimal score of the metric can only be
obtained over the network as a whole. The information about how placement of vertices affects the community structure is lost using these
current measurements. In order to include this important information, we have introduced a vertex-centric metric called {\it permanence}
\cite{Chakraborty_kdd}.

The key idea in formulating permanence is as follows. Most optimization metrics are based on the total internal and external connections of the vertex. We posit that the distribution of  the external connections of a vertex is equally important.  In particular, our vertex assignment decisions are based not on the total number of external connections but on {\it the maximum number of external connections 
to any single neighboring community}.  If vertex $v$ is in community $S$ and vertex $u$ is in community $T$ and there exists an
edge $(v,u)$, then $S$ and $T$ are neighboring communities of each other.

To the best of our knowledge, we are the first to make this distinction between the total external connections and their distribution. Permanence of a vertex thus  quantifies its propensity  to remain in its assigned community and the extent to which it is ``pulled'' \cite{Chakraborty_kdd} by the neighboring communities.

Permanence provides a {\em heterogeneous} vertex-centric measure (in the range 1 to -1) of the extent to which a 
vertex belongs to its community. A value of 1 indicates that a vertex is placed correctly in its community, and a value of -1 indicates the
vertex does not belong to its assigned community. The permanence of the network is given by the average permanence of all the vertices of a
network. It is easy to see that the more correctly the vertices are placed in their communities, the higher the overall permanence. Using
this concept we propose a new community detection algorithm, {\it MaxPerm} based on optimizing the permanence of the
network~\cite{Chakraborty_kdd}.

In this paper, after providing background information on the related work in community detection (Section \ref{related_work}) and the
datasets that we used in our experiments (Section \ref{dataset}), we explain the rationale behind creating the permanence metric (Section
\ref{permanence_def}). We show how change in permanence is more commensurate to perturbation of ground-truth communities as compared to 
other competing metrics (Section \ref{parturbation}) and  present a community detection, {\em MaxPerm}, based on maximizing permanence
(Section~\ref{comm}) . We extend this earlier work with the following new contributions:
\begin{itemize}
\item {\it In-depth Study of Network Parameters Affecting the Value of Permanence.} We study how the distribution of different network parameters, such as connectivity and clustering coefficient affect the value of  permanence (Sections \ref{comp} and \ref{parturbation}).
\item {\it Use of Permanence to Understand the Structure of the Network.} We show that permanence can provide a better understanding of the
structure of the network, such as how to strengthen the communities and revealing the core-periphery based on the community structure, as
well as its use in applications such as  vertex persistence in communities of evolving networks and identifying effective initiators for
message spreading (Section~\ref{sec:significance}).
\item {\it  In-depth Analysis of Communities Obtained using MaxPerm.} The communities obtained by MaxPerm are in general of a smaller size
than those obtained by other community detection methods. We study the structure of the communities obtained and show that these communities
are actually well defined sub-communities (Section~\ref{comm}).
\item {\it Algorithmic Factors Affecting Results of MaxPerm.} We study how different algorithmic factors such as ordering of the vertices and selection of initial seed communities affects the results of MaxPerm (Sections \ref{vertex_per} and  \ref{sec:seed}). 
\item {\it Analytical Proof on Correctness.} We provide analytical results to demonstrate how finding communities by maximizing permanence
reduces the existing limitations of community detection algorithms such as resolution limit, degeneracy of solutions and asymptotic growth
of values. (Section \ref{permanence_property}).
\end{itemize}

%
%The remainder of the paper is organized as follows; In and we provide the back ground information including an overview of t. We  In Section , we show the different factors comprising permanence affects the quantity as a whole. In   we demonstrate how In , we discuss how permanence can be used in other applications beyond community detection and in providing insights to the network structure. In  we present a community detection algorithm empirical results on how it improves community detection as compared to other algorithms. We then investigate how different algorithmic factors, such as selection of vertex ordering  and  seed community (Section) can affect the results of our algorithm. Finally, we conclude the paper with interesting insights on permanence and possible future directions in Section \ref{conc}.

%\textcolor{red}{Perhaps it is good to mention exactly what is new in the paper}

\if{0}
most of the community scoring metrics consider a community as a whole without looking much into the arrangement of nodes in the community. Here, we argue that the community membership should be {\em heterogeneous}. There should be few vertices which are highly involved into the community; others might not be that much involved. To quantify the notion of belongingness, we need an appropriate vertex-centric metric. In this article, we propose a  novel  vertex-based scoring function called {\em permanence} which precisely indicates the extent to which a vertex belongs to the community. The key idea behind formulating permanence is as follows. Most optimization metrics consider either the degree of a vertex in a community or  the total number of external connections of the vertex (i.e., those connections that are attached with the other neighbors of the vertex outside the community). We posit that the distribution of  the external connections of a vertex is equally important. For instance, in Figure \ref{example}
, even if vertex $v$ has more external connections than vertex $u$, the connections are uniformly distributed among three communities, where the distribution for vertex $u$ is non-uniform. In particular, our vertex assignment decisions are based not on the total number of external connections but on the maximum number of external connections 
to any single neighboring community.  To the best of our knowledge, we are the first to make this distinction between the total external connections and their distribution. Permanence of a vertex thus  quantifies its propensity  to remain in its assigned community and the extent to which it is ``pulled'' \cite{chakraborty} by the neighboring communities.

all networks possess community structures with equal strength. For example, a network composed of several sparsely connected dense cliques (see Figure \ref{example_diss}(a)) will have strong communities whereas a grid (see Figure \ref{example_diss}(b)) will not have any community structure at all, and between these two extremes there exist communities of different strength as per the network structure\footnote{\scriptsize{In this paper, we consider only the non-overlapping communities.}}. As of now, there is no community detection metric that can measure by what extent a vertex is a part of a community. One of the reasons for this deficiency is that the optimum value of parameters such as modularity is not  exactly related to whether the network possesses a strong community, but rather tries to identify the best community assignment, for any given 
network. For example, the highest achievable modularity in the Jazz network \cite{Newman:2006} is 0.45 \cite{chakraborty,Newman:2006} and the Western 
USA  power grid ~\cite{Watts1998} exhibits a modularity of 0.98. However, as we have seen in \cite{Newman:2006,Newman:2006}, Jazz has a much stronger community structure than the power grid. Indeed, most algorithms output a set of communities regardless of whether the network (such as a grid) possesses a community structure or not. The key reason for this is that optimization metric such as modularity frequently enforces the detection algorithm to make a choice by arbitrarily breaking ties. While this indeed increases the value of the metric, each such tie-breaking obfuscates the possibility of other community assignments. In grid-like networks, where choices can occur frequently, such tie-breaking can produce inaccurate or insignificant communities, while producing a high scoring function. Although, methods for finding consensus communities \cite{Santo} can indicate whether the communities are significant or not, these techniques are dependent on the number of algorithms used to find the consensus.

 \begin{figure}[!t]
\centering
 \includegraphics[scale=0.35]{./Fig/figure.eps}
 \caption{Toy example depicting {\em permanence} of two vertices $u$ and $v$. Even if vertex $v$ has more number of external
connections than $u$, all these six external connections are distributed equally into three neighboring communities,
resulting in the external pull proportional to 2; whereas $u$ is attached with 4 external neighbors, three of them constitute in
one community and the rest is attached with another community, resulting in the external pull proportional to 3. On the other hand, $v$ is
connected to 3 internal neighbors which are further completely connected among each other; whereas the neighbors of $u$ are partially
connected. This results in high internal pull of $v$ as compared to $v$. We capture these two notions of connectively into
the formulation of permanence.}\label{example}
\end{figure}

The sum of the  permanence of all vertices, normalized by the number of vertices provides the cumulative permanence of the network. It indicates by how much, on an average, the vertices of a network are in their correct communities. As with permanence, this value also ranges from 1 to (nearly) -1, where 1 denotes a network (such as a ring of cliques shown in Figure~\ref{example_diss} (a)) with perfect community structure, 0 (such as a grid shown in Figure~\ref{example_diss} (b)) with no community structure at all and -1 (where all vertices are incorrectly assigned). Therefore, maximizing permanence can also be used as an alternate method for identifying communities. This approach of combining the microscopic (vertex-level) information to obtain the mesoscopic (community-level) information provides a more fine-grained view of the modularity structure of the network. Specifically, permanence of a graph produces high values only if the network possesses an inherent community structure across most of its 
vertices. 
As the community structure of the networks 
degrade, so does the value of permanence of the entire network.

We conduct a set of experiments to show the effectiveness of permanence over other state-of-the-art algorithms on both synthetic and real-world networks. To start with, we analyze the effect of each individual component on the value of permanence and observe that it nicely unfolds the organization of nodes within a community. Then we study the effect of community-centric noise on the value of permanence and observe that permanence is appropriately sensitive to the noise -- it can tolerate minor noise without affecting much; however with the increase of noise, it starts deteriorating after a certain threshold. Following this, we present a bunch of results to show the implications of permanence on the real-world networks -- (i) permanence can measure the persistence of vertices within a community, (ii) permanence helps in strengthening the community structure, (iii) permanence unfolds the core-periphery organization of vertices within community, (iv)  permanence provides an inherent ranking of vertices that 
can help in different applications such as initiator selection in message passing.

Further, we propose a greedy agglomerative algorithm, called {\em MaxPerm}, that tries to maximize permanence in order to discover communities. It begins with initializing every 
vertex to a singleton community. A vertex is moved to a community only if this movement results in a net increase in the number of internal connections and/or a net decrease in the number of external connections to any of the neighboring communities. If such a move is not possible, then either the vertex remains as a singleton (such as in the case where moving to any one of the candidate communities will give equal permanence) or moves to the community where it is more tightly connected with its neighbors (this causes the vertex to have positive permanence). Experiments on a set of synthetic and real-world networks having ground-truth community structure show that our algorithm finds more meaning communities as compare to the other eight state-of-the-art baseline algorithms. Moreover, our algorithm provides a bunch of other utilities: (i) it tends to detect small-size communities (even singleton communities), which are otherwise merged into one community to form a large community, (ii) it is less 
affected by the ordering of vertices \cite{chakraborty} in detecting community structure, and (iii) both theoretical and empirical observations show that our algorithm is less prone to the standard problem of modularity maximization, such as resolution limit, degeneracy of solutions and asymptotic growth of value. Moreover, the value of permanence is relatively independent of the size of the network.       
\fi

% As we will demonstrate in this paper, the principal benefits of our approach are:
% \begin{itemize}
% \item Permanence proves to be a remarkably suitable metric for evaluating the goodness of the community structure obtained from different identification algorithms.
% \item Permanence is very sensitive to the different perturbations of the network which should be an ideal property of a community scoring metric.
% \item Maximizing permanence is more successful in finding ground-truth communities as compared to modularity-maximizing algorithms as well as other algorithms. It is also able to discover sub-communities from a large community.
% \item Community detection using maximizing permanence can overcome the resolution limit, degeneracy of solutions, in many networks. Moreover, the value of permanence is relatively independent of the size of the network.
% \end{itemize}

{ We make our experimental codes available in the spirit of reproducible research: \url{http://cnerg.org/permanence}.}

\section{Related Work}\label{related_work}

We present the ongoing research on two aspects of community detection: (i) algorithms to detect community and (ii) metrics for evaluating
the correctness of the obtained community.

\subsection{Community detection algorithms}
Most of the research in community detection algorithms are based on the idea that a community is a set of nodes that has more and/or better links between its members than with the remainder of the network. Work in this area encompasses many different approaches including, modularity optimization~\cite{blondel2008,Clauset2004,Guimera,newman03fast,Newman:2006}, spectral
graph-partitioning algorithm~\cite{Newman_13,Thomas}, clique percolation~\cite{Vicsek,PalEtAl05}, local expansion~\cite{Baumes:2005,Andrea},
fuzzy clustering~\cite{Psorakis,Sun_11}, link partitioning~\cite{AhnY2010,evans:2009}, random-walk based
approach~\cite{DeMeo:2013,JGAA-124}, information theoretic approach~\cite{rosvall2007,Rosvall29012008}, diffusion-based
approach~\cite{Raghavan:1057930}, significance-based approach~\cite{DBLP:journals} and label propagation~\cite{Raghavan-2007,Xie_11,Xie_12}.

However most of these algorithms produce different community assignments if certain algorithmic factors, 
such as the order in which the vertices are processed, changes. \cite{Santo} proposed consensus clustering by re-weighting the edges based
on how many times the pair of vertices were allocated to the same community, for different identification methods. Several pre-processing
techniques~\cite{sriram,seed-set-tr} have been developed to improve the quality of the solution. These methods form an initial estimate of
the community allocation over a small percentage of the vertices and then refine this estimate over successive steps. Recently, \cite{chakraborty} pointed out how vertex ordering influences the results of the community detection algorithms. They identified invariant groups of vertices (named as ``constant communities'') whose assignment to communities are not affected by vertex ordering. 

%Most of them are designed to detect communities from static networks. On the other hand, a large number of algorithms were proposed
%to cope with community detection on dynamically evolving networks (i.e., Internet, Online Social Networks), such as
%LabelRankT~\cite{Chen_13}, Estrangement~\cite{Kawadia_NSR2012} and intrinsically dynamic community detection
%algorithm~\cite{Bivas_11}. 

\subsection{Community evaluation metrics}
Most community detection algorithms are based on optimizing a combinatorial metric. Examples of such metrics include conductance
\cite{cond_09,Kannan:2000,Shi:2000}, cut-ratio \cite{Fortunato201075,Leskovec:2010}; however, 
the most popular and widely accepted metric is modularity~\cite{Newman:2006,NewGir04}. It is defined as the difference
(relative to the total number of edges) between the actual and the expected (in a randomized graph with the same number of 
nodes and the same degree sequence) number of edges inside a given community. Although initially defined for undirected 
and unweighted networks, the definition of modularity has been extended to capture community structure in 
weighted~\cite{PhysRevE.70} and directed~\cite{PhysRevLett.} networks. 

It was also demonstrated that modularity suffers from a {\it resolution limit}, that is, by
optimizing modularity we cannot find communities smaller than a threshold size~\cite{Barthelemy}, or weight~\cite{Berry_PRE2011}. The threshold depends on
the total number (or total weight) of edges in the network and on the degree of interconnectedness between communities. Later,
\cite{good2010} also   showed that optimizing modularity can lead to {\it degeneracy of solutions}, i.e.,  an
exponential number of high (and nearly equal)-modularity but structurally distinct solutions from a single graph. They also studied the
({\it asymptotic}) {\it growth} of modularity, showing that it depends strongly
both on the size of the network and the number of modules it contains. 

To address the resolution limit problem, multi-resolution versions of
modularity~\cite{Arenas,reichardt2006smc} were proposed to allow researchers to specify a tunable target resolution limit parameter.
\cite{Renaud} proposed different types of multi-resolution quality functions to tackle resolution limit problem. \cite{Dongxiao} considered
different community densities as good quality measures for community identification, which do not suffer from resolution limits.
Furthermore, \cite{santo_11} stated that even those multi-resolution versions of modularity are inclined to merge the smallest
well-formed communities, and to split the largest well-formed communities. Recently, \cite{Chen_2013} proposed {\it Modularity Density}
metric to solve the problems raised by \cite{santo_11}. A detailed review can be found in \cite{0002DMG16}.\\

\noindent{\bf Metrics to compare with ground-truth communities:} Although all these metrics mentioned above are useful in analytically evaluating a community structure, a stronger measure of correctness is to 
 compare the obtained
community structure with the actual known community structure (ground-truth) of a network. To compare these two community structures, different
validation metrics have been proposed, such as  Normalized Mutual Information (NMI)~\cite{danon2005ccs}, Adjusted Rand Index
(ARI)~\cite{hubert1985} and Purity (PU)~\cite{Manning}. However, \cite{Labatut} argued that these metrics are not completely relevant in the
context of network analysis, because they ignore the network structure. They proposed the  weighted versions of these measures where
misplacing a high degree vertex would incur higher penalty compared to a low degree vertex. In our experiments, we, therefore, also use the weighted versions
of these measures, namely Weighted-NMI ({\em W-NMI}), Weighted-ARI ({\em W-ARI}) and Weighted-Purity ({\em W-PU}) (we refer the reader to the
paper \cite{Labatut} for the detailed descriptions of these metrics). Note that all the metrics are bounded between 0 (no matching) and 1
(perfect matching).

% 
% In Section \ref{dataset}, we present a detailed description of the synthetic and real-world datasets used in this experiment. In Section \ref{permanence_def}, we demonstrate the motivation and the definition of our proposed measure, {\em permanence}. Following this, we describe the experimental setup in Section \ref{setup}. Next,  we show that permanence can be considered as a better community scoring function compared to the other existing metrics such as modularity, conductance and cut-ratio in Section \ref{goodness}. Section \ref{parturbation} shows that permanence is appropriately sensitive to the network perturbations. We further characterize permanence from difference perspectives in Section \ref{sec:significance}. In Section \ref{comm}, we demonstrate a greed agglomerative algorithm using maximizing permanence to detect non-overlapping communities from both synthetic and real-world networks and compare its performance with 8 baseline algorithms in Section \ref{sec:
% eval}. In Section \ref{permanence_property}, we provide theoretical results to prove that maximizing permanence can deduce the effect of resolution limit, degeneracy of solutions and asymptotic growth of values, often observed in modularity maximization. Finally, we draw the conclusion with possible future directions in Section \ref{discussion}.

\section{Network Datasets and Ground-truth Communities}\label{dataset}
We examine a set of artificially generated networks and three real-world complex networks whose underlying ground-truth community
structures are known to us. The brief description of the used datasets and their ground-truth communities are mentioned below.

\subsection{Synthetic networks}  \label{sec:synthetic_nw}
We select the LFR benchmark model \cite{Lancichinetti} to generate artificial networks with a community structure.  The model allows to
control directly the following properties: number of nodes $n$, desired average degree $k$
and maximal degree $k_{max}$, exponent $\gamma$ for the degree distribution, exponent $\beta$ for the community size distribution, and
mixing coefficient $\mu$. The parameter $\mu$ represents the desired average proportion of links between a node and the nodes located outside
its community, called \emph{inter-community links}.  Unless otherwise stated, the LFR network is generated with the number of nodes ($n$) as
1000, and $\mu$ is varied from 0.1 to 0.6. For the rest of the
parameters, we use the default value of the parameters mentioned in the 
implementation\footnote{\url{https://sites.google.com/site/santofortunato/inthepress2}} designed by \cite{Lancichinetti}.

\begin{table}[!h]
\tbl{Properties of real-world networks. $n$ and $e$ are the number of nodes and edges, $c$ is the number of communities, $<k>$ and
$k_{max}$ its average and maximum degree, $n_c^{min}$ and $n_c^{max}$ the sizes of its smallest and largest
communities.\label{tab:dataset}}{
\begin{tabular}{l|c|c|c|c|c|c|c}
\hline
Network            & $n$ & $e$ & $<k>$ & $k_{max}$ & $c$ & $n_c^{max}$ & $n_c^{min}$\\
\hline\hline
Football           & 115 & 613 & 10.57 & 12 & 12 & 13 & 5  \\
Railway           & 301 & 1,224 & 6.36 & 48 & 21 & 46 & 1 \\
Coauthorship     & 103,677 & 352,183 & 5.53 & 1,230 & 24 & 14,404 & 34 \\
\hline
\end{tabular}}
\end{table} 

\subsection{Real-world networks}\label{real_nw}
We use three real-world networks mentioned below whose ground-truth community structures are known a priori. The
properties of these dataset are summarized in Table~\ref{tab:dataset}. 

\noindent \textbf{Football network} \cite{GN} contains the network of American football games between Division IA colleges during the 
regular season of Fall 2000. The vertices in the graph represent teams (identified by their college names) and edges represent regular-season games
between the two teams they connect. The teams are divided into conferences (indicating communities) containing around 8-12 teams each. 

% Games are more frequent
% between members of the same conference than between members of
% different conferences, with teams playing an average of about seven intra-conference games and four inter-conference games in the 2000
% season. Inter-conference play is not uniformly distributed; teams that are geographically close to one another but belong to different
% conferences are more likely to play one another than teams separated by large geographic distances.

\noindent \textbf{Railway network} \cite{Ghosh} consists of nodes representing stations, where two
stations $s_i$ and $s_j$ are connected by an edge if there exists at least one train-route such that both $s_i$ and $s_j$ are scheduled halts on that route.  Here the communities
are states/provinces of India since the number of trains within each state is much higher than the trains in-between
two states. 

\noindent \textbf{Coauthorship network} \cite{asonam}  is derived from the citation dataset\footnote{\url{http://cnerg.org/}}.
 Here each node represents an author and an undirected edge between authors is drawn if the two authors
collaborate at least once via publishing a paper. The communities are marked by the research fields since authors have a tendency to
collaborate with other authors within the same field. Besides the aggregated network, we also create some intermediate networks mentioned in
Table \ref{asymptotic} by cumulatively aggregating all the vertices and edges over each year, e.g., 1960-1971, 1960-1972, ..., 1960-1980.

\if{0}

%
%{\color{red} A legitimate concern is whether the used data sets are suitable for testing community detection in general, and the question is whether the assignment to ground-truth communities corresponds to the community structure of a network. For that, we use internal density~\cite{Ciglan} (ratio between number of actual connections and the possible number of connections in a community) of the ground-truth communities and compare it to internal density of a same-sized group of nodes in a random graph with identical number of network elements (nodes and edges). The results are presented in last column of Table~\ref{tab:dataset} and show that the average internal density of ground-truth communities is high, orders of magnitude higher than in the case of random graphs. The conclusion is that the ground-truth communities used in the experiments are related to the informal notion of a community as a densely connected subgraph.}

\vspace{-3mm}
\subsection{Scoring Functions for Evaluating Community Structure}\label{subsec:eval}
The goodness of a community is often measured by how well certain scoring functions are optimized. Here we compare the optimal value of
permanence for the obtained communities versus three popular scoring functions, namely modularity (Mod)~\cite{Newman:2006},
conductance (Con)~\cite{cond_09} and cut-ratio (Cut)~\cite{Leskovec:2010}. In order to make the higher the better, we measure (1-Con) and (1-Cut) for conductance and cut-ratio respectively. 

%We run each algorithm discussed in subsection~\ref{algorithms} on all the datasets mentioned in section~\ref{sec:dataset}, and for every output, we compute all four community scoring functions. 
%Then we rank the algorithms based on each of the community scoring functions separately. Table~\ref{table1} shows the scores and the ranks (within parenthesis) for the football network (columns 2-5) using different community scoring functions. The lower the value of conductance and cut-ratio, the better the quality of
%the communities. 

\subsection{Metrics to Compare with Ground-truth}\label{metrics}
A stronger test of the correctness of the community detection algorithm, however, is by comparing the obtained community with a given ground-truth structure.
We use three standard validation metrics, namely Normalized Mutual Information (NMI)~\cite{danon2005ccs}, Adjusted Rand Index (ARI)~\cite{hubert1985} and Purity (PU)~\cite{Manning} to measure the accuracy of the detected communities with respect to the ground-truth community structure. \cite{Labatut} argues that these measures have certain drawbacks in that they ignore the connectivity of the network. We therefore also use the weighted versions of these measures, namely Weighted-NMI (W-NMI), Weighted-ARI (W-ARI) and Weighted-Purity (W-PU) as proposed in~\cite{Labatut}. Note that, all the metrics are bounded between 0 (no matching) and 1 (perfect matching).

%are not completely relevant in the context of network analysis, because they
%ignore the network structure. He proposed the modified versions of these measures where
%misplacing a high degree vertices would incur higher penalty compared to a low degree vertices. The modified versions of NMI, ARI and Purity are %We refer the reader to the paper~\cite{Labatut} for the detailed descriptions of these metrics. We obtain the ranking with respect to each of the six validation metrics. The ranking for football network is mentioned within parenthesis in the corresponding cells (columns 6-13 in Table~\ref{table1}). Average ranking of normal and weighted metrics are mentioned in columns 9 and 13 respectively. 

\subsection{Community Detection Algorithms}\label{algorithms}
We use the following community detection algorithms for comparison with our proposed algorithm discussed in
Section~\ref{comm}:\\
\textbf{(i) Modularity-based:}  FastGreedy \cite{newman03fast}, Louvain \cite{blondel2008} and CNM \cite{Clauset2004}.\\
%Modularity identifies groups of nodes as communities if they have more internal connections than they would have if they were connected by chance~\cite{Newman:2006}. We use three different modularity optimization algorithms, namely 
\textbf{(ii) Random walk-based:} WalkTrap \cite{JGAA-124}. \\
%These algorithms identify communities based on the concept that for a tightly connected community, random walk through a network will get trapped into a community. We use the random walk-based 
\textbf{(iii) Compression-based:} InfoMod \cite{rosvall2007} and InfoMap~\cite{Rosvall29012008}.

\fi

\begin{figure}[!t]
\centering
 \includegraphics[scale=0.4]{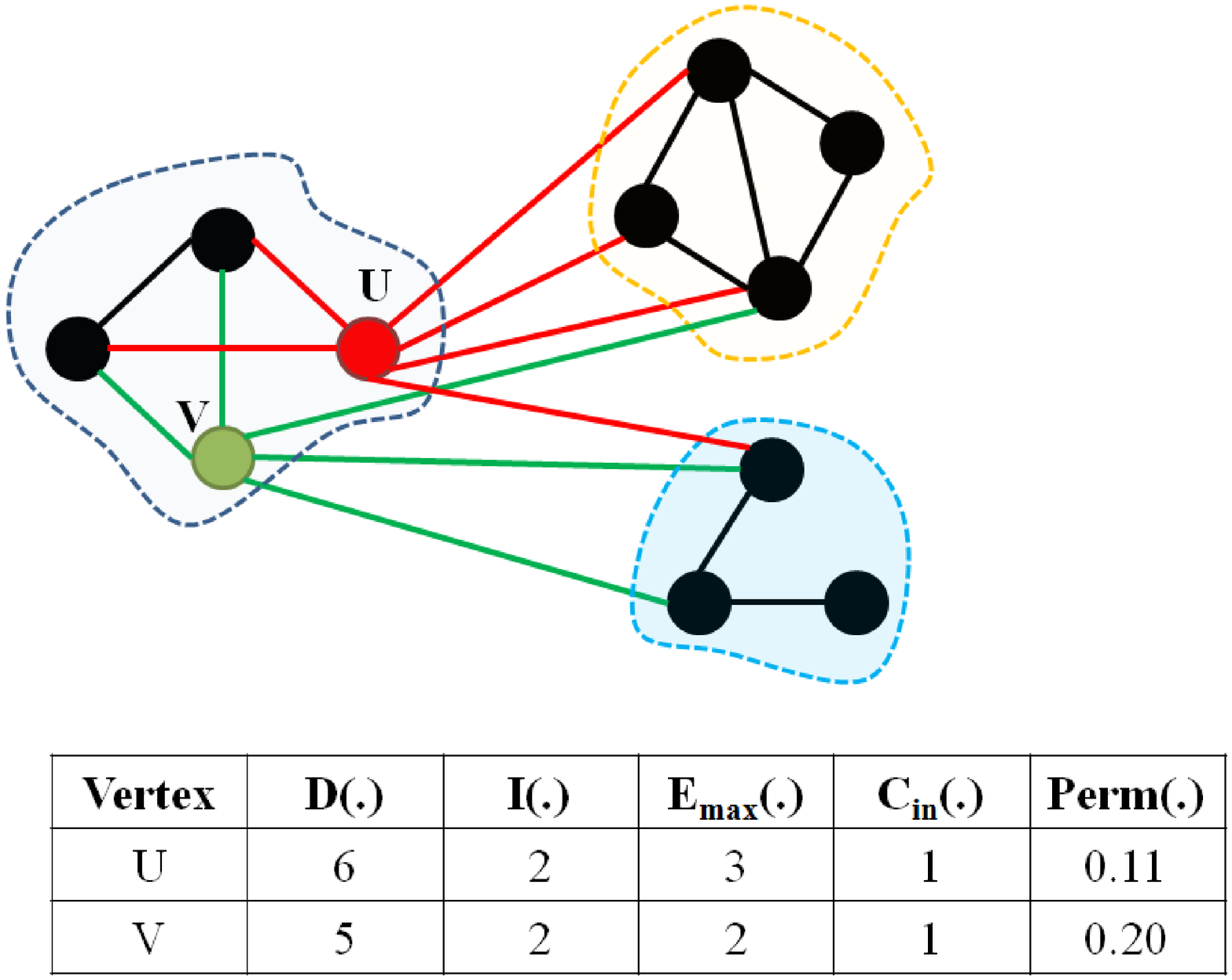}
 \caption{(Color online) Toy example depicting {\em permanence} of two vertices $u$ and $v$. The communities are represented
by broken lines. \label{example}}
%   Although vertex $v$ has more number of external
%connections than $u$, all these six external connections are distributed equally into three neighboring communities,
%resulting in an external pull proportional to 2. On the other hand, $u$ is attached with 4 external neighbors, three of them constitute in
%one community and the remaining one is attached with another community, resulting in an external pull proportional to 3. In the case of internal connections, $v$ is connected to 3 internal neighbors which are connected with each other; whereas the neighbors of $u$ are partially
%connected. This results in high internal pull of $v$ as compared to $u$.  Permanence captures both these notions of connectedness.
\end{figure}

\section {Defining Permanence}\label{permanence_def}

In this section, we describe the permanence metric and the two primary concepts behind its formulation. \\

\noindent{\bf  Concept I:} {\em A vertex should have more number of internal connections than the number of connections to any of the external
neighboring communities.} 

Most optimization metrics consider the {\it total number of external neighbors} of the vertex. However, in our
earlier experiment \cite{Chakraborty_kdd,1742,0002KGMB16}, we empirically demonstrated that a group of vertices are likely to be placed
together so long as the
number of internal connections is {\it larger} than the number of connections to {\it any one single external community}. In other words,
a vertex which has connections to some external communities, experiences a {\it separate} ``pull'' from each of these external communities. In formulating permanence we consider the maximum pull, which is 
proportional to the maximum number of connections to an external community (see Figure \ref{example}).  \\
% 
% \begin{figure}[!h]
% \begin{center}
%  \begin{tabular}{c c}
% %  \multicolumn{2}{c}{\includegraphics[scale=0.2]{./Fig/example}}  \\
% %   \multicolumn{2}{l}{\scriptsize{(a) Example demonstrating the importance of the distribution of} }\\
% %   \multicolumn{2}{l}{\scriptsize{external connections.} }\\
% %   \multicolumn{2}{c}{\includegraphics[scale=0.20]{./Fig/ratio.eps}}  \\
% %   \multicolumn{2}{l}{\scriptsize{(b) Fraction of vertices versus the ratio between the number of }}\\
% %   \multicolumn{2}{l}{\scriptsize{ total ($E_{sum}$) and maximum ($E_{max}$) external connections.}}\\
%       \multicolumn{2}{c}{\includegraphics[scale=0.3]{./Fig/demo1.eps} } \\
% %   \multicolumn{2}{l}{\scriptsize{(c) Two networks with same modularity, conductance and cut-ratio,}}\\
% %    \multicolumn{2}{l}{\scriptsize{but the left one has more prominent community structure.} }\\
% % %     \multicolumn{2}{c}{\includegraphics[scale=0.18]{./Fig/c_in_bin.eps}}  \\
% %   \multicolumn{2}{l}{\scriptsize{(d) Fraction of vertices with a specific range of internal clustering} }\\
% %   \multicolumn{2}{l}{\scriptsize{ coefficient ($c_{in}$) in LFR and real-world networks.} }\\\\
% 
%  \end{tabular}
% \end{center}
%  \caption{(Color online) Two networks with same modularity, conductance and cut-ratio, but the left one has more prominent community
% structure.  }\label{examplefig}
% \end{figure}

\if{0}
that in permanence we consider the distribution of external connections of the vertex to its neighboring communities. A vertex that has
equal number of connections to all its external communities (e.g., a vertex with total 6 external connections with 2 to each of 3
neighboring communities) has equal ``pull" from each community whereas a vertex with more external connections to one  particular community 
(e.g., a vertex with total 6 external connections with 1 connection each to two neighboring communities and 4 connections to the third
neighboring community), will experience more ``pull" from that community due to large number of external connections to it.

This property is demonstrated by a toy example in Figure~\ref{examplefig}(a). If the edge $(a,b)$ is deleted and the edge $(a,c)$ is added, then the number of external connections remains the same, 
and the value of modularity, conductance and cut-ratio are also the same for this change. However, in the initial graph, vertex $a$ had more
``pull" from the community of $b$, in fact proportional to the number of its internal connections, whereas in the modified version $a$ has
equal pull from both the communities of $b$ and $c$. Our  permanence formula, defined in Section~\ref{sec:perm}, takes this distinction
into account. 

Figure~\ref{examplefig}(b) shows a histogram of the fraction of
vertices versus the ratio between the number of total ($E_{sum}$) and maximum ($E_{max}$) external connections for two representative
networks. We notice that -- (i) very few vertices have (closely) similar values of $E_{sum}$ and $E_{max}$ (i.e., ratio=1); the majority 
have
significantly different $E_{sum}$ and $E_{max}$ (ii) the ratio between these two quantities  is not constant;  it is spread over
a wide range of values.  Therefore, we cannot estimate  $E_{max}$ from the value of $E_{sum}$. Consequently,  metrics that are  based on
total number of external connections lack the information as to what extent a vertex may be ``pulled'' by the neighboring communities which
can better estimated by $E_{max}$.  Using $E_{sum}$ can potentially result in frequent ties that need to be arbitrarily resolved by the community detection algorithms based on such metrics.

{\em When computing permanence, we use the maximum number of external connections, i.e., the maximum pull, to any one external community,
instead of combining all the external connections.}

\fi

\noindent{\bf Concept II:} {\em Within the substructure of a community, the internal neighbors of the vertex should be highly connected
among each other.}

Most optimization metrics only consider the internal connections of a vertex within its own community.  
However, how strongly a vertex is connected also depends on whether its internal neighbors are connected with each other.  To measure this connectedness of a vertex, we
compute the clustering coefficient of the
vertex with respect to its internal neighbors. For a vertex $v$ belonging to community $c$, it is measured by the ratio between
the actual number of edges among the neighbors (which also belong to $c$) of $v$ and the total number of possible edges among the neighbors
\cite{Holland1971}. The higher this internal clustering coefficient, the more tightly the vertex is connected to
its community (see Figure \ref{example}).

\if{0}
As an empirical study, we further obtain the internal clustering coefficient per vertex of the benchmark networks for their ground-truth
communities. Figure~\ref{examplefig}(d) shows a histogram of the internal clustering coefficient versus the number of vertices corresponding
to a specific range of internal clustering coefficient. As can be seen from the histogram, for most vertices the internal clustering
coefficients are generally towards the high range. However, for LFR ($\mu$=0.6) there is a reverse trend. In this network,
there are more vertices with lower internal clustering coefficient. This network by construction has a weaker community structure than the
other networks in the set, and thus quite a few of its vertices are loosely connected internally (see more in Section~\ref{comm}).

{\em To represent whether vertices are tightly connected within their communities, we include the internal clustering coefficient  as a factor in computing permanence.}

\fi

We combine these two criteria to formulate permanence of a vertex $v$, as follows: 
\begin{equation}\label{perm}
Perm(v)= \Big[  \frac{I(v)}{E_{max}(v)} \times  \frac{1}{D(v)}\Big] - \Big[1-c_{in}(v)\Big]
\end{equation}

\noindent where $I(v)$ is the number of internal (in its own community) neighbors of $v$,  $E_{max}(v)$ is the maximum number of connections of $v$ to
any one of the external communities, $D(v)$ is the degree of $v$ and $c_{in}(v)$ is the clustering coefficient among the internal neighbors
of $v$. Figure \ref{example} presents a toy example to calculate the permanence of a vertex. %For detailed explanation, the interested readers are encouraged to read our earlier paper \cite{Chakraborty_kdd}.

For vertices that do not have any external connections, $Perm(v)$ is considered to be equal to the internal clustering coefficient (i.e.,
$Perm(v) = c_{in}(v)$). If the number of internal connections, $I(v)$, is less than 2, we set the internal clustering coefficient, $c_{in}(v)$, to be 0. 
Therefore, for a vertex in a singleton
community, $Perm(v)=0$. 

The maximum value of $Perm(v)$ is 1 and is obtained when vertex $v$ is an internal node and part of a clique. The
lower bound of $Perm(v)$ is  close to -1. This is obtained when $I(v) \ll D(v)$, such that $\frac{I(v)}{D(v)E_{max}(v)} \approx 0$ and  $
c_{in}(v)=0$. Therefore for every vertex $v$, $-1 < Perm(v) \leq 1$. The permanence of a graph $G(V,E)$, where $V$ is the set of vertices
and $E$ is the set of edges, is given by $Perm(G)=\frac{1}{|V|}\sum_{v \in V}Perm(v)$. For a graph $G(V,E)$, the range is $-1 < Perm(G) \le
1$. $Perm(G)$ will be closer to 1 if majority of vertices have high permanence, that is the vertices are in well-defined communities. This can
happen only if the network inherently possesses a strong community structure. 

%If multiple community assignments are possible in the network, such as for a grid network, each individual vertex will have either high
% external pull or low internal clustering coefficient, and this will reduce the value of the overall permanence of the graph.
% The maximum value obtained is when $G$ consists of a series of disconnected cliques. If there is a vertex bridging between two cliques,
%then
% the highest overall permanence will be obtained if each clique acts as a separate community and bridging vertex forms a singleton
%community.
% For a grid, the best value of $Perm(G)$ will be zero, i.e., each vertex is assigned to a singleton community. 

\if{0}
% where $I(v)$ is the number of internal neighbors of $v$, $D(v)$ is the degree of $v$, $E_{max}(v)$ is the number of connections of $v$ to
that external community which has the maximum number of (external) neighbors 
% of $v$, and $c_{in}(v)$ is the clustering coefficient of $v$ with respect to its internal neighbors only. For a vertex in a singleton
community, $Perm(v)=0$. 
Figure~\ref{example} depicts a toy example for measuring permanence of two vertices $u$ and $v$. Note that, this formula actually
differentiates between the two cases in Figure~\ref{examplefig}(a) with higher permanence value for the case where the external pull is
uniform. Similarly, the formula differentiates between the two cases in Figure~\ref{examplefig}(c) by imposing more penalty on the network
that has a less tightly knit internal substructure.

\subsection* {Boundary conditions of permanence:}
For vertices that do not
have any external connections, $Perm(v)$ is considered to be equal to the internal clustering coefficient (i.e., $Perm(v) = c_{in}(v)$). The
maximum value of $Perm(v)$ is 1 and is obtained when vertex $v$ is an internal node and part of a clique. The lower bound of $Perm(v)$ is 
close
to -1. This is obtained when $I(v) \ll D(v)$, such that $\frac{I(v)}{D(v)E_{max}(v)} \approx 0$ and  $ c_{in}(v)=0$. Therefore for every
vertex $v$, $-1 < Perm(v) \leq 1$. The permanence of a graph $G(V,E)$, where $V$ is the set of vertices and $E$ is the set of edges,
is given by $Perm(G)=\frac{1}{|V|}\sum_{v \in V}Perm(v)$. For a graph $G(V,E)$, the range is $-1 < Perm(G) \le 1$. 

$Perm(G)$ will be closer to 1 as more vertices have high permanence, that is more vertices are in well-defined communities. This can happen
only if the network has a strong community structure. 
%If multiple community assignments are possible in the network, such as for a grid network, each individual vertex will have either high
external pull or low internal clustering coefficient, and this will reduce the value of the overall permanence of the graph.
The maximum value obtained is when $G$ consists of a series of disconnected cliques. If there is a vertex bridging between two cliques, then
the highest overall permanence will be obtained if each clique acts as a separate community and bridging vertex forms a singleton community.
For a grid, the best value of $Perm(G)$ will be zero, i.e., each vertex is assigned to a singleton community.

\fi

\if{0}
\subsection{Advantages of Permanence over Modularity } Permanence has two significant advantages over modularity -- (i) the value of
permanence is not as much affected by the symmetric growth of
the network size, and (ii) the value tends to fall sharply as the community structure deteriorates. Table~\ref{change} supports both these 
points. If we increase the size of the LFR network (see in Section \ref{dataset}) keeping $\mu$ (average ratio between the external
connections of a node to its degree) constant, the change in permanence is much less compared to the change in modularity (details in
Section \ref{permanence_property}, Table~\ref{asymptotic}). On the other hand, given a fixed network size,
if we worsen the quality of the community structure by increasing $\mu$, the value of permanence decreases significantly. The negative value
of permanence clearly indicates the less stringent community structure in LFR ($\mu$=0.6) network. Therefore, the value of permanence is a strong indicator of the quality of the community structure of a network, and  this value is very sensitive to changes
 in the community structure.    

\vspace{-4mm}
\begin{table}[!h]
\caption{Change in modularity and permanence with the increase of $\mu$ and $n$ in LFR network.}\label{change}
\begin{center}
\scalebox{0.8}{
  \begin{tabular}{|c|c|c|c|c|c|c|}
  \hline
  & \multicolumn{3}{c|}{Modularity} & \multicolumn{3}{c|}{Permanence} \\\hline
 \diaghead{\theadfont $NNN\mu$} {$\mu$}{n}      & 1000 & 3000 & 6000 & 1000 & 3000 & 6000\\\hline
 0.1   & 0.85 & 0.88 & 0.89 & 0.56 & 0.57 & 0.57 \\\hline
 0.3   & 0.66 & 0.68 & 0.69 & 0.26 & 0.27 & 0.27 \\\hline
 0.6   & 0.46 & 0.48 & 0.49 & -0.13 & -0.13 & -0.13 \\\hline
 \end{tabular}}
 \end{center}
\end{table}

\fi

\begin{figure}[!ht]
\centering
  \includegraphics[width=\columnwidth]{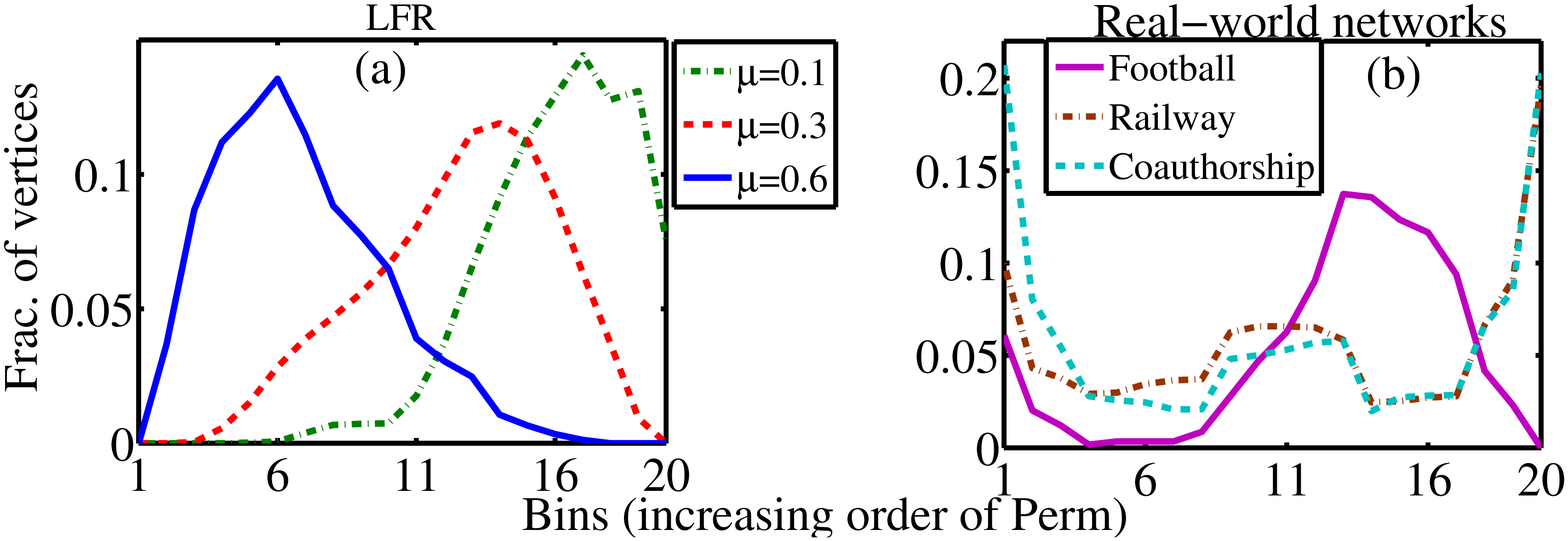}
 \caption{(Color online) Distribution of the values of Perm  for different networks. The value of Perm  of vertices is equally divided into
20 buckets indicated in x-axis (bin 1: $-1 \leq Perm <-0.9$, ..., bin 20: $0.9 \leq Perm \leq 1$).}\label{perm_dist}
 \includegraphics[width=\columnwidth]{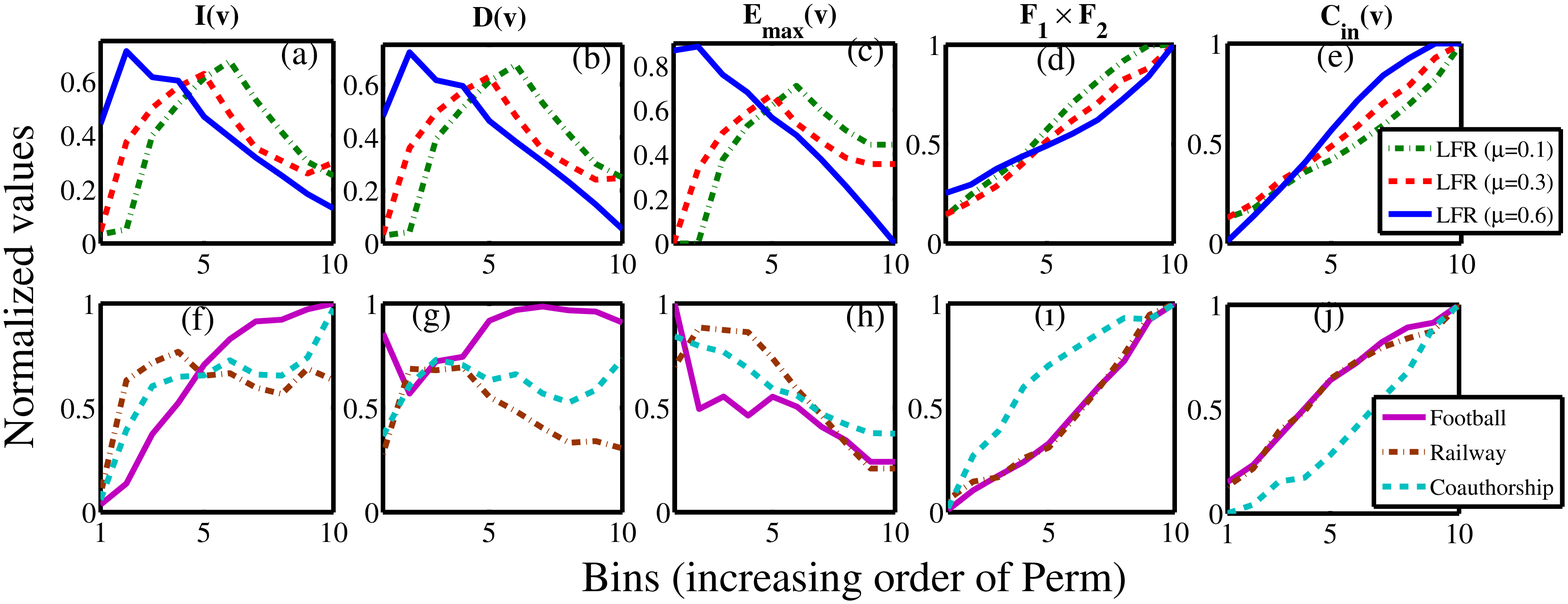}
 \caption{(Color online) The relation of average Perm  with (a),(f) $<I(v)>$, average internal neighbors per vertex, (b),(g) $<D(v)>$, average degree per vertex, (c),(h) $<E_{max}(v)>$, average maximum connections to an external community per vertex, (d),(i) average value of $F_1\times F_2=\frac{I(v)}{D(v)}\times \frac{1}{E_{max}(v)}$ per vertex, (e),(j) $<C_{in}(v)>$, average internal clustering coefficient per vertex for LFR (upper panel) and real-world (lower panel) networks.}\label{dist}
\end{figure}

\section{Effect of individual components on permanence}\label{comp}
In this section, we study the distribution of permanence values corresponding to the vertices in the graph based on their communities.  We first compute the permanence  of each vertex based on the ground-truth communities of the benchmark networks. We divide the permanence values ranging from $-1$ to $1$ into 20 bins where the low (high) numbered bins contain nodes with lower (higher) permanence. We plot the bins
on x-axis, and for each bin, on the y-axis, we plot the fraction of vertices whose permanence value falls in that bin. We observe in Figures \ref{perm_dist}(a) that this curve follows a Gaussian-like distribution, i.e., there are few vertices with very high or very low permanence values with a peak at the intermediate values.  The peak shifts from left to right with the decrease of $\mu$ value in the LFR network (keeping the other parameters of LFR
constant). The shift in the peak shows that as the communities get more well-defined with the decrease of $\mu$, most vertices move towards higher permanence. Figure \ref{perm_dist}(b) shows that Football network also follows similar kind of behavior, where most of the vertices fall in medium Perm  range. However, for Railway and Coauthorship networks, the curve follows ``U-shaped'' pattern, indicating maximum vertices falling in either very low or very high permanence buckets. %This indicates that these two networks have fuzzier communities with many vertices being in the boundary of multiple communities.

This phenomenon indicates that the community structure is not very clear in these networks. Recall that we have computed
communities from ground-truths. The high proportion of lower values indicates that the networks contain entities that are not easy to
classify, such as railway stations that are at the border of the states or  authors who publish in multiple fields.

To understand the dependence on each component of the permanence equation (Equation \ref{perm}), we further plot permanence with respect to each individual component, i.e., $I(v)$, $D(v)$, $E_{max}(v)$, $C_{in}(v)$ and their combination in Figure \ref{dist}. Figure \ref{dist}(a) shows a decreasing trend of $I(v)$ with the increase of permanence for LFR, where the pattern is completely opposite for real-world networks as shown in Figure \ref{dist}(f).  The trend is almost similar for the relation between $D(v)$ and permanence in Figures \ref{dist}(b) and \ref{dist}(g). However, here most of the real-world networks except Football show similar pattern with that of LFR, where high degree nodes tend to exhibit low or medium permanence value. Further, we plot the relation between permanence and $E_{max}(v)$ in Figures \ref{dist}(c) and \ref{dist}(h) and observe that while all the real-world networks show an inverse relation,
 for two LFR networks ($\mu=0.3$ and $\mu=0.6$) it initially increases and then starts decreasing. From these observations, one can not find any universal relation as such among different networks. However, once we combine these factors together and plot the dependence between permanence and $\frac{I(v)}{D(v)}\times \frac{1}{E_{max}(v)}$ in Figures \ref{dist}(d) and \ref{dist}(i), we observe a consistent behavior for all the networks in that the value of the combination tends to increase almost linearly with permanence. A similar trend is followed in Figures \ref{dist}(e) and \ref{dist}(j) where the value of internal clustering coefficient tends to increase with the increase of permanence. These results show that permanence depends on two factors -- (i) the combined effect of  $\frac{I(v)}{D(v)}\times \frac{1}{E_{max}(v)}$, and (ii) the value  of $C_{in}(v)$.

\section{Effect of perturbations on ground-truth communities}\label{parturbation}
One of the crucial measures for an effective community scoring metric is how it behaves under different
perturbations of the ground-truth community structure \cite{Yang:2012}. The metric should be robust to small perturbations of 
the ground-truth communities, such as when groupings of nodes that differ very slightly from the original ground-truth grouping.
Furthermore, the metric should 
also be sensitive to large perturbations. If the change is so large that the ground-truth structure dissolves to a random set of nodes, then
the value of the scoring function should be low.
In this section, we compare the change in value of permanence with three other community scoring metrics and demonstrate that among them permanence is both robust to noise as well as sensitive to large changes in the network.

\subsection{Community scoring metrics}\label{subsec:eval} We consider the following community scoring metrics:

\begin{itemize}
 \item {\bf Modularity (Mod):} Modularity \cite{Newman:2006} is  defined by the fraction of the edges that fall within the given groups minus the expected such fraction if edges were
distributed at random. Formally, for a given graph $G(V,E)$, it is quantified as follows:
 \begin{equation}
  Q=\frac{1}{2m}\sum_{u,v \in V} [A_{uv} - \frac{k_uk_v}{2m}]\delta(c_u,c_v) 
 \end{equation}
 where $m=|E|$, $A_{uv}$ is the $(u,v)$ entry of the adjacency matrix $A$, $k_u$ is the degree of vertex $u$, $c_u$ is the community of
vertex $u$, and $\delta(c_u,c_v)=1$ if $u$ and $v$ are in the same community or $0$ otherwise.

 \item {\bf Conductance (Con):} Conductance \cite{cond_09} is the ratio between the number of edges inside the cluster and the number of edges leaving the cluster \cite{Kannan:2000,Shi:2000}. More
formally, conductance $\Phi(S)$ of a set of nodes $S$ is defined as follow:
 \begin{equation}
  \Phi(S)=\frac{C_S}{min(Vol(S),Vol(V \setminus S))}
 \end{equation}
 where $C_S$ denotes the size of the edge boundary, $C_S=|{(u,v)}: u \in S, v \notin S|$, and $Vol(S)=\sum_{u\in S} d_u$ where $d_u$ is the
degree of vertex $u$.

 \item {\bf Cut-ratio (Cut):} Cut-ratio is a standard metric in graph clustering \cite{Fortunato201075,Leskovec:2010}, which is defined as
the fraction of all possible edges leaving the cluster $S$. Formally, given an undirected graph $G(u,v)$, the cut-ratio $\theta(S)$ of a set
of nodes $S$ is defined as follow:
 \begin{equation}
  \theta(S)=\frac{C_S}{n_S(n-n_S)}
 \end{equation}
 where $C_S$ is defined earlier, and $n_S=|S|$. 
 \end{itemize}
 
 Note that the higher the value of modularity, the better the quality of the community structure; however for conductance and
cut-ratio, the opposite argument is applicable. Therefore, to make these two measures comparable to modularity and permanence, we measure
(1-Con) and (1-Cut) for conductance and cut-ratio respectively.

\subsection{Perturbation strategies}
Given a graph $G=<V,E>$  and  \emph{perturbation intensity}  $p$, we  restructure the ground-truth community by applying a perturbation
strategy.  We experiment with the three perturbation strategies as proposed in \cite{Yang:2012}. We designate a given ground-truth
community as $S$ and the rest of the network as $S'$.

\begin{enumerate}

\item {\bf Edge-based perturbation:} We select an inter-community edge $(u,v)$ where $u \in S$ and $v \in S'$ (where $S \neq S'$) and assign
$u$ to $S'$ and $v$ to $S$. We continue this process for $p \cdot |E|$ iterations. This strategy preserves the size of $S$, but certain
vertices within the ground-truth community may become disconnected. 

\item {\bf Random perturbation:}  We pick two random nodes $u \in S$ and $v \in S' $ (where $S \neq S'$) that may not be connected by an edge and then swap their memberships. We continue this process for $p \cdot |V|$ iterations.  Random perturbation maintains the size of $S$, but the community may have disconnected vertices.
%Random perturbations degrade the quality of $S$ faster than edge-based strategy, since edge-based strategy only affects the boundaries of the
%community.

\item {\bf Community-based perturbation:} This perturbation is similar to the edge-based strategy. However, each community $S$ is perturbed one by one for
 $p \cdot |S|$, until the nodes of the community are swapped with nodes outside the community. This process is repeated for
all the communities separately. 
%This perturbation decreases the quality of the ground-truth communities the fastest  among the three since the number of swaps is much higher than the others.
\end{enumerate}

We perturb networks using these perturbation strategies for values of $p$ ranging between 0.01 to 0.5.   We compute how the perturbations,
as given by the value of $p$, affect the values of
 modularity, permanence, 1-Con and 1-Cut. For small values of $p$, small
change of the scoring function is
desirable. This indicates that the scoring function is robust to noise. For high perturbation, that is larger values of $p$,  the communities become more random. Therefore, the values should drop significantly. \\

% This behavior would indicate
% that the community scoring function is sensitive to the change in ground-truth communities.

%
%For each of the community scoring functions namely modularity, permanence, conductance and cut-ratio, we
%measure the values for perturbation intensity $p$ ranging between 0.01 and 0.5. For small $p$, small degradation of the original value is
%desirable since it indicates that the scoring function is robust to noise. For high perturbation intensities $p$, high degradation is
%preferred because this suggests that the community scoring function is sensitive, i.e., as the community becomes more ``random'' we want the
%scoring function to significantly drop. LFR ($\mu$=0.1 and 0.3), the curves for conductance and modularity are nearly aligned which indicates their similar response to perturbation. Moreover, in football network, the cut-ratio seems to be even more sensitive than modularity. However, for

\begin{figure*}
\centering
\includegraphics[width=\columnwidth]{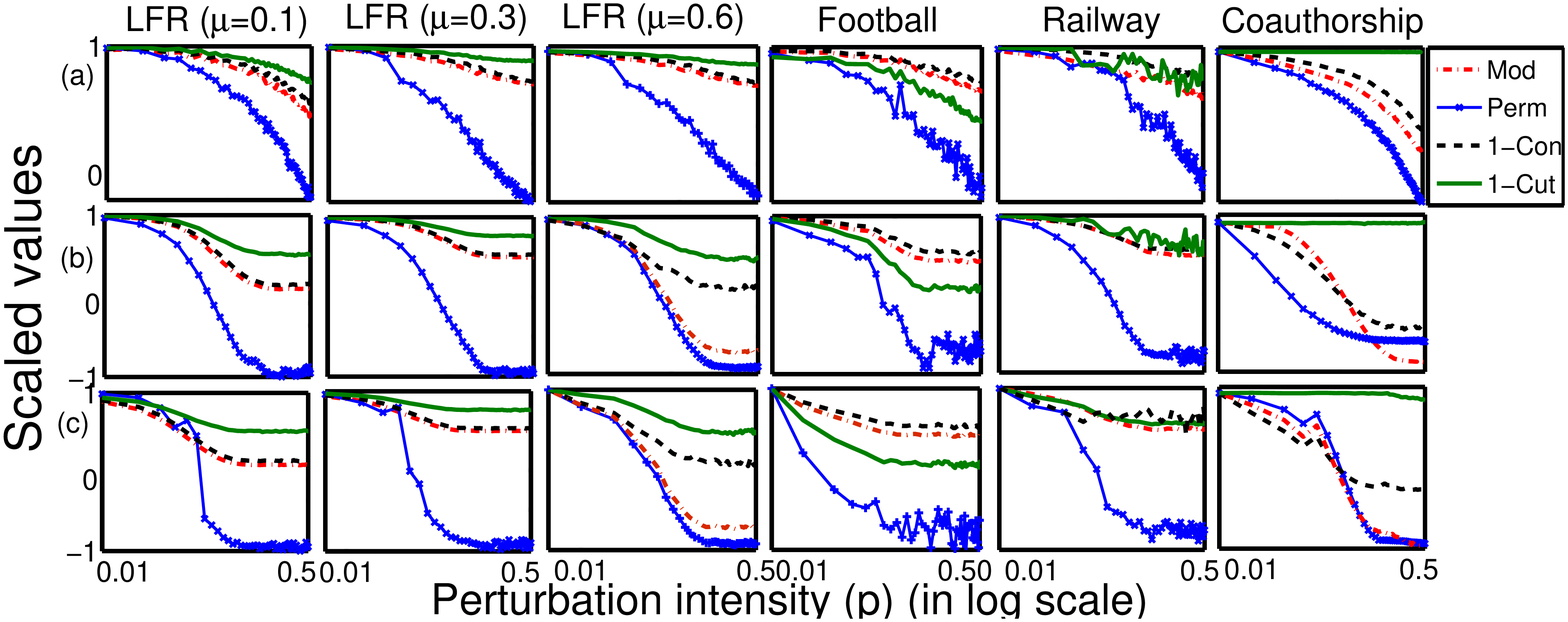}
\caption{(Color online) Change in the value of the scoring functions with the increase of perturbation intensity $p$ in (a) edge-based,
(b) random and (c) community-based perturbation strategies. The values are normalized by the maximum value obtained from each function. }
\label{fig:perturbation}
\end{figure*}

\subsection{Experimental results} Figure~\ref{fig:perturbation} shows the  results of our experiments. To compare equitably across the different scoring functions,  we
scale the values of each parameter by normalizing  with the maximum value obtained from that function. For all three strategies,
 the scoring function values  decrease with the increase of $p$. Of the three methods Community-based produces the fastest degradation, followed by Random perturbation and Edge-based perturbation is the slowest. However, once $p$ has reached a certain threshold, the decrease is much faster in permanence, while other scoring functions do not always show this sensitivity to large perturbation. 
% On more careful inspection, we find that this happens because the internal structure of a community completely breaks down if 
% perturbation is taken beyond a point and thus has an avalanche effect on the value of the clustering coefficient ($c_{in}(v)$ in equation
%(\ref{perm})). This in turn quickly pulls the value of permanence down. 
% Summarizing, the results  indicate that permanence is a better measure for distinguishing true communities from randomized sets of nodes
%than the other parameters. 

\begin{figure}[!ht]
\centering
\includegraphics[scale=0.3]{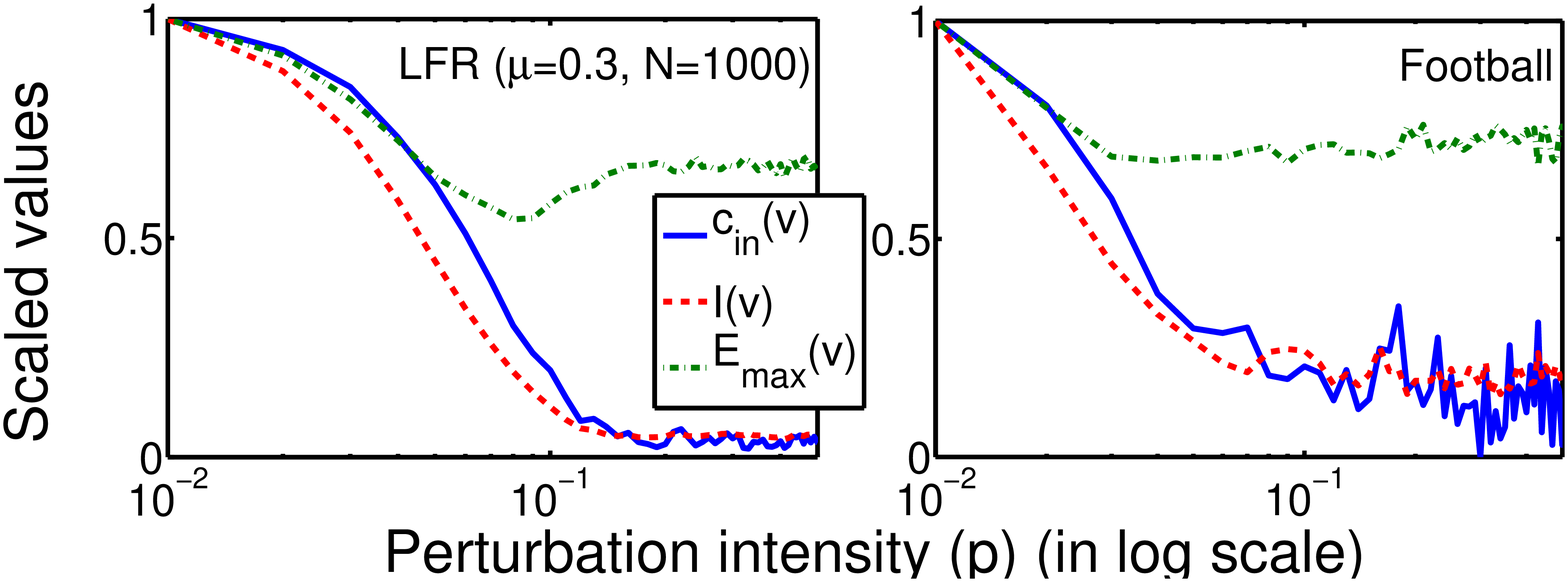}
\caption{(Color online) Change in the average values of internal degree $I(v)$, maximum external connections $E_{max}(v)$ and internal clustering-coefficient $c_{in}(v)$ of vertices of two representative networks with the increase of perturbation intensity in random perturbation strategy. }
\label{fig:perturbation_comp}
\end{figure}

In order to further observe how perturbation affects each of the three major components of the permanence metric, namely the internal degree
$I(v)$, the maximum external connections $E_{max}(v)$ and the internal clustering-coefficient $c_{in}(v)$, we further measure the change of
their individual values as a function of $p$. Figure~\ref{fig:perturbation_comp} shows the rate of these changes for random perturbation.
The most sensitive components of permanence are the internal degree and the average internal clustering-coefficient of vertices. These
values tend to be comparatively stable for small perturbations, but degrades significantly as $p$ increases. This provides
another justification for
incorporating the internal clustering-coefficient as a penalty factor in the formulation of permanence.  

%In random and community-based
%strategies, we observe a sudden fall in permanence value after a certain number of perturbations which indicates the degree of
%sensitivity of permanence for the corresponding networks.
\if{0}
\fi

\section{Implications of permanence}\label{sec:significance}
We have shown in \cite{Chakraborty_kdd},  using a rank correlation approach, that permanence is a better quality
scoring metric as compared to modularity, conductance and cut-ratio. In Section \ref{parturbation}, we demonstrated how
permanence is robust to small perturbations and sensitive to larger ones.
In this section, we analyze the characteristics of permanence from different perspectives of community structure, based on their known
ground-truth structure. 

%Since the ground-truth community of each network is known to us, a set of following experiments are conducted on the ground-truth structure to exploit the notion of permanence in real graphs.

\subsection{Measuring persistence of a vertex in its community}
We observe, using the metadata of the co-authorship networks, that the permanence of a vertex is proportional to its persistence, i.e., how
long a vertex remains in an evolving community.
In the original publication dataset \cite{asonam} as mentioned in Section \ref{dataset}, each scientific article is categorized into one of
the 24 research fields (such as Algorithms, Programming Languages, AI etc.). We tag each author by the field in which she has published
maximum papers. Each field corresponds to a community \cite{asonam}. Essentially, we intend to measure the persistence of an author in her
own community in terms of her {\em research age}. For this, we define {\em research age} ($\xi$) of an author in a field/community in two
different ways as follows.\\

\noindent{\bf Definition 1: Collective research age ($\xi_c$):} The collective research age $\xi_c^f(a)$ of an author $a$ in a field/community $f$ is defined by the total number of distinct years author $a$ has published at least one paper in the field $f$.

\noindent{\bf Definition 2: Discounted research age ($\xi_d$):} The discounted research age $\xi_d^f(a)$ of an author $a$ in a
field/community $f$ is defined by the total number of distinct years author $a$ has published at least one paper in the field $f$, where
each year is linearly penalized by the number of its immediate consecutive preceding years when she has not published a single paper on $f$.
The linear penalty is introduced to bring in the effect of ``consistency break'' in the publication career of an author. Ideally,  an author
who is publishing consistently in  a field should be more persistent than an author publishing in stretches with intermediate gaps. The
penalty is used to significantly put more emphasis on the former case than the latter case.

\begin{figure}[!h]
\centering
 \includegraphics[width=\columnwidth]{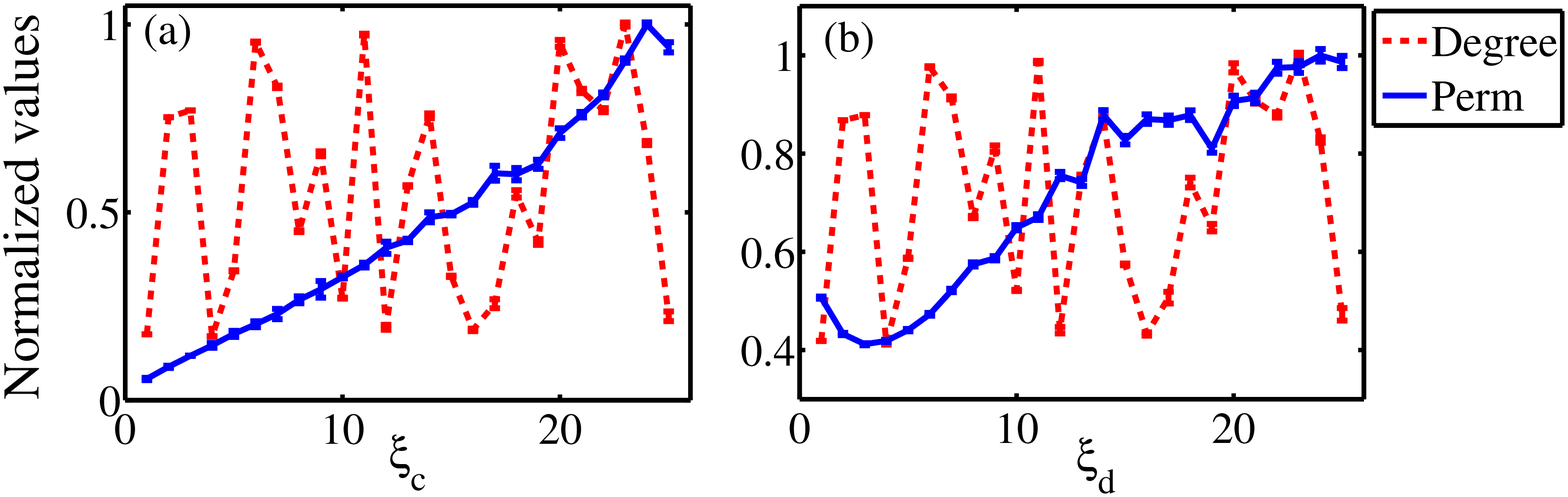}
 \caption{Changes of average degree and average permanence (with variance) of authors with the increase in (a) collective research age $\xi_c$, and (b) discounted research age $\xi_d$.  }\label{age}
\end{figure}

For example, let us assume that an author $a$ has published papers in the following years: 1960, 1965, 1966, 1967, 1970. Therefore,  $\xi_c^f(a)=5$ and $\xi_d^f(a)=(1+1/4+1+1+1/2)=3.75$ (year 1965 is penalized by its previous 4 consecutive unproductive years, similarly for 1970). We then plot the average degree and the average permanence of authors against two types of research ages in Figure \ref{age}. We observe that while there is almost no correlation between the average degree and the research age of an author, the permanence value of an author is almost linearly correlated with the research age. This evidence essentially leads to the 
the following conclusions: (i) permanence of a vertex is a suitable way of representing its persistence  in its own community, which can not be derived from the degree of a vertex,  (ii) since the result in Figure \ref{age} is reported over all the authors in different communities, we can also compare the extent of persistence of two vertices belonging to two different communities, 
i.e., same permanence value 
of two vertices in two different communities indicates the equal extent of persistence in their corresponding communities.

\subsection{Strengthening the community structure}
The value of permanence of a vertex signifies its propensity to remain in its own community. Therefore, 
vertices having low permanence in a community are loosely connected to the community. We explore whether we can strengthen the community
structure by deleting vertices with low permanence. Note that when a vertex is deleted from its community, it would also affect the
permanence value of the remaining vertices. Therefore, we rank\footnote{We use dense ranking scheme to rank the authors.} the vertices at
the very beginning based on permanence and do not consider further changes in permanence during the deletion. Then in each step, we measure
the quality of the cluster by {\em edge-density} (ratio between the actual number of edges and the expected number of edges in that
cluster). 

\begin{figure}[!h]
\centering
 \includegraphics[width=\columnwidth]{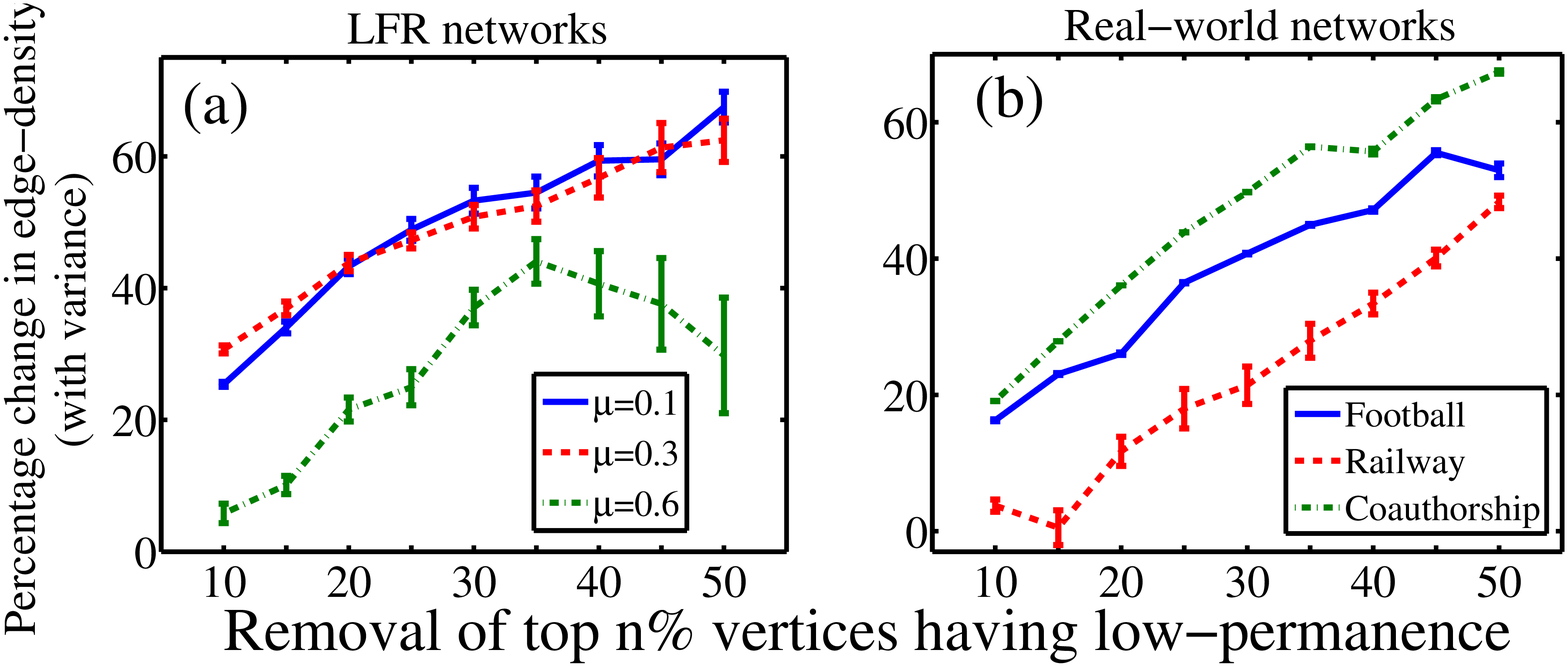}
 \caption{(Color online) Percentage change in edge-density after removing top $n\%$ (varies from 10\% to 50\%) low-ranked vertices based on
permanence. Each point in this plot is averaged over all the communities, and therefore, the variance is also plotted.}\label{edge_density}
\end{figure}

For each community, we remove top $n\%$ low ranked vertices based on permanence and measure the percentage change of edge-density (averaged over all the communities) due to this removal.  One can observe in Figure \ref{edge_density} that the edge-density increases with the increase in $n$. Although  
overall there is increase in edge density, we notice that in LFR ($\mu=0.6$), the edge-density starts decreasing after removing 35\% of
vertices. The reason might be described as follows. In LFR ($\mu=0.6$) vertices generally exhibit small permanence value due to
the overall inferior community structure. The range of permanence values of vertices in each community of LFR ($\mu=0.6$) is also not high
as compared to the same for LFR ($\mu=0.1$). Therefore, removing 35\% of vertices from LFR ($\mu=0.6$) might result in the removal of
vertices having relatively high ranking based on permanence. On the other hand, since the range is high for  LFR ($\mu=0.1$) and LFR
($\mu=0.3$), the same extent of deletion of vertices might not affect the high-ranked vertices in the community. However for these networks,
such decrease in edge-density can also be observed for higher extent of deletion (usually beyond 50\%, not shown in Figure
\ref{edge_density}).

\subsection{Heterogeneity and core-periphery organization of community structure}
Although it is implicitly assumed that all the constituent members of a community belong to the community equally, this is not true in
reality. Within a community, the extent of involvement and activity may not be same for all members -- permanence can capture this {\em
heterogeneity}. The permanence of a node $v$ belonging to a community $c$ indicates the extent to which the node belongs to the community.
With this value several inferences can be drawn about the communities present. For instance, it inherently creates a gradation/ranking of
the constituent vertices in a community. This ranking may be important in many cases -- for example in exploring the core-periphery
structure of a community. 

\begin{figure}[!h]
\centering
 \includegraphics[width=\columnwidth]{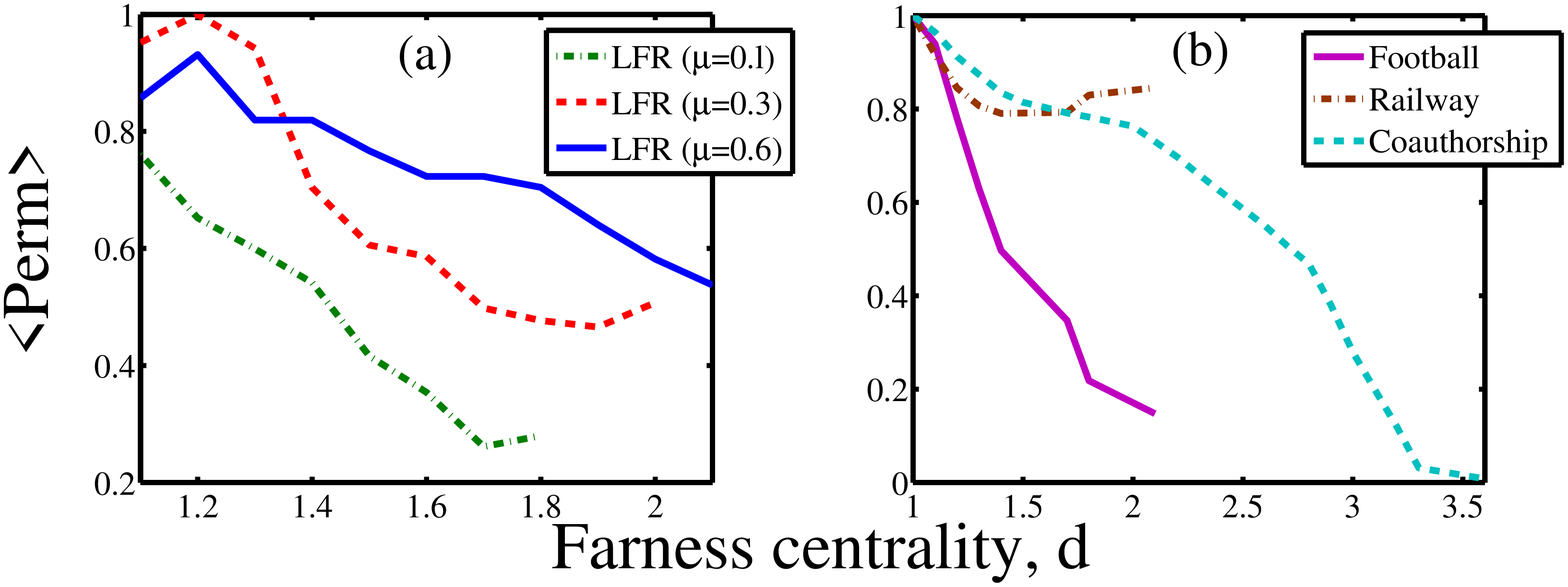}
 \caption{(Color online) Community-wise average permanence, $<Perm>$ of vertices as a function of farness centrality $d$ for LFR and real-world networks.}\label{fig:farness}
\end{figure}

To explore the relation of permanence of a vertex with its position vis-a-vis core of a community, we use {\em farness centrality} ($d$) proposed in \cite{Yang14} as a measure to locate the position of a vertex within a community. In order to measure farness centrality for each community, we construct the induced subgraph constituting all the nodes in the community and measure average shortest path for each vertex within this subgraph. Thus, the lower the value of $d$ for a vertex, the closer the vertex is to the core part of the community\footnote{Farness centrality is just the reverse of closeness centrality in a connected component.}.  We plot average permanence of vertices as a function of farness centrality in Figure \ref{fig:farness}. We observe that for both LFR and
real-world networks, average permanence decreases with the distance from the center of the community. Therefore,  the value of permanence can act as a strong indicator of the position of the vertex in the  
community.

The next investigation reveals the manner in which 
the permanence value of vertices decreases from the core. A smooth decrease in value  would indicate 
that the nodes in a community are arranged in layers with each layer of vertices roughly having similar permanence. 
In order to understand the mixing pattern of vertices 
we measure permanence-based assortativity ($r$)\footnote{Assortativity ($r$) lies between -1 and 1. When $r$ = 1, the network is said to have perfect
assortative patterns, when $r$ = 0 the network is non-assortative, while at $r$ = -1 the network is completely disassortative.}
 \cite{Newman-assort-2003} to observe the  preference for a network's nodes in a community
$c$ to attach to other nodes that have nearly similar permanence. We divide the permanence values into 20 bins so that nodes within a bin are considered to have equivalent permanence values, and then measure $r$ of a network. For comparison, we also measure degree-based assortativity of vertices in each network.
We observe in Table \ref{assor} that  both synthetic and real-world networks are highly assortative in terms of permanence, rather than in terms of degree. This result indeed indicates that in general, a community is organized into several layers, 
where each layer is composed of vertices exhibiting similar permanence, and vertices  tend to be highly connected {\em within each layer} than across different layers.   

\begin{table}[h!]
\caption{Average of the assortativity scores, $<r>$ (degree-based and $Perm$-based) of the communities per network.}\label{assor}
\centering
\scalebox{1}{
 \begin{tabular}{|c|c|c|c|}
\hline
$<r>$ & LFR ($\mu=0.1$) & LFR ($\mu=0.3$) & LFR ($\mu=0.6$) \\\hline
Degree-based  &   0.088              &      0.108           &      0.082           \\\hline
$Perm$ based &    0.520         &    0.551             &   0.430 \\\hline
 \end{tabular}}

\scalebox{1}{
 \begin{tabular}{|c|c|c|c|}
\hline
$<r>$         & Football & Railway & Coauthorship \\\hline
Degree-based  &    -0.105         &  0.153      &  0.155     \\\hline
$Perm$ based &  0.747        &  0.531     &   0.489 \\\hline
 \end{tabular}}

\end{table}

\subsection{Initiator selection for message spreading}
Message spreading is one of the challenging problems in complex networks and
distributed systems \cite{ChierichettiLP10}. Starting with one source node/initiator having a message, the protocol proceeds in a sequence of
synchronous rounds with the goal of delivering it to every node in the network. At every time step, each node in the system having the  message communicates with one node (not having the message)  in its neighborhood and transfers the message. The algorithm terminates when all the nodes in the system have received the message. A fundamental issue in message spreading is the selection of initiators.  Selecting initiators based on the degree leads to faster spreading (requires less steps in average) of message than the random node selection \cite{Demers:1987}. Since vertices with higher permanence form the core of the community, we posit that initiator selection based on  permanence would  help in 
faster dissemination of the message.  Note the  message spreading algorithms are based on only the local view of the vertices,
therefore global methods such as those described in~\cite{Kempe:2003} will not be applicable under this formulation.

\begin{figure}[!h]
\centering
\includegraphics[width=0.8\columnwidth]{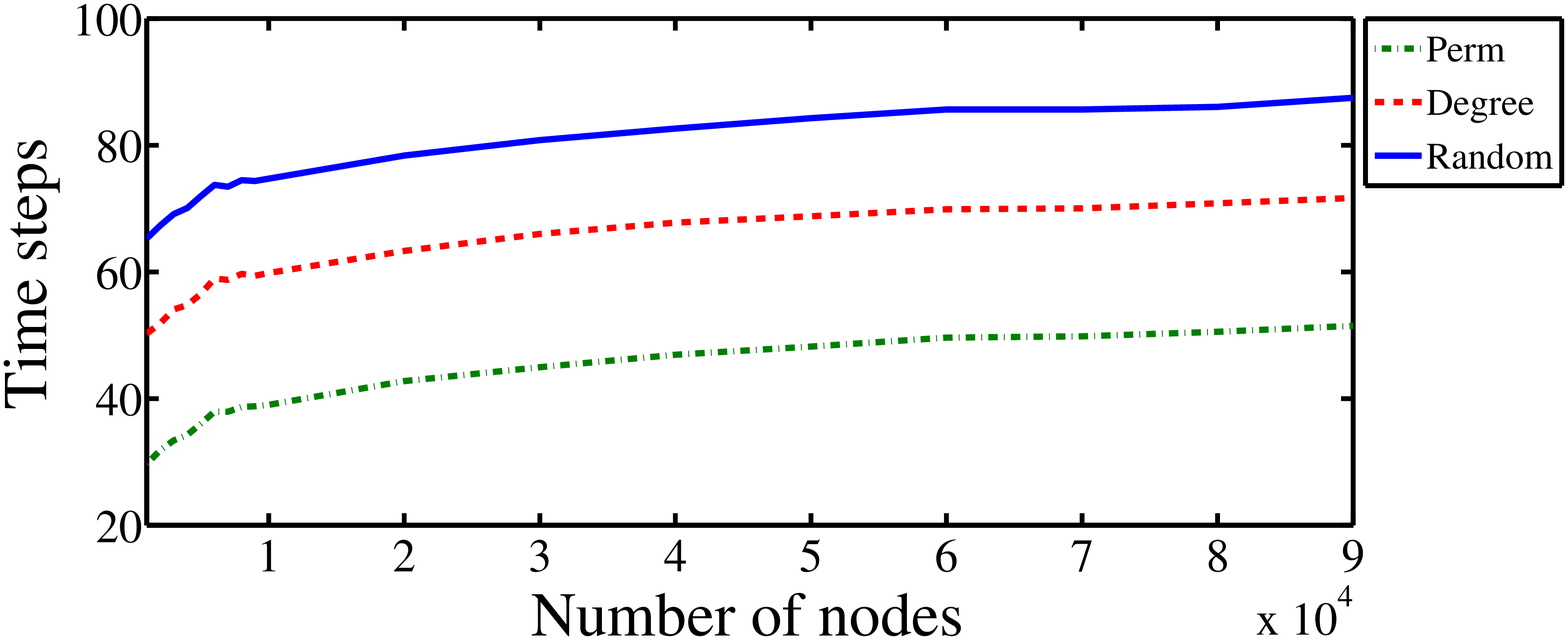}
\caption{(Color online) Number of time steps required to broadcast a message in LFR network by varying the number of nodes.}\label{spreading}
\label{farness}
\end{figure}

To validate this hypothesis, we consider the LFR network and vary the number of nodes from 10,000 to 90,000, keeping the other parameters constant (see Section \ref{dataset}). We  
select multiple initiators by picking one node per community present in the ground-truth structure based on the following criteria separately: (i) random, (ii) highest degree, (iii) highest permanence. For each network configuration, we run 500 simulations and report in Figure \ref{spreading} the average number of time steps required for the message to reach all the nodes in the network. We observe that the permanence based initiator selection from ground-truth communities requires minimum time steps to spread the message compared to the degree-based selection. 
%These results thus highlight the importance of Perm based ranking within a community.

\if{0}
\subsection{Dependence on three components}
In equation \ref{perm}, we notice that  permanence $Perm(v)$ of a vertex $v$ is dependent on the four basic components: (i) $d(v)$, degree
of $v$, (ii) $I(v)$, internal degree of $v$ in its own community, (iii) $E_{max}(v)$, maximum connection to any one of the external
communities and (iv) $c_{in}$, clustering coefficient among the internal neighbors of $v$. Here we exhaustively analyze the dependence of
$Perm(v)$ on each of these ingredients on different networks. We observe that the internal degree of a vertex in its community is almost
proportional to its total degree, and therefore, permanence behaves equally with internal degree and total degree.
A general observation on analyzing the ground-truth communities of both synthetic and real-world networks is that nodes having moderate internal degree (and degree) and high internal clustering coefficient exhibit maximum permanence when they experience almost zero pull from external communities. In the rest of this subsection, we discuss more on these dependencies individually for different networks. 

\subsubsection{Dependence on $I(v)$ and $c_{in}(v)$} In Figure \ref{scatter}(a), we observe that for most of the networks, the value of permanence decreases with the increase of $I(v)$. The reason is that with the increase of the number of internal neighbors, the chance of getting higher internal clustering coefficient tends to decrease, resulting in low permanence. The range of $I(v)$ also tends to get reduced with the increase in $\mu$ in LFR networks; however the average value of $c_{in}(v)$ decreases as well. In football network, most the the vertices exhibit high $I(v)$ and $c_{in}(v)$, resulting hight permanence and prominent community structure. However, in football and coauthorship networks, $I(v)$ is respectively low which in turn results higher $c_{in}$. However, we shall see later in Figure \ref{scatter}(b) that for these two networks, $E_{max}(v)$ tends to be higher as well.

\subsubsection{Dependence on $I(v)$ and $E_{max}(v)$} Figure \ref{scatter}(b) shows how permanence varies with the changes in $I(v)$ and $E_{max}(v)$. Here, we notice that although the range of $E_{max}$ expands with the increase in $\mu$, resulting in the decrease of permanence. On the other hand, most of the points in football network are clubbed in high $I(v)$ and medium $E_{max}(v)$ zone; while in railway network the cluster of points almost covers  the entire diagonal region ($I(v):$ 20-40, $E_{max}(v):$ 10-20). Surprisingly , in coauthorship network the value of permanence tends to increase almost linearly with the increase in $I(v)$ and $E_{max}$.

\subsubsection{Dependence on $c_{in}(v)$ and $E_{max}(v)$} Figure \ref{scatter}(c) depicts the dependence of permanence on $c_{in}(v)$ and $E_{max}(v)$. In general, the pattern is almost similar for all the networks that for a certain value of $E_{max}(v)$, higher $c_{in}(v)$ tends to increase the value of permanence. In football and railway networks, most of the points lie in low $E_{max}(v)$ and high $c_{in}(v)$ range, where the pattern for coauthorship network is quite similar to the corresponding figure in Figure \ref{scatter}(a) in that most of the vertices tend to have low $E_{max}(v)$, but cover entire range of $c_{in}(v)$.

\begin{figure}[!t]
\centering
 \includegraphics[width=\columnwidth]{./Fig/scatter.eps}
 \includegraphics[width=\columnwidth]{./Fig/scatter1.eps}
 \caption{Changes in permanence with three combinations of its ingredients: (a) $I(v)$ and $c_{in}(v)$, (b) $I(v)$ and $E_{max}(v)$, and
(c) $E_{max}(v)$ and $c_{in}(v)$ for different networks.}\label{scatter}
\end{figure}
\fi

\section{Community detection using permanence maximization}\label{comm}
We now present an algorithm for detecting communities by maximizing the permanence of the network. Our algorithm, {\bf MaxPerm} (pseudocode in Algorithm~\ref{algo}), finds high permanence partitions of large networks using a greedy agglomerative approach, similar to the methods used in \cite{blondel2008,Clauset2004}.

Initially the vertices are assigned to random connected subgraphs as their initial seed community. At each iteration, a vertex is moved from one community to another if its permanence increases. This process is continued for several iterations  until the value of the permanence of the network is unchanged. Although convergence to a fixed value is not theoretically guaranteed, we have observed that on test cases the algorithm converges within ten iterations. As in the case of
modularity-maximization methods, 
we observe that creation of appropriate seed communities can help improve the quality of community
detection. We shall discuss this issue later in Section \ref{sec:seed}.

%The accuracy of the results depend on the 
%
%Similar to the greedy approach of modularity maximization algorithms,  
%
%our algorithm  strives to obtain a high  value of
%permanence. In this algorithm, we begin with initializing every 
%vertex to a singleton community. 
%A vertex is moved to a community only if this movement results in a net increase in the number of internal connections and/or a net decrease
%in the number of external connections to any of the neighboring communities. If such a move is not possible, then either the vertex remains
%as a singleton 
%(such as in the case where moving to any one of the candidate communities will give equal permanence) or moves to the community where it is
%more tightly connected with its neighbors (this causes the vertex to have positive permanence). This process is repeated for each vertex
%and the entire relocation of all vertices is repeated over several iterations until the permanence value converges. However, the convergence
%is
%not theoretically guaranteed, thus is worth studying since it could give good heuristics to enhance the computation time. Preliminary
%results on several test cases seem to indicate that  the algorithm converges with high probability.

 \begin{algorithm}[!t]
 \KwData{A graph $G(V,E)$}
 \KwResult{Permanence of $G$; Detected communities}
  Each vertex is assigned to its seed community\;
  Set value of maximum iteration as $maxIt$ \;
  $vertices \gets |V|$\;
  $Sum \gets  0 $\;
  $Old\_Sum \gets  -1 $\;
  $Itern \gets 0 $\;
 \While{$Sum \neq Old\_Sum$ and $Itern < maxIt$}{
   $Itern \gets Itern+1$\;
   $Old\_Sum \gets  Sum $\;
   $Sum \gets  0 $\;
   \ForAll{$v \in V $}{
       \tcp{\color{blue}{ Compute current permanence of $v$}}
       $cur\_p \gets Perm(v)$\;
       \If{$cur\_p ==1$}{
           $Sum \gets  Sum+cur\_p $\;
           {\bf continue};\;
       }
       $cur\_p\_neig \gets 0$\;
       $Neig(v)$=set of neighbors of $v$\;
       \ForAll{$u \in Neig(v)$}{
            \tcp{\color{blue}{ Compute current permanence of $u$}}
            $cur\_p\_neig \gets cur\_p\_neig + Perm(u)$\; 
       }
       
     \tcp{ \color{blue}{$Comm(v)$ is the set of neighboring communities of $v$}}
      \ForAll {$C \in Comm(v)$}{
            Move $v$ to community $C$\;
            \tcp{ \color{blue}{Compute permanence of $v$ in community $C$}}
            $n\_p \gets Perm(v)$ \;
            \tcp{\color{blue}{Neighbors of $v$ are affected for this movement}}
            $n\_p\_neig \gets 0$\;
            \ForAll{$u \in Neig(v)$}{
                \tcp{\color{blue}{ Compute new permanence of $u$}}
                  $n\_p\_neig \gets n\_p\_neig + Perm(u)$\; 
            }
            \eIf {($cur\_p < n\_p)$ and ($cur\_p\_neig< n\_p\_neig)$}{
                $cur\_p \gets n\_p$\;
            }{
                Replace $v$ to its original community\;
            }
      }
      $Sum \gets  Sum+cur\_p $\;
  }
}
$Netw\_perm=Sum/vertices$ \tcp*{\color{blue}{Permanence of $G$}}
{\bf Return} {$Netw\_perm$}\;
       
 \caption{MaxPerm: Community Detection using {\bf Max}imizing {\bf Perm}anence}\label{algo}
\end{algorithm}

\subsection{Computational complexity and strategies for improvement} The computational complexity of the algorithm is as follows. The most
expensive part in computing the permanence of a vertex $v$ is the internal clustering co-efficient, $c_{in}(v)$. Given the degree of vertex
$v$ is $D(v)$, computing $c_{in}(v)$ takes time $O(D(v)^2)$. For each vertex we compute the permanence for its own and each of its
neighboring communities. Let the number of neighboring communities of vertex $v$ at iteration $k$ be $C_k(v)$.  Let the total number of
iterations required by the algorithm be $maxIt$. Therefore, the time to execute ${MaxPerm}$ is: 
$\sum_{k=1}^{k=maxIt} \sum_{v=1}^{v=|V|} (C_k(v) O(D(v)^2))$.  

Let $d_{max}$ be the maximum degree of the network. We also note that the maximum number of communities that a node can belong to is $d_{max}+1$.  The upper bound for ${MaxPerm}$ is $O(maxIt \cdot |V| \cdot d_{max}^3)$. Since only a few nodes of the network has
the highest degree, in practice the time is much lower than the value given by this upper bound.

The execution time can be further reduced by a few simple strategies. First, instead of recomputing $Perm(v)$, for each community, we can
store the number of edges each vertex has in each of its neighboring communities. We update these values only when a vertex or its neighbor
changes communities. Second, since we want the permanence to increase if a vertex changes communities, the only communities to consider for
relocation are those with both high internal degree and high clustering coefficient. By computing permanence for only communities that
satisfy these criteria we can reduce the computation time. Both these strategies require us to keep track of  the communities for neighbors
of the vertex, for each vertex. Together, this requires extra storage of order $O(|V| D(v)) \approx O(E)$.

%%NG - Till here
\subsection{Baseline community detection algorithms}\label{algorithms}
There exist numerous community detection algorithms, which differ in the way they define the community structure. Here we select the
following  set of algorithms and categorize them according to the principle they use to identify communities as per~\cite{Labatut}.

\begin{enumerate}
 \item  {\bf Modularity-based approaches:} We select three modularity optimization algorithms, namely {\bf FastGreedy}
approach~\cite{newman03fast}, {\bf Louvain}~\cite{blondel2008} and {\bf CNM}~\cite{Clauset2004}, which differ in the way they perform this
optimization.
 
 \item {\bf Node similarity-based approaches:} This category deals with the notion that a community is viewed as a group of nodes which are
similar to each other, but dissimilar from the rest of the network. It includes {\bf WalkTrap}~\cite{JGAA-124} which is built on the notion
that random walks tend to get trapped into a community.
 
 \item {\bf Compression-based approaches:} These approaches assume the community structure as a set of regularities in the network topology,
which can be used to represent the whole network in a more compact way than the whole adjacency matrix. The best community structure is
supposed to be the one maximizing compactness while minimizing information loss. The quality of the representation is assessed through
measures derived from information theory. Two popular such algorithms are {\bf InfoMod}~\cite{rosvall2007} and {\bf
InfoMap}~\cite{Rosvall29012008}.
 
 \item {\bf Significance-based approaches:} According to these approaches, a community structure can be expected under certain
circumstances, but groups of densely connected nodes can also appear only by chance. {\bf Order Statistics Local Optimization Method
(OSLOM)}~\cite{DBLP:journals} is a local optimization method applied to measure the statistical significance of individual communities.
 
 \item {\bf Diffusion-based approaches:} These approaches rely on the assumption that information is more efficiently exchanged between
nodes of the same community. In {\bf Community Overlap Propagation Algorithm (COPRA)} \cite{Raghavan-2007}, the information takes the form
of a label, and the propagation mechanism relies on a vote between neighbors. Communities are then obtained by considering groups of nodes
with the same label.
\end{enumerate}

Each algorithm is used with its default parameters. Although the algorithms OSLOM and COPRA are suitable for overlapping community
detection, these are used here to detect mutually exclusive communities (i.e., non-overlapping communities) by setting a priori the number of overlapping nodes as zero.

\subsection{Validation measures}\label{metrics}
A stronger test of the correctness of the community detection algorithm, however, is by comparing the obtained community with a given
ground-truth structure.
We use three standard validation metrics, namely Normalized Mutual Information (NMI)~\cite{danon2005ccs}, Adjusted Rand Index
(ARI)~\cite{hubert1985} and Purity (PU)~\cite{Manning} to measure the accuracy of the detected communities with respect to the ground-truth
community structure. \cite{Labatut} argue that these metrics 
ignore the network structure and propose the  weighted versions of these measures where
misplacing a high degree vertices would incur higher penalty. We therefore also use the weighted versions
of these measures, namely Weighted-NMI ({\em W-NMI}), Weighted-ARI ({\em W-ARI}) and Weighted-Purity ({\em W-PU}). All these metrics are
bounded between 0 (no matching) and 1
(perfect matching).

 \begin{table*}[!ht]
\caption{Differences of  MaxPerm with the other algorithms for different networks. Each value is obtained by averaging the
values of all six validation metrics. The expanded results are shown in Appendix. Positive differences indicate the improvement of our
algorithm. }\label{avg_Improvement}
\centering
 \scalebox{0.75}{
\begin{tabular}{|c|c|c|c|c|c|c|c|c|}
\hline
{\bf Networks}        &  {\bf Louvain} & {\bf FastGreedy} & {\bf CNM}   & {\bf WalkTrap} & {\bf Infomod} & {\bf Infomap} & {\bf COPRA} &
{\bf OSLOM}\\\hline\hline
LFR ($\mu$=0.1)       &  0.00    & 0.00	& 0.14	& 0.00 	   & 0.06    & 0.00    & 0.11  & 0.00 \\\hline	
LFR ($\mu$=0.3)       &  0.00	   & 0.87	& 0.40  & 0.00	   & 0.08    &0.00     & 0.02  & 0.00 \\\hline
LFR ($\mu$=0.6)       &  -0.75   & 0.02	& -0.13 & -0.50    & -0.20   & -0.72   & -0.09 & -0.68 \\\hline
Football              &  0.02    & 0.01       & 0.30  & 0.02     & 0.01    & 0.00    & 0.03  & 0.01 \\\hline
Railway               &  0.14    & 0.37       & 0.20  & 0.02     & 0.19    & 0.02    & 0.01  & 0.11\\\hline
Coauthorship          &  0.00    & 0.14       & 0.05  & 0.02     & -0.04   & -0.02   & 0.09  & 0.09\\\hline
                          
 \end{tabular}}
\end{table*}

\subsection{Performance analysis}\label{sec:eval}
Table~\ref{avg_Improvement} shows results of the improvement  of our method (as differences between the NMI of a baseline algorithm to
MaxPerm) compared to the algorithms
given in Section~\ref{algorithms} and averaged over all the validation metrics. The detailed results of improvement in terms of
each validation measure separately are shown in Table \ref{Improvement} of Appendix.

For LFR ($\mu$=0.1) network, MaxPerm is as efficient  as
Louvain, WalkTrap, Infomap and OSLOM (and achieves an average accuracy of 0.95), which is followed by FastGreedy, Infomod, COPRA and CNM.
For
LFR ($\mu$=0.3) network, MaxPerm once again seems to be comparable in performance to  Louvain, WalkTrap, Infomap and OSLOM (and achieves average accuracy of
0.86), which is followed by COPRA, Infomod, CNM and FastGreedy. 
However, MaxPerm does not work well for the LFR ($\mu=0.6$) network. In this case, Louvain outperforms other competing algorithms with the average accuracy of 0.53, which is followed by Infomap, OSLOM, WalkTrap, Infomod, CNM, MaxPerm and FastGreedy. We hypothesize that maximizing permanence performs better at identifying communities from networks
which actually possess modular structure. If the communities are not that well-separated as in LFR ($\mu=0.6$), more singleton communities
are formed and the permanence value tends to degrade. We shall address this issue further at the end of this section.

For {\it Football} network, MaxPerm achieves highest average accuracy of 0.86 with Infomap, followed by FastGreedy, Infomod and OSLOM at the
second position, Louvain and WalkTrap at the third position, COPRA and CNM at the forth and fifth positions respectively.  For {\it Railway}
network, MaxPerm completely dominates others with the average accuracy of 0.78, which is followed by COPRA, Infomap, WalkTrap, OSLOM,
Louvain, CNM and FastGreedy. Once again, MaxPerm shows moderate performance for {\it Coauthorship} network, and seems to be as good as
Louvain
(achieves average accuracy of 0.34). Although MaxPerm seems to be superior than FastGreedy, CNM, WalkTrap, COPRA and OSLOM, they are
dominated by two information-theoretic approaches (Infomod and Infomap). 

To summarize our algorithm is competitive with other standard
algorithms, except for  LFR($\mu=0.6$) and coauthorship networks. In order to understand this behavior we further look at the community
structure of these two networks individually.

\subsection{Analyzing the community structure of LFR ($\mu=0.6$)}
To understand why MaxPerm is not as competitive for  LFR ($\mu=0.6$), we study the quality of the ground-truth communities. 
We observe that the average internal clustering coefficient in the network decreases with increase in $\mu$. The value is 0.78 for
 LFR ($\mu=0.1$), it reduces to 0.36 for LFR ($\mu=0.6$). Moreover, 97\% of vertices in ground-truth communities of LFR
($\mu$=0.6) have less internal connections than the external connections. In contrast, LFR ($\mu$=0.1) and LFR ($\mu$=0.3) have almost no
such nodes.
 This indicates that the LFR ($\mu=0.6$) network does not have modular structure. 

To further validate this hypothesis, we measure the similarity of the communities obtained by different community detection algorithms using the validation measures. 
The results in Table~\ref{similarity} clearly show that the values of the validation metrics decrease 
 with the increase in $\mu$. This is because as $\mu$ increases, the communities in LFR
network become more fuzzy and the consensus between the outputs of different algorithms dilutes. The results
of a good community detection algorithm should reflect such absence of modular structure in the network (hence show poor performance). In the absence of a modular structure, permanence-based algorithm tends to detect more singleton communities rather than arbitrarily assigning
vertices into communities.

\subsection{Analyzing the community structure of coauthorship network}
To explain the results of MaxPerm obtained from coauthorship network, we analyze the meta data of 
the communities. The titles and the abstract written by the authors in each community obtained by MaxPerm show that 
our method splits large ground-truth communities into denser submodules. 

This phenomenon is more prominent in older research areas such as Algorithms and Theory, Databases etc. These submodules are actually the
subfields (sub-communities) of a field (community) in computer science domain. Few examples of such sub-communities obtained from our
algorithm are noted in Table~\ref{subfield}. Thus, our algorithm, in addition to identifying well-defined communities, is also able to
unfold the hierarchical organization of a network (see Section \ref{sec:comp} for more discussion).

\begin{table}[!t]
\caption{Average pairwise similarities between outputs of the community detection algorithms on different LFR
networks.}\label{similarity}
\centering
 \scalebox{0.75}
 {
\begin{tabular}{|c|c|c|c|}
\hline
Validation & LFR  & LFR  & LFR \\
measure  &  ($\mu$=0.1)                &      ($\mu$=0.3)            &       ($\mu$=0.6)        \\\hline
NMI  & 0.95 & 0.82 & 0.53 \\\hline
ARI   &  0.98 & 0.79 & 0.48 \\\hline
PU   & 0.99  & 0.85 & 0.56 \\\hline
W-NMI & 0.94 & 0.85 & 0.54 \\\hline
W-ARI  & 0.97  & 0.78 & 0.50\\\hline
W-PU   & 0.98 & 0.83 & 0.57 \\\hline
\end{tabular}}
\end{table}

\begin{table}
\centering
\caption{Size of the largest communities obtained from different community detection algorithms and their similarities with the ground-truth
structure.}\label{max_comm}
 \scalebox{0.7}
 {
\begin{tabular}{|c|r|r|r|r|r|r|}

 %& \multicolumn{2}{c|}{Largest community size}& \multicolumn{2}{c|}{Similarity} \\\cline{2-5}
 \multicolumn{7}{c}{{\bf Size of the large communities}} \\\hline
 &  LFR ($\mu=0.1$) &  LFR ($\mu=0.3$) &  LFR ($\mu=0.6$)  & Football & Railway  & Coauthorship \\\hline
Ground-truth & 63 & 49 & 42 & 12 & 13 & 12674 \\\hline
Louvain & 65 & 62 & 57 & 24 &  17&  1254\\\hline
FastGreedy & 78 & 95 & 76 & 18 & 4 & 9875  \\\hline                                    
CNM  & 91 & 86 & 72 & 32 & 32 &  11251\\\hline
Walktrap  & 71 & 83 & 65 & 15 & 14 & 8620 \\\hline
Infomod & 65 & 61 & 46 & 16  & 4 &  324 \\\hline
Infomap &  65 & 59 & 48 & 16  & 4 & 357\\\hline
COPRA   & 54  & 56 & 76 & 20  & 10 &  465\\\hline
OSLOM  &  57 & 42   & 87 & 18  &12 &  732\\\hline
MaxPerm & 60  & 49 & 40 &  13 &  13 & 318  \\\hline

\multicolumn{7}{c}{{ }}\\
\multicolumn{7}{c}{{\bf Similarity of largest community obtained from the algorithm with that of the ground-truth}} \\\hline
 &  LFR ($\mu=0.1$) &  LFR ($\mu=0.3$) &  LFR ($\mu=0.6$)  & Football & Railway  & Coauthorship \\\hline
Louvain &  0.89  & 0.70 & & 0.41 & 0.87 & 0.70 \\\hline
FastGreedy & 0.51& 0.32 & 0.39 & 0.65 & 0.52 & 0.39  \\\hline                                    
CNM  &  0.82 & 0.52 & 0.76 & 0.31 & 0.71 &  0.66\\\hline
Walktrap  & 0.88 & 0.51 & 0.73 & 0.57 & 0.75 & 0.64\\\hline
Infomod & 0.90 & 0.79 & 0.82 & 0.86  & 0.84 & 0.78 \\\hline
Infomap & 0.90  & 0.74 & {\bf 0.83} & 0.86 & 0.85 & 0.78\\\hline
COPRA   & 0.79  & 0.67 & 0.70 & 0.78  & 0.52 & 0.59 \\\hline
OSLOM  & 0.81  & 0.81   & 0.73  & 0.72  & 0.68&0.61  \\\hline
MaxPerm & {\bf 0.95}  & {\bf 1} & {\bf 0.83} &  {\bf 0.92} &  {\bf 0.87} & {\bf 0.79 } \\\hline

\end{tabular}}
\end{table}

\begin{figure}[!ht]
\centering
\includegraphics[width=\columnwidth]{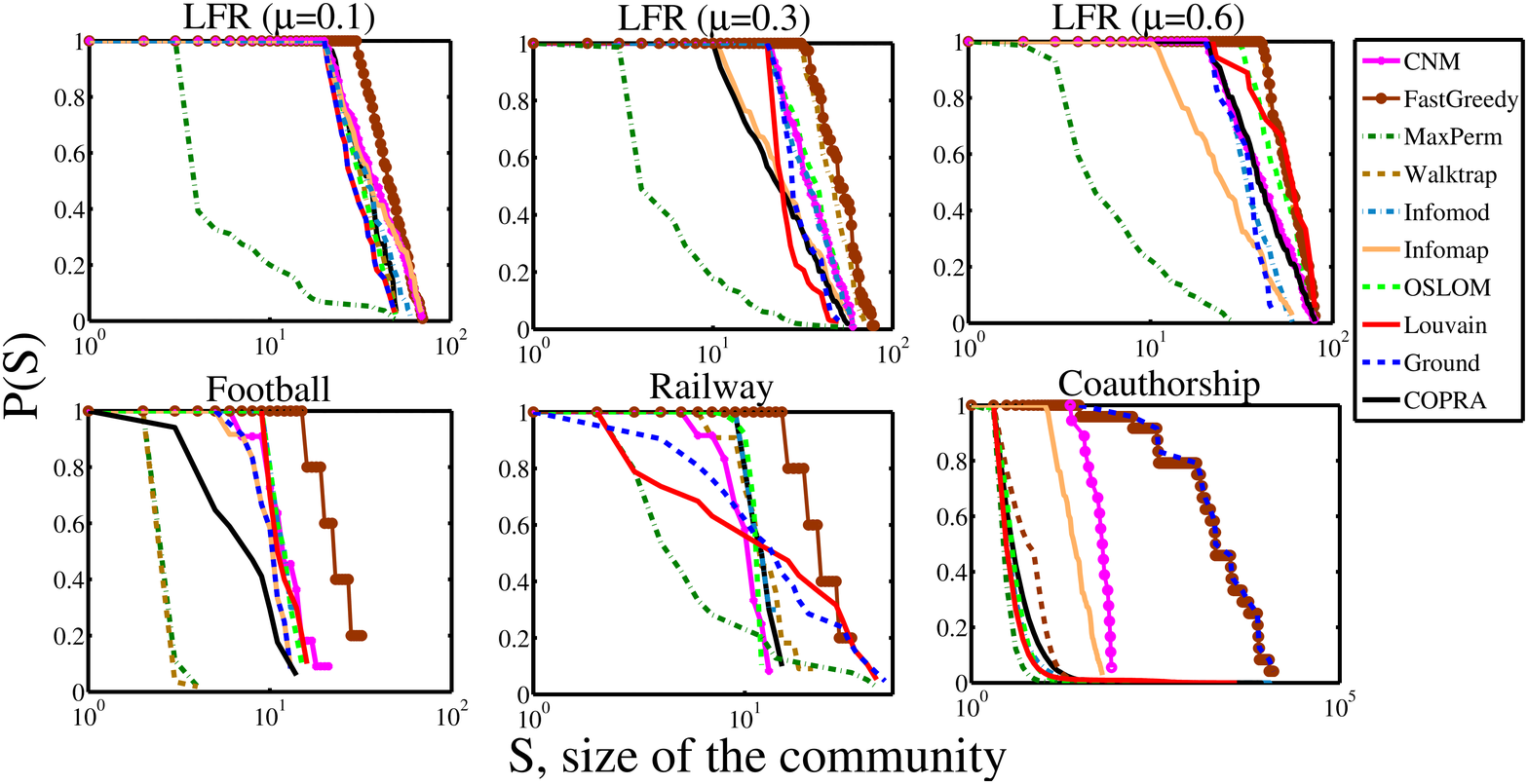}
\caption{(Color online) Distribution of the community size obtained from the ground-truth structure vis-a-vis other community detection algorithms.}
\label{fig:size_dist}
\end{figure} 

\subsection{Detection of small sized communities}\label{sec:comp}
Many optimization algorithms tend to ignore smaller size communities and combine them to produce larger communities. This phenomenon is known
as the ``resolution limit'' problem. Here we provide experimental results to show that MaxPerm can mitigate the effects of resolution limit
(analytical proof is given in Section \ref{permanence_property}).

In our test suite, we observe that all the
competing algorithms produce larger sized communities as compared to those obtained by permanence. 
Figure \ref{fig:size_dist} shows the community size distribution of the ground-truth structure, and that obtained from other community detection algorithms. We observe that with the increase of $\mu$ in LFR network, the size distributions obtained from ground-truth and Louvain respectively start separating, while the pattern obtained from MaxPerm remains almost same in that most of the communities are small in size. Interestingly, for coauthorship network even if most of the communities in ground-truth are large in size, the communities obtained from Louvain and MaxPerm are almost of similar size. 
%\textcolor{red}{The Figure shows opposite, Louvain matches the ground truth community sizes, whereas MaxPerm separates}

We therefore investigate whether these smaller size communities are arbitrary or actually represent different sub-communities of a large
community present in the ground-truth structure. As shown in Table \ref{subfield}, for the co-authorship network the smaller communities
actually represented sub-disciplines of a larger academic field. However, such meta data is not available for all networks. To
answer this
question, we construct a bipartite network consisting of the communities obtained from the algorithm $A$  (denoted as set $C_A$) as vertices in
one partition and the ground-truth communities (denoted as set $C_G$) as vertices in another partition. We create the edges $C_A \times C_G$ with edge weights derived as follows: the weight of the edge connecting $c_a\in C_A$ and $c_g\in C_G$ is measured by the fraction of vertices in $c_a$ that are also part of $c_g$.  
We only consider edges with non-zero weights. If a detected community is mostly subsumed by one ground-truth community, it produces very
high edge-weight (ranging from 0.8-1) and very small-edge-weight (0-0.2); whereas medium edge-weight (0.4-0.7) indicates that the
detected community is equally absorbed in multiple ground-truth communities. For example, assume that $c_a$ contains 10 nodes which are distributed into two ground-truth communities $c_g^1$ and $c_g^2$ in two different ways: ({\em Case 1.}) 8 vertices of $c_a$ are in $c_g^1$ and 2 are in $c_g^2$, ({\em Case 2.}) 4 vertices of  $c_a$ are in $c_g^1$ and 6 are in $c_g^2$. Therefore, in Case 1 the edge weights are 0.8 and 0.2 for  
($c_a \rightarrow c_g^1$) and   ($c_a \rightarrow c_g^2$) respectively; whereas in Case 2 the edge weights are 0.4 and 0.6 for  ($c_a
\rightarrow c_g^1$) and   ($c_a \rightarrow c_g^2$) respectively.

We construct such weighted bipartite graphs separately for all the algorithms. In  Figure \ref{bucket}, we divide the edge-weights into 10
buckets such that bucket 1 corresponds to higher-edge weight. Then in y-axis,  we plot the fraction of edges falling in each bucket. We
observe that while the proportion of edges for baseline algorithms is higher in medium weight zone, for MaxPerm most of the edges either
fall in higher weighted buckets or lower-weighted buckets. This indicates that the communities obtained by  MaxPerm are indeed subgroups
within one larger community, rather than being scattered across multiple communities. 
%corroborates our earlier observation that MaxPerm is able to detect more fine-grained sub-communities of a large ground-truth community.           

We also observe that despite finding small communities, the largest-size community obtained by MaxPerm best corresponds to the largest
ground-truth community.
In Table~\ref{max_comm}, we show for all the networks that the size of the largest communities detected by the other algorithms is much larger than
 the size of the largest community present in the ground-truth structure. We also measure the maximum similarity (using Jaccard
coefficient)
between the largest-size community detected by each algorithm with the communities in ground-truth structure and notice that MaxPerm
is able to detect largest size community which is most similar to the ground-truth structure (see Table~\ref{max_comm}). These experimental results indicate that MaxPerm is more effective in reducing the  effect of resolution limit.

\begin{table}[!ht]
\centering
\scalebox{0.8}{
\begin{tabular}{|c|c|}
\hline
{\bf Communities} & {\bf Sub-communities}\\\hline\hline
Algorithms & Theory of computation; Formal methods; \\
and Theory & Information \& coding theory; \\ 
           & Computational geometry; Data structure;\\\hline
Databases & Models; Query optimization; Database \\
          & languages; storage; Performance;\\
          & security, and availability \\\hline
\end{tabular}}
\caption{Example of communities and sub-communities obtained from coauthorship network using MaxPerm algorithm. }\label{subfield}

\end{table}

\begin{table}
\centering
 \includegraphics[width=1\textwidth]{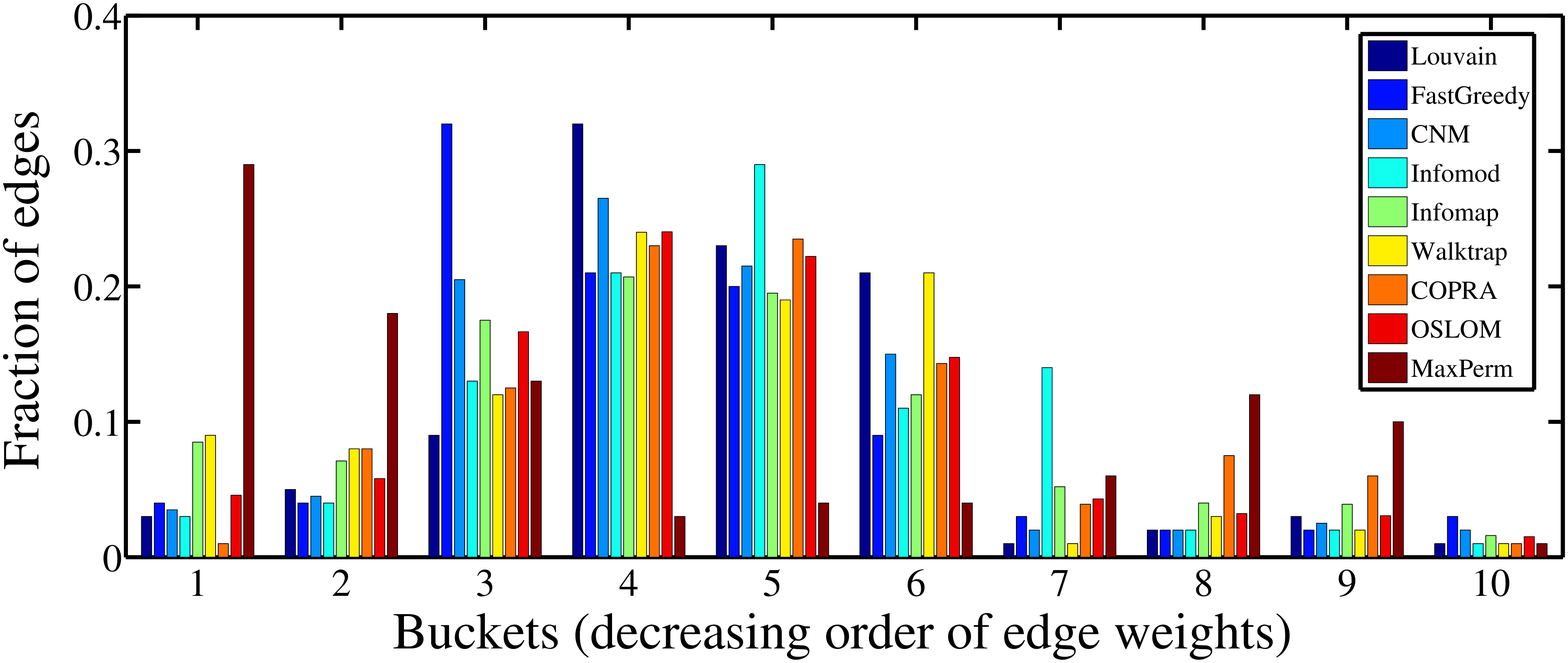}
\captionof{figure}{ (Color online) Fraction of edges of the bipartite network with a specific weight. The range of the edge-weight
is divided into ten buckets (bucket 1: 0.9-1, bucket 2: 0.8-0.89 and so on). The bipartite network corresponds to the coauthorship network.}\label{bucket}
\end{table}

\section{Effect of vertex ordering}\label{vertex_per}
Most of the community detection algorithms attempt to optimize certain functions (such as modularity) and therefore are heavily
dependent on the order in which vertices are processed. This is an important source of concern among researchers on how to reconcile results where  
the final outcome can change due to the mere change in  vertex ordering \cite{sjri2012,lf2012,dyb2010,lln2010,chakraborty}. In
this direction, \cite{chakraborty} show that despite such fluctuations in the final outcome, there exist few invariant groups of vertices
in  a network that always remain together, and they are known as ``constant communities''. Further, they study the change in community
structure based on the number of constant communities by a metric, called \textit{sensitivity} ($\phi$), which is measured as the ratio of
the number of constant communities to the total number of vertices. For a particular network, if the value of $\phi$ for an algorithm
remains consistent over different vertex orderings, the algorithm would be qualified to be resilient to the effect of vertex ordering.

\begin{figure}[!ht]
\centering
 \includegraphics[scale=0.35]{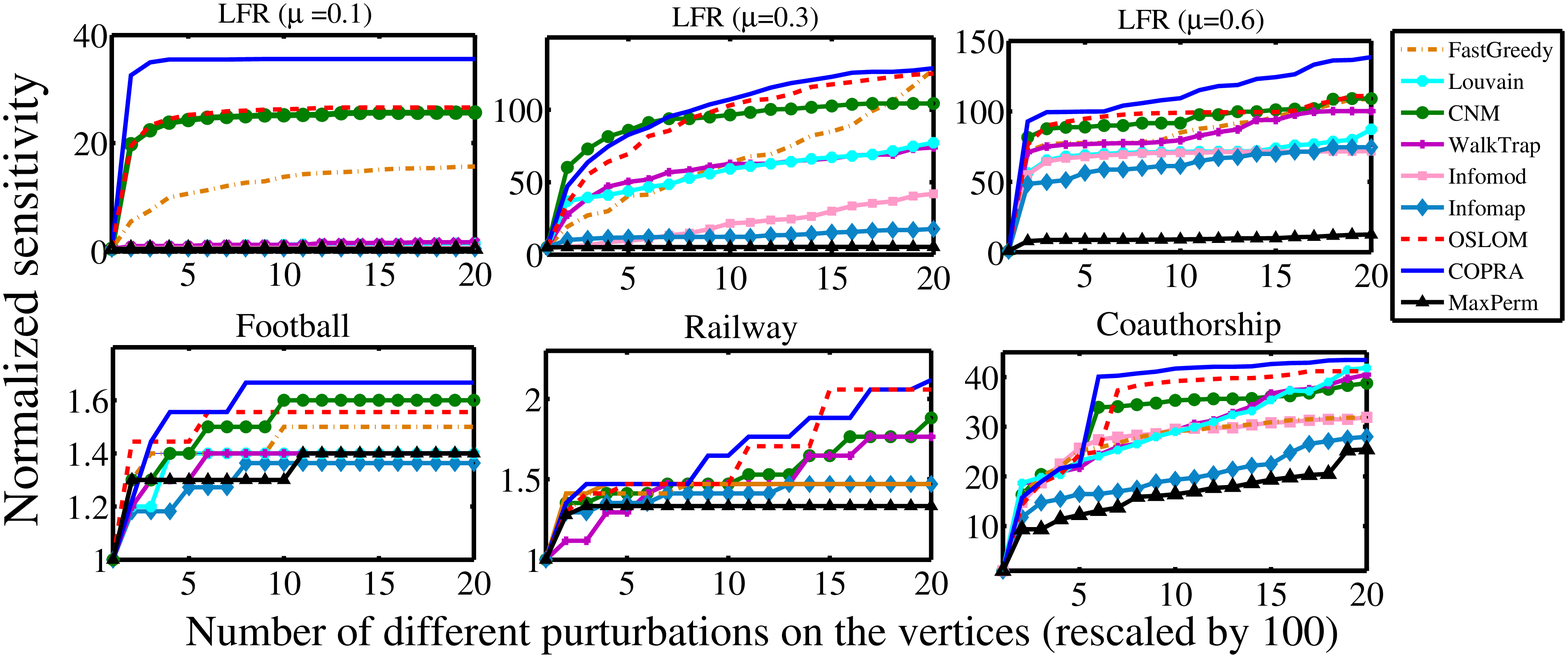}
 \caption{Sensitivity of each algorithm across 2000 permutations. The x-axis is rescaled by a constant factor of 100. Y-axis indicates the value of sensitivity as it changes over the permutations. For better visualization, we rescale the value of sensitivity with the minimum value for each algorithm so that the sensitivity of all the algorithms always starts from 1.}\label{constant_comm}
\end{figure}

We plot the value of sensitivity over different vertex orderings for each algorithm in Figure \ref{constant_comm}. In x-axis, we plot the
number of different permutations of the vertices. For fair comparison, we normalize the sensitivity values by the minimum value for each
algorithm and plot it in y-axis so that the sensitivity profiles of all the algorithms start from 1. For a particular network, the lower the
value of sensitivity of an algorithm across different perturbations, the better the algorithm resilient to the initial vertex ordering.
There are two consequences of this result: (i) For a specific  LFR network, say LFR ($\mu=0.3$), we can observe that MaxPerm remains almost
consistent in terms of sensitivity over different iterations, which is in most cases followed by Infomod, Infomap and Louvain. COPRA and
OSLOM perform worst among the others. (ii) Across different LFR networks, we observe  that with the increase of $\mu$, the performance of
all the algorithms 
start deteriorating. The reason could be that with the increase of $\mu$, communities in LFR network become fuzzier, and therefore multiple community partitions can be equally good.
 Similar result is observed across different real-world networks where the algorithms tend to be largely
 insensitive for coauthorship network due to the lack  of clear separation between communities.

\begin{figure}[!t]
\centering
\scalebox{0.3}{
\includegraphics{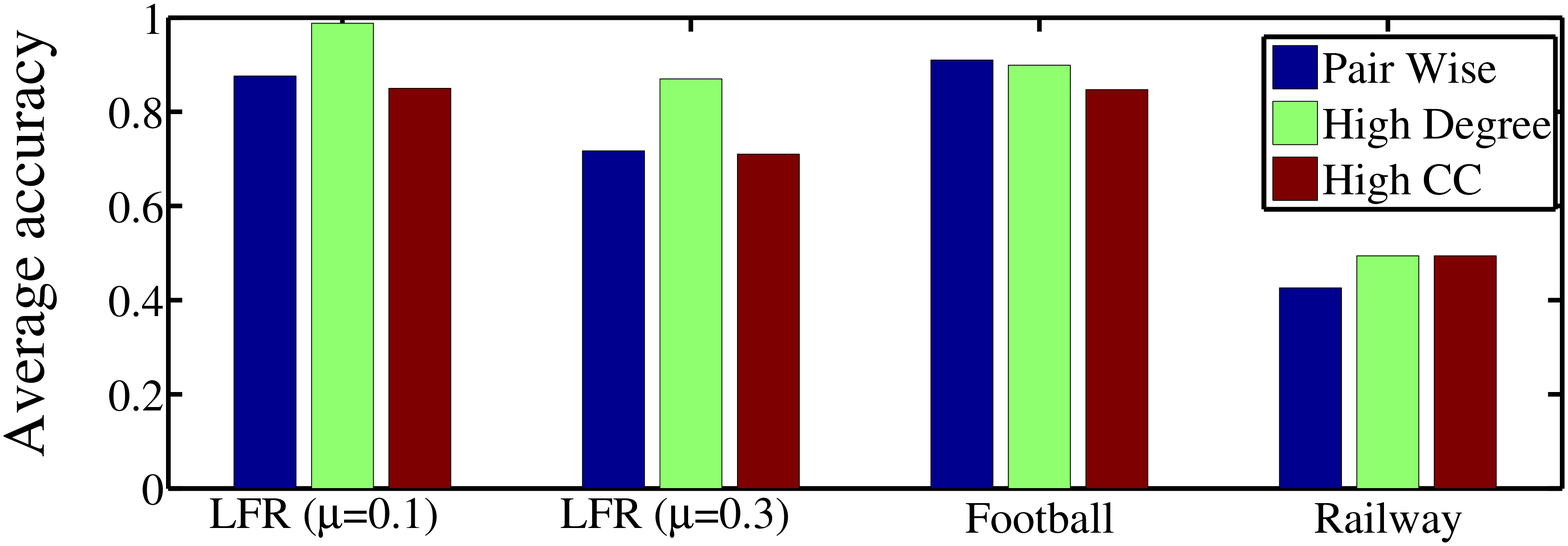}}
\caption{Average accuracy of community detection results with different seeding techniques. Seeding based on high degree gives the highest
accuracy.} \label{seed}
\end{figure}

\section{Effect of seed community selection}\label{sec:seed}
The first step before starting the iterations to maximize permanence is to place the vertices in initial (seed) communities. The importance of seed communities in detecting communities has been studied for other metrics as in ~\cite{seed-set-tr}. In this section, we explore how it affects the MaxPerm algorithm.

We note that vertices in seed communities should consist of connected subgraphs. If the vertices in the communities are not connected 
to each other, then their initial permanence will be zero (because $I(v)$ is zero), and moving to a neighboring community 
does not improve this value. We consider three seed selection strategies as follows:
\begin{itemize}
\item {\bf Pair Wise:} Two vertices are assigned to the same seed community if they are connected by an edge. If we encounter a vertex whose neighbors have all been already assigned to communities, that unmatched vertex is kept as a singleton. This is the fastest out of the three seeding methods.

\item {\bf High Degree:} We first order the vertices in the decreasing order of degree. The vertex with the highest degree and its neighbors are assigned to the same community. We continue combining each unassigned vertex in the sorted list and its unassigned neighbors into a community. This seeding is based on maximizing $I(v)$ for the high degree vertices.

\item {\bf High CC:}  We order the vertices in the decreasing order of clustering coefficient, and similar to the {\em high degree}, combine
the vertices with high clustering coefficient and their neighbors in a community. This seeding is based on maximizing $c_{in}(v)$ and is the
most expensive of the three methods.

\end{itemize}

We test the seeding strategies on the four networks (LFR with $\mu=0.1, 0.3$, Football and Railway) on which MaxPerm consistently
outperformed the other algorithms. As can be seen in Figure~\ref{seed}, the best accuracy comes from using the high degree strategy. The
reason for this is that Pair Wise depends on the vertex ordering, and the pairings can change depending on how vertices are numbered. High
CC is too restrictive, because once a vertex is in a tightly coupled group, there is less chance for it to migrate to a larger (if slightly)
less tightly coupled group. Thus maximizing permanence using high clustering coefficient  tends to fall into local minima. In High Degree,
the groupings are less random compared to Pair Wise, but they also provide more flexibility for vertices to migrate between communities as 
compared to High CC. We believe that this is the reason behind  High Degree providing the best accuracy.

%Jason Riedy, David A Bader, Karl Jiang, Pushkar Pande, and Richa Sharma. Detecting communities from given seeds in social networks. 2011.

\section{Handling limitations of modularity maximization algorithms}\label{permanence_property}
In this section we analytically show how finding communities by maximizing permanence can reduce the effect of some of the common issues in community detection
including (a) resolution limit, (b) degeneracy of solution
and (c) dependence on the size of the graph. %We now discuss how each of these problems are ameliorated by maximizing permanence.  
 
\begin{figure}[!ht]
\centering
 \begin{tabular}{l}
\scalebox{0.25}{ \includegraphics{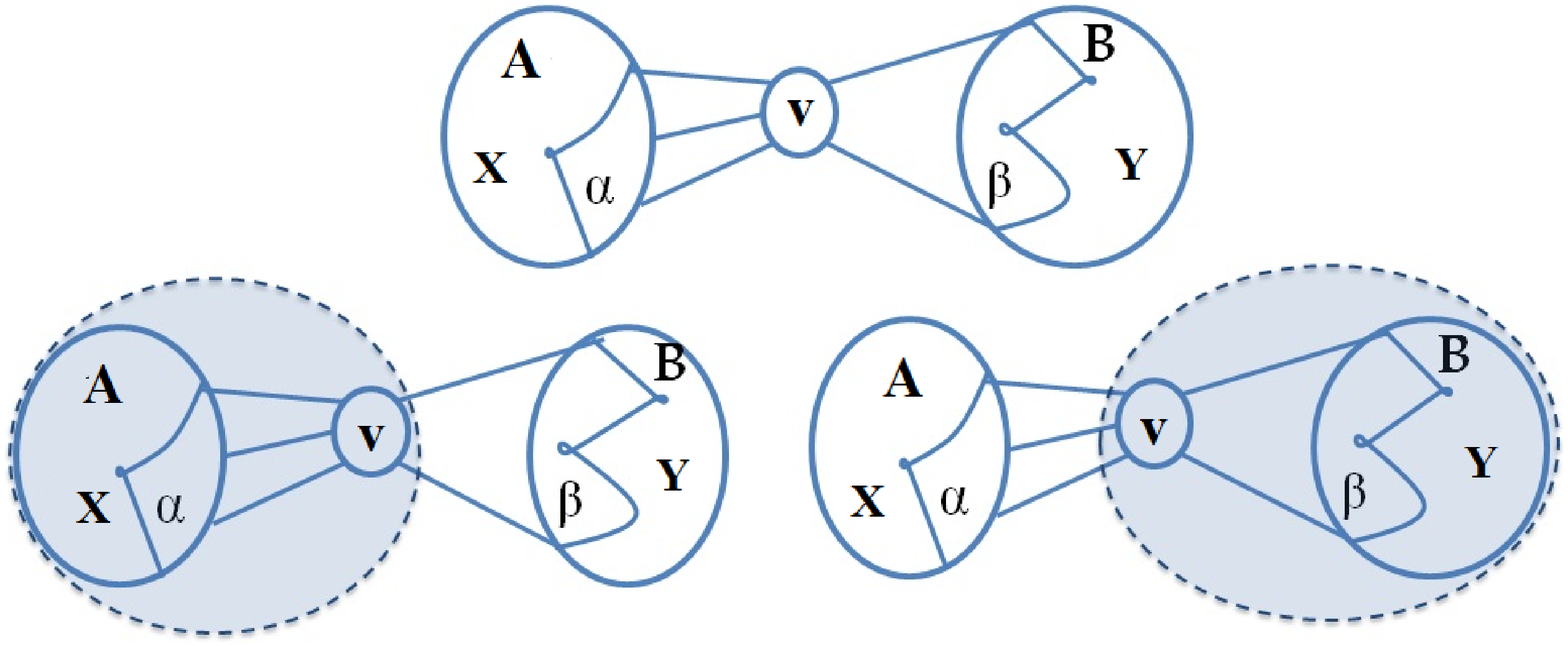}} \\

 \scriptsize{Case 1: [$(A+v):B$] \hspace{10 mm} Case 2: [$A:(B+v)$] } \\

 \scalebox{0.25}{ \includegraphics{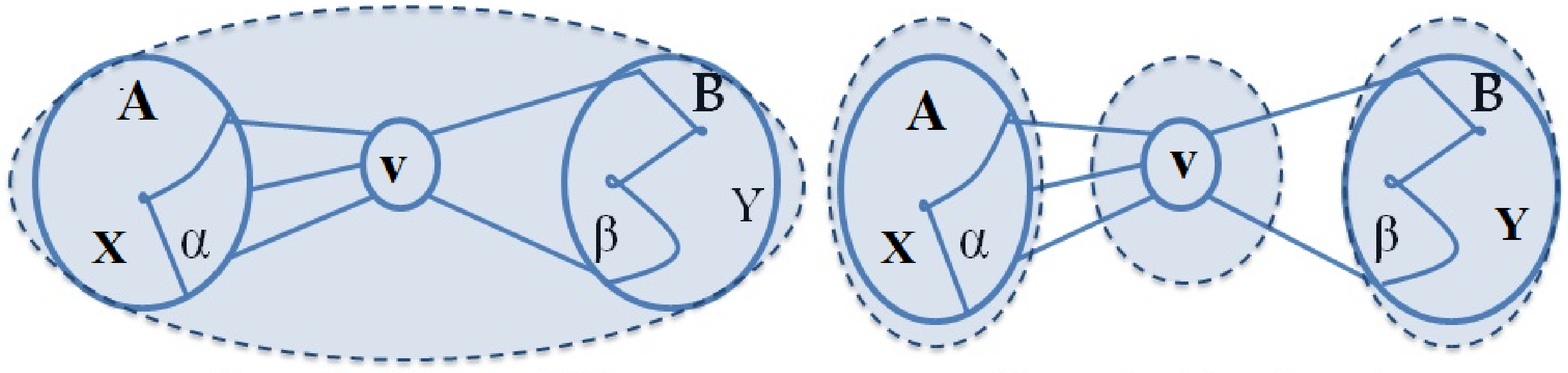}} \\
 
     \scriptsize{Case 3: [$(A+v+B)$] \hspace{10 mm} Case 4: [$A:B:v$] } \\

 \end{tabular}
 \caption{(Color online) Illustrative example explaining four possible cases of community assignment of vertex $v$.}\label{example1}
\end{figure}

\begin{figure}
\centering
\begin{minipage}{.5\columnwidth}
  \centering
  \includegraphics[width=.4\linewidth]{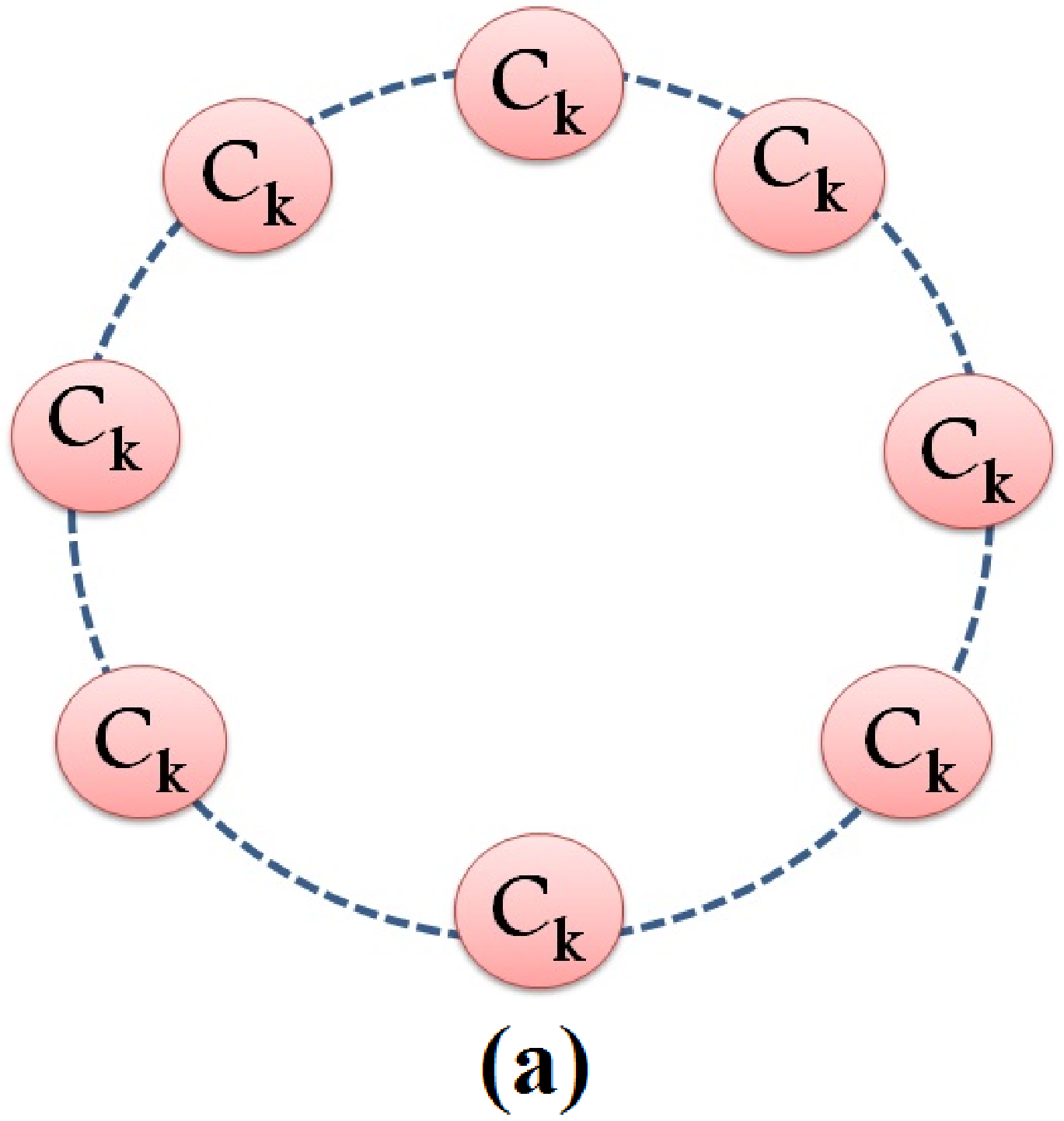}
  
\end{minipage}%
\begin{minipage}{.6\columnwidth}
  \centering
  \includegraphics[width=.3\linewidth]{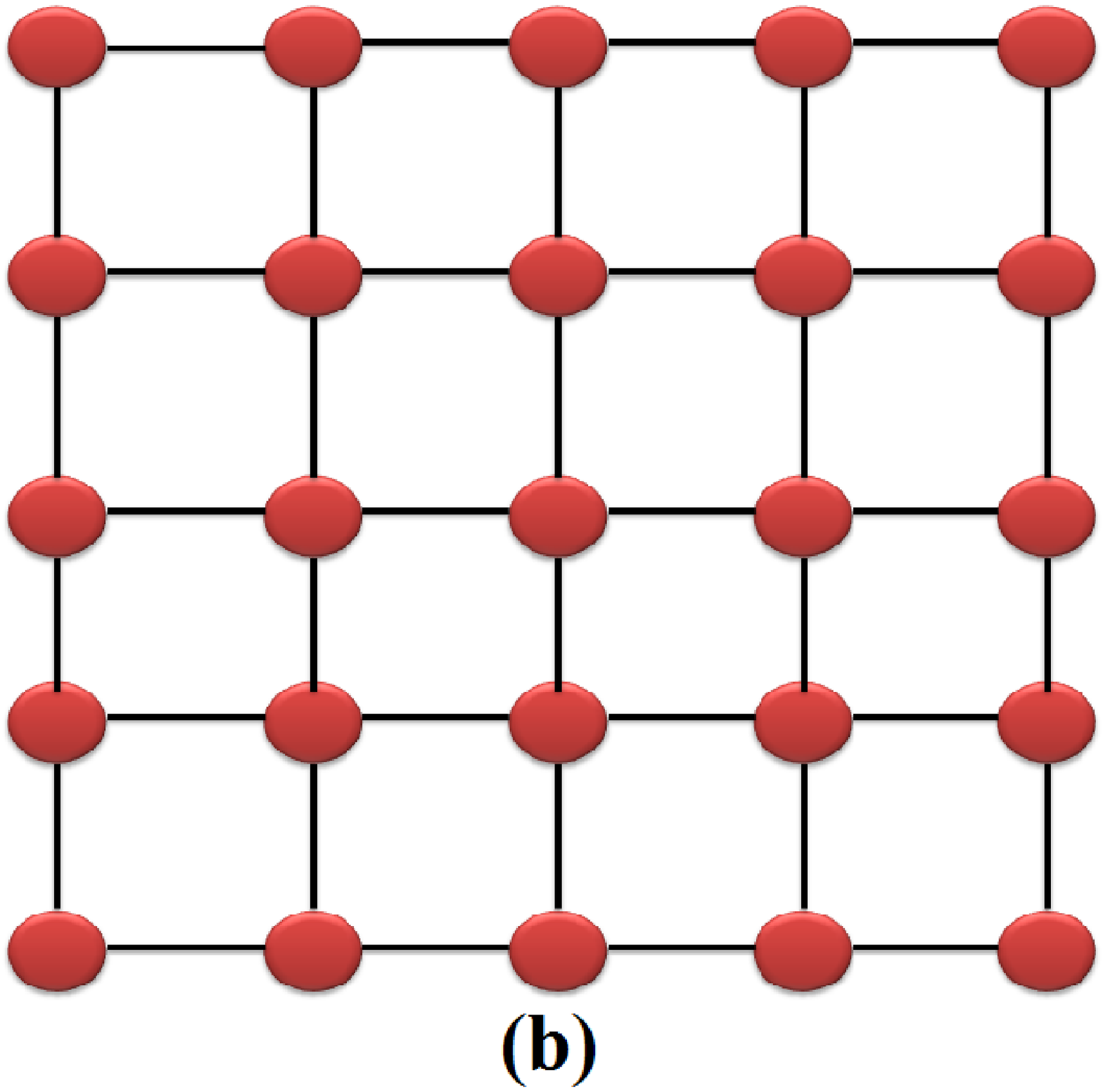}
\end{minipage}
\caption{(a) A cycle of $m$ identical $k$-cliques each having $k$ vertices and connected by single edges; (b) a $5\times 5$ grid network.}\label{example_diss}
\end{figure}

 We illustrate our proof using a simple example of two communities $A$ and $B$ connected by one
vertex $v$ (as shown in Figure~\ref{example1}). There is no edge between the communities $A$ and $B$, except through the vertex $v$.
This simple example covers many of the scenarios where problems due to degeneracy of solutions or resolution limits arise. For example, by
considering $A$ and $B$ to be cliques, and $v$ to be a vertex within the clique $A$, we can form a subgraph from the circle of cliques as
shown in Figure~\ref{example_diss}(a). This figure is a common example to show the existence of resolution limit \cite{Barthelemy}. In a
similar manner, by considering $A$ and $B$ as single vertices, we can obtain the subgraphs of a grid as shown in
Figure~\ref{example_diss}(b). A grid is an example of a network where multiple solutions  can occur.

 %single nodes themselves we can create any subgraph from a grid as shown in Figure~\ref{

%We shall show how the assignments differ when maximizing modularity versus maximizing permanence. 

\subsection{Terminology and Theorems}

Let vertex $v$ be connected to $\alpha$ ($\beta$) nodes in community $A$ ($B$), and these $\alpha$ ($\beta$) nodes form the set $N_\alpha$
($N_\beta$). The number of vertices in community $A$ is ($x+ \alpha$), and the number of vertices in community $B$ is ($y + \beta$).
 Let the average internal degree (connections to internal neighbors only) of a vertex $a\in N_\alpha$ and a vertex $b\in N_\beta$, before
$v$ is
added, be $I_{\alpha}$ and $I_{\beta}$ respectively. Let the average internal clustering coefficient\footnote{{\scriptsize Note that
internal clustering
coefficient of $v$ is obtained by considering the ratio of the existing connections and the total number of possible connections among the
{\em internal neighbors} of $v$.}} of the neighboring nodes in communities $A$ and $B$ be $C_A$ and $C_B$ respectively.

If $v$ is added to
communities $A$ ($B$) then the average internal 
 clustering coefficient of $v$ becomes $C^{v}_{A}$ ($C^{v}_{B}$) respectively, and the average internal clustering coefficients of the nodes
in $N_\alpha$($N_\beta$) become $C^{\alpha}$ ($C^{\beta}$). We will use these average values to approximate the permanence measure. 
 
 We assume that the communities $A$ and $B$ are tightly connected internally such that the values of $C_A$ and $C_B$ are very high (at least
greater than 0.5). We note that the values $C^{\alpha}$ ($C^{\beta}$) will depend on the connections of $v$ to the communities
and the connections of the vertices in $N_\alpha$ and $N_\beta$. 

To simplify this, we will consider two special cases. One case is when the nodes in the community are
tightly connected and adding $v$ does not significantly change the internal clustering coefficient. In this case, we assume $C^\alpha=C^A$ and $C^\beta =C^B$. The other case is when $v$ is added that no new  connections are formed among the neighbors of $v$, but the internal degree increases by one. Therefore,  $C^{\alpha}=C_A\frac{(I_{\alpha}-1)}{(I_{\alpha}+1)}$ (similarly,
$C^{\beta}=C_B\frac{(I_{\beta}-1)}{(I_{\beta}+1)}$ ).

%where the vertex $v$ is loosely connected most of the nodes in $A$ ($B$) and these nodes themselves are connected to each other. Because
%the value of $C_A$ ($C_B$) is already high, we assume that in this case, including $v$ does not significantly change the clustering
%coefficient and $C^{\alpha}$=$C_A$ ($C^{\beta}$=$C_B$). 
%The second case is when

 %%%%%%%%%
 %%%%%%%%%%%%%
 %%%%%%%%%%

 The combination of communities $A$, $B$ and the vertex $v$ can have four cases (see Figure~\ref{example1}) as follows:
\begin{itemize}
 \item {\bf Case 1.} $v$ joins with community $A$ only. We denote this configuration as $[(A+v):B]$, and its total permanence as
$P_{(A+v):B}$. We assume that the combined permanence of all nodes $x \not \in (N_{\alpha} \cup N_{\beta} \cup v)$ as $P_x$. This value will
not be
affected due to the re-assignments. Therefore, the total permanence is the sum of the following factors: $P_x$, $[\alpha C^\alpha]$ (for the
nodes in $N_{\alpha}$ connected to $v$), $[\frac{\alpha}{(\alpha+\beta)\beta}-(1-C^v_A)]$ (for vertex $v$) and
$[\beta(\frac{I_{\beta}}{I_{\beta}+1}
-(1-C_B))]$ (for the nodes in $N_{\beta}$).\\ 

$ P_{(A+v):B} =  P_x+\alpha C^\alpha+  \frac{\alpha}{(\alpha+\beta)\beta}-(1-C^v_A)+\beta(\frac{I_{\beta}}{I_{\beta}+1} -(1-C_B))$\\

 \item {\bf Case 2.} $v$ joins with community $B$ only. We denote this configuration as $[(A:(v+B)]$, and its total permanence as
$P_{A:(v+B)}$. The values of this total permanence is  the sum of the following factors: $P_x$, $[\alpha(\frac{I_{\alpha}}{I_{\alpha}+1}
-(1-C_A))]$
(for the nodes in $N_{\alpha}$), $[\frac{\beta}{(\alpha+\beta)\alpha}-(1-C^v_B)]$ (for vertex $v$) and $[\beta C^\beta]$ (for the nodes in
$N_{\beta}$
connected to $v$).\\
 
 $ P_{A:(v+B)} =  P_x+ \alpha(\frac{I_{\alpha}}{I_{\alpha}+1} -(1-C_A)) + \frac{\beta}{(\alpha+\beta)\alpha}-(1-C^v_B) + \beta C^\beta$\\
 
 \item {\bf Case 3.} $A$, $B$ and $v$ merge together. We denote this configuration as $[(A+v+B)]$, and its total permanence as
$P_{(A+v+B)}$.  The values of this total permanence is  the sum of the following factors: $P_x$, $[\alpha C^\alpha]$ (for the nodes in
$N_{\alpha}$), $[\frac{\alpha(\alpha-1)C^v_A+\beta(\beta-1)C^v_B}{(\alpha+\beta)(\alpha+\beta-1)}]$ (for vertex $v$) and $[\beta C^\beta]$
(for the
nodes in $N_{\beta}$ connected to $v$).\\
 
 $P_{(A+v+B)} = P_x + \alpha C^\alpha+ \frac{\alpha(\alpha-1)C^v_A+\beta(\beta-1)C^v_B}{(\alpha+\beta)(\alpha+\beta-1)} + \beta C^\beta$\\
 
 \item{\bf Case 4.} $A$, $B$ and $v$ remain as separate communities. We denote this configuration as $[(A:v:B)]$, and its total permanence
as $P_{(A:v:B)}$. The values of this total permanence is  the sum of the following factors: $P_x$, $[\alpha(\frac{I_{\alpha}}{I_{\alpha}+1}
-(1-C_A))]$ (for the nodes in $N_{\alpha}$), 0 (for vertex $v$) and  $[\beta(\frac{I_{\beta}}{I_{\beta}+1} -(1-C_B))]$ (for the
nodes in $N_{\beta}$).\\
 
 $P_{(A:v:B)} = P_x + \alpha(\frac{I_{\alpha}}{I_{\alpha}+1} -(1-C_A)) + \beta(\frac{I_{\beta}}{I_{\beta}+1} -(1-C_B))$\\
 
 \end{itemize}
 
 We present a set of theorems as to when these conditions will occur. By using these theorems we can analytically show that degeneracy of solutions and resolution limit is reduced when maximizing permanence. 

{\bf Lemma 1}
{\em Given $C^{\alpha}=C_A$ and $C^{\beta}=C_B$, let $Z1=\frac{\alpha -\beta}{\alpha \beta} +\left ( C^{v}_A -C^{v}_B \right)+\left(
\frac{\alpha}{I_{\alpha}+1} - \frac{\beta}{I_{\beta}+1}  \right ) $. 
The assignment $[(A+v):B]$ will have a higher permanence than $[A:(v+B)]$, if $Z1>0$  and a lower permanence if $Z1<0$. \\

Given  $C^{\alpha}=C_A\frac{(I_{\alpha}-1)}{(I_{\alpha}+1)}$ and $C^{\beta}=C_B\frac{(I_{\beta}-1)}{(I_{\beta}+1)}$, let $Z2=\frac{\alpha -\beta}{\alpha \beta} +\left ( C^{v}_A -C^{v}_B \right)+\left(
 \frac{\alpha( C_A+1)}{I_\alpha+1} -  \frac{\beta (C_B+1)}{I_\beta+1} \right)$. 
The assignment $[(A+v):B]$ will have a higher permanence than $[A:(v+B)]$, if $Z2>0$  and a lower permanence if $Z2<0$.} \\

{\bf Proof. } Here we are comparing between Case 1 and Case 2. The difference in total permanence between these two assignments by assuming
$C^\alpha=C_A$ and $ C^\beta=C_B$ is:
\begin{equation*}
\begin{split}
P_{(A+v):B}-P_{A:(v+B)}&  =  \frac{\alpha}{(\alpha+\beta)\beta}+C^v_A+\beta(\frac{I_{\beta}}{I_{\beta}+1}-1) \\
				      & -  (\alpha(\frac{I_{\alpha}}{I_{\alpha}+1}-1) + \frac{\beta}{(\alpha+\beta)\alpha}+C^v_B)\\
&= \frac{\alpha-\beta}{\alpha\beta}+ (C^v_A-C^v_B)
+(\frac{\alpha}{I_{\alpha}+1}-\frac{\beta}{I_{\beta}+1})  
\\				       
\end{split}
\end{equation*}

The difference in total permanence between these two assignments by assuming
$C^\alpha=C_A\frac{(I_{\alpha}-1)}{(I_{\alpha}+1)}$ and $ C^\beta=C_B\frac{(I_{\beta}-1)}{(I_{\beta}+1)}$ is:
\begin{equation*}
\begin{split}
P_{(A+v):B}-P_{A:(v+B)}&  =  \frac{\alpha}{(\alpha+\beta)\beta}+C^v_A+\beta(\frac{I_{\beta}}{I_{\beta}+1}-1) \\
				      & -  (\alpha(\frac{I_{\alpha}}{I_{\alpha}+1}-1) + \frac{\beta}{(\alpha+\beta)\alpha}+C^v_B)\\
&= \frac{\alpha-\beta}{\alpha\beta}+ (C^v_A-C^v_B)
+( \frac{\alpha( C_A+1)}{I_\alpha+1} -  \frac{\beta (C_B+1)}{I_\beta+1} )\\				       
\end{split}
\end{equation*}

If this difference is greater than zero then $[(A+v):B]$ will have a higher permanence. If the difference is less than zero then $[A:(v+B)]$
will have higher permanence. \\

{\bf Lemma 2}
{Joining $v$ to community $A$  gives higher permanence than merging the communities $A$, $B$ and $v$, if  (i) $C^{\beta}= C_B$, and  $X >0$ and (ii) if $C^{\beta}= C_B\frac{I_{\beta}-1}{I_{\beta}+1}$ and $X+\beta C_B\frac{2}{I_\beta+1} > 0$; where $X= \frac{\alpha}{(\alpha+\beta)\beta}-\frac{{\beta}}{I_{\beta}+1}-1+\frac{\beta(\beta-1)(C^v_A+C^v_B)}{(\alpha+\beta)(\alpha+\beta-1)}+\frac{2\alpha\beta C^v_A}{(\alpha+\beta)(\alpha+\beta-1)} $

{\bf Proof. } We are comparing Case 1 and Case 3 and in this case $C^{\beta}= C_B$. The difference in total
permanence
is:
\begin{equation*}
\begin{split}
P_{(A+v):B}-P_{(A+v+B)} &  = \frac{\alpha}{(\alpha+\beta)\beta}-1+C^v_A+\beta(\frac{I_{\beta}}{I_{\beta}+1}-1+C_B) \\
					& -(\frac{\alpha(\alpha-1)C^v_A+\beta(\beta-1)C^v_B}{(\alpha+\beta)(\alpha+\beta-1)} +\beta C^\beta)\\
					&\mbox{ Substituiting $C_B$ with $C^{\beta}$}\\
					&  = \frac{\alpha}{(\alpha+\beta)\beta}-1+C^v_A-\frac{{\beta}}{I_{\beta}+1} \\
					& -(\frac{\alpha(\alpha-1)C^v_A+\beta(\beta-1)C^v_B}{(\alpha+\beta)(\alpha+\beta-1)} )\\	
					&  = \frac{\alpha}{(\alpha+\beta)\beta}-1-\frac{{\beta}}{I_{\beta}+1} \\
					& +\frac{\beta(\beta-1)(C^v_A+C^v_B)}{(\alpha+\beta)(\alpha+\beta-1)}+\frac{2\alpha\beta C^v_A}{(\alpha+\beta)(\alpha+\beta-1)}\\		
\end{split}
\end{equation*}

Now we consider the case where $C^{\beta}= C_B\frac{I_{\beta}-1}{I_{\beta}+1}$. 
The difference in total permanence is:
\begin{equation*}
\begin{split}
P_{(A+v):B}-P_{(A+v+B)} &  = \frac{\alpha}{(\alpha+\beta)\beta}-1+C^v_A+\beta(\frac{I_{\beta}}{I_{\beta}+1}-1+C_B) \\
					& -(\frac{\alpha(\alpha-1)C^v_A+\beta(\beta-1)C^v_B}{(\alpha+\beta)(\alpha+\beta-1)} +\beta C^\beta)\\
					&\mbox{ Substituiting $C^{\beta}= C_B\frac{I_{\beta}-1}{I_{\beta}+1}$}\\
					&  = \frac{\alpha}{(\alpha+\beta)\beta}-1-\frac{{\beta}}{I_{\beta}+1} +\beta C_B\frac{2}{I_\beta+1}\\
					& +\frac{\beta(\beta-1)(C^v_A+C^v_B)}{(\alpha+\beta)(\alpha+\beta-1)}+\frac{2\alpha\beta C^v_A}{(\alpha+\beta)(\alpha+\beta-1)}\\	
\end{split}
\end{equation*}

{\bf Lemma 3}
{\em If  $C^{\alpha}= C_A$ and $C^{\beta}= C_B$  then the 
communities will merge  (i.e., $[(A+v+B)]$), rather than remain separate (i.e., $[A:B:C]$). \\
If $C^{\alpha}=
C_A\frac{(I_{\alpha}-1)}{(I_{\alpha}+1)}$
$C^{\beta}=
C_B\frac{(I_{\beta}-1)}{(I_{\beta}+1)}$ and  then the communities will merge if:\\
$ \frac{\alpha(\alpha-1)C^v_A+\beta(\beta-1)C^v_B}{(\alpha+\beta)(\alpha+\beta-1)} >\alpha\frac{(2C_A-1)}{I_{\alpha}+1} +\beta\frac{(2C_B-1)}{I_{\beta}+1}$.}\\}

{\bf Proof:} We are comparing Case 3 and Case 4, and the case $C^{\alpha}= C_A$ and $C^{\beta}= C_B$
 The difference in total
permanence is:
\begin{equation*}
\begin{split}
P_{(A+v+B)}-P_{(A:v:B)} & =  \alpha C^\alpha+ \frac{\alpha(\alpha-1)C^v_A+\beta(\beta-1)C^v_B}{(\alpha+\beta)(\alpha+\beta-1)} + \beta
C^\beta \\
                                         & -(\alpha(\frac{I_{\alpha}}{I_{\alpha}+1} -(1-C_A)) + \beta(\frac{I_{\beta}}{I_{\beta}+1} -(1-C_B)))\\
                                         &= \frac{\alpha(\alpha-1)C^v_A+\beta(\beta-1)C^v_B}{(\alpha+\beta)(\alpha+\beta-1)}\\
                                         &+\frac{\alpha}{I_{\alpha}+1} +\frac{\beta}{I_{\beta}+1} \\
                                           \end{split}
\end{equation*} 
This value is always positive so the communities will merge.\\

We now consider the case where $C^{\alpha}=C_A\frac{(I_{\alpha}-1)}{(I_{\alpha}+1)}$
$C^{\beta}=
C_B\frac{(I_{\beta}-1)}{(I_{\beta}+1)}$
permanence is:
\begin{equation*}
\begin{split}
P_{(A+v+B)}-P_{(A:v:B)} & =  \alpha C^\alpha+ \frac{\alpha(\alpha-1)C^v_A+\beta(\beta-1)C^v_B}{(\alpha+\beta)(\alpha+\beta-1)} + \beta
C^\beta \\
                                         & -(\alpha(\frac{I_{\alpha}}{I_{\alpha}+1} -(1-C_A)) + \beta(\frac{I_{\beta}}{I_{\beta}+1} -(1-C_B)))\\
                                       &  = \frac{\alpha(\alpha-1)C^v_A+\beta(\beta-1)C^v_B}{(\alpha+\beta)(\alpha+\beta-1)} \\
                                       &- \alpha\frac{(2C_A-1)}{I_{\alpha}+1} -\beta\frac{(2C_B-1)}{I_{\beta}+1}\\
 \end{split}
\end{equation*}

{\bf Lemma 4}
If  $C^{\alpha}= C_A$ and $C^{\beta}= C_B$  then the
communities will remain separate (i.e., $[A:v:B]$) rather than $v$ joining with community $A$ (i.e., $[(A+v):B]$), if
$\alpha(\frac{1}{I_{\alpha}+1}+\frac{1}{(\alpha+\beta)\beta}) +(C^v_A-1) <0$.\\

Otherwise, if  $C^{\alpha}= C_A\frac{(I_{\alpha}-1)}{(I_{\alpha}+1)}$; $C^{\beta}= C_B\frac{(I_{\beta}-1)}{(I_{\beta}+1)}$ and  then the
communities will remain
separate if $\alpha(\frac{1-2C_A}{I_{\alpha}+1}+\frac{1}{(\alpha+\beta)\beta}) +(C^v_A-1) <0$.

%$\alpha(\frac{2C_A-1}{I_{\alpha}+1}) +(1-C^v_A) \ge \frac{\alpha}{(\alpha+\beta)\beta}$}\\
{\bf Proof:} We are comparing Case 1 and Case 4 for the case $C^{\alpha}= C_A\frac{(I_{\alpha}-1)}{(I_{\alpha}+1)}$; $C^{\beta}=
C_B\frac{(I_{\beta}-1)}{(I_{\beta}+1)}$. The difference in total permanence is:
\begin{equation*}
\begin{split}
P_{(A+v):B}-P_{(A:v:B)} & = \alpha C^\alpha+  \frac{\alpha}{(\alpha+\beta)\beta}-(1-C^v_A)\\
                                         &-(\alpha(\frac{I_{\alpha}}{I_{\alpha}+1} -(1-C_A)) \\
                                  &=\alpha(\frac{1-2C_A}{I_{\alpha}+1})+ \frac{\alpha}{(\alpha+\beta)\beta}+(C^v_A-1)\\
		\end{split}
\end{equation*}

 %This value will be negative (favor merge) if: $\alpha(\frac{2C_A-1}{I_{\alpha}+1}+(1-C^v_A)) >  \frac{\alpha}{(\alpha+\beta)\beta}$\\
  
  If we consider the case $C^\alpha=C_A$ and $C^\beta=C_B$, then
  \begin{equation*}
\begin{split}
  P_{(A+v):B}-P_{(A:v:B)} & = \alpha C^\alpha+  \frac{\alpha}{(\alpha+\beta)\beta}-(1-C^v_A)\\
                                         &-(\alpha(\frac{I_{\alpha}}{I_{\alpha}+1} -(1-C_A)) \\
                                            &=\alpha(\frac{1}{I_{\alpha}+1}+ \frac{1}{(\alpha+\beta)\beta})+(C^v_A-1)\\
                                \end{split}
\end{equation*} 

 \subsection {Mitigation of issues in modularity maximization}
 Using the above lemmas, we can determine the conditions for which
a particular assignment (of the four possible ones) will give the highest permanence. Using these conditions, we shall show how permanence
overcomes three major shortcomings of modularity maximization.

(i) {\bf Degeneracy of solution} is a problem where a community scoring metric (e.g., modularity) admits multiple distinct high-scoring
solutions and typically lacks a clear global maximum, thereby, resorting to tie-breaking~\cite{good2010}.  Consider the case where the vertex $v$ has equal connections to groups $A$ and $B$, therefore $\alpha$ = $\beta$, and the neighbors of $v$ are not connected to each other, i.e. $C^{\alpha}= C_A\frac{(I_{\alpha}-1)}{(I_{\alpha}+1)}$; $C^{\beta}=C_B\frac{(I_{\beta}-1)}{(I_{\beta}+1)}$. Since the neighbors are not connected therefore  $C^v_A=C^v_B=0$.

In this case the condition in Lemma 4 becomes; $P_{(A+v):B}-P_{(A:v:B)}= \alpha\frac{1-2C_A}{I_{\alpha}+1}+ \frac{1}{2\alpha} -1$
Because the values of $C_A$ range from $0$ to $1$, the values of $(1-2C_A)$ is negative. Moreover $\frac{1}{2\alpha}$ is less than 1.  Therefore the value of  $P_{(A+v):B}-P_{(A:v:B)}$ is negative, indicating that permanence is higher if $v$, $A$ and $B$ form separate communities.

According to Lemma 3, the communities will merge if:\\
$ \frac{\alpha(\alpha-1)C^v_A+\beta(\beta-1)C^v_B}{(\alpha+\beta)(\alpha+\beta-1)} >\alpha\frac{(2C_A-1)}{I_{\alpha}+1} +\beta\frac{(2C_B-1)}{I_{\beta}+1}$.  Since $C^v_A=C^v_B=0$, the left hand side is 0.  Therefore permanence is higher if $v$, $A$ and $B$ form separate communities.

Therefore, {\em if $v$ has equal number of connections to each community,  and the neighbors  of $v$ are not connected then $v$ will remain as singleton, rather than arbitrarily joining any of its neighbor groups.}

%In our example, if
%$\alpha$ = $\beta$, modularity maximizing algorithm will assign $v$ arbitrarily to $A$ or $B$. However, in the case of permanence, such ties
%are often resolved because $v$ will remain as a separate community so long as the following condition is maintained:
%
%{\em Condition 1. 
%If $\alpha=\beta$, $C^{\beta}= C_B\frac{I_{\beta}-1}{I_{\beta}+1}$, then communities $A$, $B$ and $v$ will merge  if 
% $\alpha(\frac{2C_A-1}{I_{\alpha}+1}) +(1-C^v_A) \ge \frac{1}{2\alpha}$.}
%
%We can see that this condition is true by applying Lemma 4 and then substituting the appropriate values of the quantities.
%We also observe that when $\alpha=\beta=1$, then $C^v_A=0$ and the communities will always remain separate. Furthermore, as $\alpha$ increases,
%the left-hand of the equation is going to be larger than the right, thus increasing the chance of separate communities. In the example of a $5\times5$ complete grid (Figure~\ref{example_diss} (b)), we observe that optimizing permanence  generates
%only one solution by assigning each vertex into separate singleton communities; whereas modularity produces multiple  solutions by combining
%two or more vertices. 
%

%However, permanence
%does not completely eliminate the solution degeneracy problem. In certain cases, higher permanence is obtained by moving $v$ to either $A$
%or to $B$. 

(ii) {\bf {Resolution limit}} is a problem where communities of certain small size are merged into larger ones~\cite{Barthelemy,good2010}.
One of the classic examples where modularity cannot identify communities of small size is a cycle of $m$ cliques (see
Figure~\ref{example_diss} (a)). Here maximum modularity is obtained if two neighboring cliques are merged. 

 In the case of permanence, we can determine that whether two communities $A$ and $B$ would merge (as in modularity) or whether $v$ would
join community $A$ (we select $A$ as the community to explain the case, but similar analysis can also be done for the case when $v$ joins
$B$).

We assume that the communities $A$ and $B$ are tightly connected, such that $ C_A >.5$ and $C_B>.5$.  We assume $v$ is tightly connected to group $A$, such that $C^v_A \approx 1$ and connected by one edge to group $B$ ($\beta=1$), such that $C^\beta=C_B\frac{I_{\beta}-1}{I_{\beta}+1}$.

From Lemma 2 we have;
$P_{(A+v):B}-P_{(A+v+B)}  = \frac{\alpha}{(\alpha+1)}-1-\frac{{1}}{I_{\beta}+1} +\frac{2C_B}{I_\beta+1} +\frac{2}{\alpha+1}$

which is equal to $\frac{1}{(\alpha+1)} +\frac{2C_B-1}{I_{\beta}+1}$. Since $C_B >.5$ therefore the value is positive. This indicates that  permanence is higher if $v$ joins group $A$ rather than if the three groups merge.

% 
%{\em Condition 2. 
%Joining $v$ to community $A$  gives higher permanence than merging the communities $A$, $B$ and $v$ if $C^{\beta}= C_B$, and
%$\frac{\gamma}{(\gamma+1)\beta} +\frac{C^v_A(2\gamma+1)-C^v_B}{(\gamma+1)^2} -\frac{\beta}{I_{\beta}+1}>1$;  where $\gamma = \alpha/\beta$
%and also if $C^{\beta}= C_B\frac{I_{\beta}-1}{I_{\beta}+1}$, and $\frac{\gamma}{(\gamma+1)\beta}
%+\frac{C^v_A(2\gamma+1)-C^v_B}{(\gamma+1)^2}
%+\frac{\beta(2C_B-1)}{I_{\beta}+1}>1$.  }
%  
% We can see that this condition is true by applying Lemma 2. 
% The clique example follows a special case where $v$ is connected by one edge to community $B$ and is connected to all nodes in community
%$A$.  Therefore $\beta=1$ and adding $v$ decreases the internal clustering coefficient of $B$ ($C^{\beta}=
%C_B\frac{I_{\beta}-1}{I_{\beta}+1}$) and $C^v_B=0$ (we set permanence of vertices with less than two neighbors as zero). Moreover, when $A$
%is a clique, and $v$ is connected to all the nodes of $A$, $C^v_A$ becomes $1$. By substituting these values in  Condition 2, we see, as per Lemma 2, that
%higher permanence is obtained by joining $v$  to community $A$ and the neighboring communities should not be merged.
% 
 Note that, this result is independent of the size of the communities $A$ and $B$. This phenomenon highlights that in general, {\em if $v$ is very tightly connected to a community and very loosely connected to another
community, highest permanence is obtained when $v$ joins the community to which it is more connected.}

\begin{table*}[!t]
\caption{Change in modularity, permanence and the other network parameters with the (near-)symmetric growth of coauthorship network as
discussed in Section~\ref{real_nw}. $N$: number of nodes, $C$: number of communities, $I$: internal degree, $D$: degree, $c_{in}(v)$:
clustering coefficient of $v$ with respect to its internal neighbors, $E_{max}(v)$: maximum external connectivity of $v$, $CD$: average
intra-community density (number of edges normalized by the number of nodes). The consistency of four network parameters indicates symmetric
growth of the network in different instantiations. }\label{asymptotic}
\centering
 \scalebox{0.68}{
\begin{tabular}{|c|c|c|c|c|c|c|c|c|c|c|c|c|}
\hline
\multirow{8}{*}{\begin{sideways}Coauthorship\end{sideways}} & \multirow{5}{*}{Network } &  $N$ & 964 & 1515 & 1991 & 2681 & 3386 & 4836 &
6284 & 7814 & 9001 & 10386 \\\cline{3-13}
			      & \multirow{5}{*}{properties} & $C$ & 24 & 24 & 24 & 24 &24 &24 &24 &24 &24 &24 \\\cline{3-13}
			      & &  $\frac{I}{D}$ &  0.082 & 0.095 & 0.093 & 0.091 & 0.089 & 0.104 & 0.111 & 0.112 & 0.115 & 0.113
\\\cline{3-13}
			      & &   $\frac{1}{E_{max}(v)}$ ($\times 10^{-4}$) & 3.8 & 3.2 & 2.9 & 3.9 & 2.8 & 2.11 & 2.39 & 2.92 & 2.69 &
3.22\\\cline{3-13}
			      & & $1-c_{in}(v)$ & 0.239 & 0.248 & 0.246 & 0.251 & 0.251 & 0.260 & 0.265 & 0.269 & 0.270 &
0.274\\\cline{3-13}
			      & & $CD$ & 74.30 & 80.30 & 90.34 & 98.18 & 102.68 & 118.68 & 118.72 & 123.29 & 110.22 & 123.292\\\cline{2-13}
			      & \multicolumn{2}{c|}{Modularity} & 0.369 & 0.374 & 0.395 & 0.392 & 0.421 & 0.422 & 0.465 & 0.471 & 0.493 &
0.501\\\cline{2-13}
			      & \multicolumn{2}{c|}{Permanence} & 0.094 & 0.092 & 0.092 & 0.096 & 0.095 & 0.095 & 0.097 & 0.097 & 0.097 &
0.098 \\\hline

\end{tabular}}
\end{table*}

(iii) {\bf Asymptotic growth of value} of a metric implies a strong dependence on both the size of the network and the number of modules the
network contains~\cite{good2010}. Rewriting equation~\ref{perm}, we get the permanence of the entire network $G$ as follows:
{\scriptsize$Perm(G)=\frac{1}{|V|} \sum_{v \in V}\left[\frac{I(v)}{D(v)E_{max}(v)}\right]-\frac{1}{|V|}\sum_{v \in
V}\left[(1-c_{in}(v))\right]$}. We can notice that most of the parameters in the above formula are independent of the symmetric growth of
network size and the number of communities. 
Table~\ref{asymptotic} illustrates the property from a real-life example of coauthorship network where the modularity increases with
increase in the size of the network, while permanence remains almost constant.

\section{Conclusion and Future Work}\label{conc}
In this paper, we present a new vertex-centric community quality metric, called {\em permanence}, that unlike other metrics considers both the connection density among internal neighbors and the distribution of external connectivity of a vertex. We empirically demonstrated on synthetic and real-world networks that permanence is an effective community evaluation metric compared to other well-known approaches such as modularity, conductance and cut-ratio. We also showed how permanence is appropriately sensitive to the fluctuations of community structure. Further experiments on characterizing permanence revealed that -- (i) permanence can measure the persistence of a vertex in its own community, and (ii) one can strengthen the community structure by suitably removing nodes with low permanence value. Finally, we developed a new community detection algorithm by maximizing permanence  - MaxPerm that has a much superior performance compared to state-of-the art algorithms 
on most datasets. Moreover, MaxPerm detects more efficient and realistic 
community structure -- (i) the obtained communities are highly connected, irrespective of the size of the communities, as a result of which one is
able to detect small and even singleton communities; (ii) the communities obtained by MaxPerm are less affected by the initial vertex ordering.  

The proposed metric calls for deeper levels of investigation. More algorithms and datasets from diverse areas need to be selected to reinforce the robustness of our proposed metric. Since permanence is a local metric, one immediate direction would be to discover local community boundary for a particular seed node. We intend to extend permanence metric to enable evaluation of
the quality of overlapping community structures and to weighted and directed networks. 
%With the gradual increase of network size in time-varying environment, we are interested to see how permanence captures the movement of a vertex in different communities. 
Overall, we believe that this metric will help in formulating a strong theoretical foundation in the identification and evaluation of various types of community strictures where the ground-truth is not known.

\appendix
\section*{APPENDIX}
 In this appendix, we expand Table \ref{avg_Improvement}. In this table, the differences between the results obtained from
MaxPerm and all the other algorithms are shown in terms of six validation measures for all the networks.

\begin{sidewaystable} 
\centering
\caption{Differences of {\bf MaxPerm} with the other algorithms in terms of the validation metrics. Positive differences
indicate the improvement of our algorithm. The rows indicated by ``L'' show values obtained from the LFR
graphs with $\mu=$0.1, 0.3 and 0.6 respectively (from left to right and separated by semicolons). The rows indicated by ``R'' show values 
for football, railway and
coauthorship networks (from left to right and separated by semicolons). The average improvements over different validation
measures are shown in rows 8 and 15  for LFR and real-world networks respectively.}\label{Improvement}

\scalebox{0.7}{
\begin{tabular}{|c|c|c|c|c|c|c|c|c|c|}
\hline
Type & Validation& Louvain & FastGreedy & CNM & WalkTrap & Infomod & Infomap & COPRA & OSLOM\\
 & metrics        &         &             &        &      &        &         &       &        \\\hline\hline

 \multirow{6}{*}{L} & NMI & 0.14; 0.00; -0.78 & 0.00; 0.81; -0.02 & 0.07; 0.24; -0.25 & 0.00; 0.00; -0.13 & 0.04; 0.05; -0.78  &  0.00;
0.01;
0.12 & 0.08; 0.09; -0.78 & 0.00; 0.01; 0.12 \\\cline{2-10}
                   & ARI & 0.00; -0.02; -0.76& 0.00; 0.98; 0.03 & 0.24; 0.59; -0.10 & 0.00; 0.01; -0.52 & 0.11; 0.13; -0.05 & 0.00; -0.01;
-0.95 & 0.16; 0.01; 0.02& 0.00; -0.01; -0.86\\\cline{2-10}
                   & PU & 0.00; 0.00; -0.72 & 0.00; 0.86; 0.04 & 0.12; 0.41; -0.13 & 0.00; -0.01; -0.58 & 0.08; 0.09; -0.11 & 0.00; 0.00;
-0.83 & 0.09; -0.01; 0.06 & 0.00; 0.01; -0.81\\\cline{2-10}

                    & W-NMI & 0.00; 0.00; -0.78 & 0.00; 0.81; 0.01& 0.06; 0.23; -0.10 & 0.00; 0.00; -0.58 & 0.02; 0.04; -0.08 & 0.00;
0.00;
-0.94  & 0.07; 0.03; 0.04 & 0.00; 0.00; -0.88 \\\cline{2-10}

                    & W-ARI & 0.00; 0.02; -0.72 & 0.00; 0.93; 0.07 & 0.23; 0.54; -0.04 & 0.00; -0.01; -0.52& 0.05; 0.08; -0.02 & 0.00;
-0.01; -0.88 & 0.16; -0.01; 0.10 & 0.00; -0.01; -0.82\\\cline{2-10}

                    & W-PU & 0.00; 0.00; -0.79& 0.00; 0.86; 0.00 & 0.11; 0.39; -0.20 & 0.00; 0.00; -0.67 & 0.05; 0.09; -0.18 & 0.00; 0.00;
-0.89 & 0.09; 0.00; 0.01& 0.00; 0.00; -0.88\\\cline{2-10}

 &{\bf Avg.} &{\bf 0.02; 0.00; -0.75} &{\bf 0.00; 0.87; 0.02} &{\bf 0.14; 0.40; -0.13} &{\bf 0.00; 0.00; -0.50} &{\bf 0.06; 0.08;
-0.20} & {\bf 0.00; 0.00; -0.72} & {\bf 0.11; 0.02; -0.09} & {\bf 0.00; 0.00; -0.68} \\\hline\hline

  \multirow{6}{*}{R}  & NMI & 0.01; 0.37; 0.03  & 0.00; 0.14; 0.13  & 0.22; 0.07; 0.02  & 0.01; 0.11; 0.01  & 0.01; 0.25;
-0.02 &  0.01; 0.06; -0.06  & 0.02; 0.05; 0.14 & 0.03; 0.16; 0.09 \\\cline{2-10}

		& ARI & 0.00; 0.04; -0.08 & 0.00; 0.36; 0.13  & 0.42; -0.09; 0.02 & 0.03; -0.07; 0.08 & 0.03; 0.05; -0.01 & 0.04; -0.06;
-0.05 & 0.09; -0.06; 0.12 & -0.06; 0.01; 0.05\\\cline{2-10}

		    & PU& -0.01; 0.08; -0.10 & -0.01; 0.41; 0.13 & 0.26; 0.11; 0.12 & 0.05; 0.07; 0.05 & 0.02; 0.13; -0.01 & -0.02; 0.06;
-0.06 & 0.04; 0.06; 0.07 & -0.04; 0.11; 0.06 \\\cline{2-10}

		       & W-NMI & 0.07; 0.27; 0.05 & 0.02; 0.56; 0.18 & 0.37; 0.10; 0.05 & 0.05; 0.14; 0.07 & 0.05; 0.43; -0.06 & 0.03; 0.17;
-0.01 & 0.04; 0.14; 0.02 & 0.03; 0.28; 0.12\\\cline{2-10}

		      & W-ARI & 0.02; 0.05; 0.05 & 0.00; 0.31; 0.11 & 0.34; -0.15; 0.02  & 0.02; -0.11; 0.05 & 0.02; 0.17; -0.06 & 0.00;
-0.07; -0.01  & 0.01; -0.10; 0.09 & 0.00; 0.05; 0.11\\\cline{2-10}

		      & W-PU & 0.01; 0.05; 0.02 & 0.06; 0.44; 0.14 & 0.21; -0.06; 0.06 & -0.05; -0.04; -0.15 & -0.05; 0.12; -0.12 & -0.06;
-0.05; 0.05 & 0.00; -0.05; 0.08 & -0.06; 0.00; 0.10\\\cline{2-10}

 & {\bf Avg.} & {\bf 0.02; 0.14; 0.00} &{\bf 0.01; 0.37; 0.14} &{\bf 0.30; 0.00; 0.05} &{\bf 0.02; 0.02; 0.02} &{\bf 0.01; 0.19; -0.04}
&{\bf 0.00; 0.02; -0.02}  &{\bf 0.03; 0.01; 0.09} & {\bf -0.01; 0.11; 0.09} \\\hline

\end{tabular}}
\end{sidewaystable}

\bibliographystyle{ACM-Reference-Format-Journals}
\bibliography{sigproc}


\begin{thebibliography}{00}

%%% the Web address.

\ifx \showCODEN    \undefined \def \showCODEN     #1{\unskip}     \fi
\ifx \showDOI      \undefined \def \showDOI       #1{{\tt DOI:}\penalty0{#1}\ }
  \fi
\ifx \showISBNx    \undefined \def \showISBNx     #1{\unskip}     \fi
\ifx \showISBNxiii \undefined \def \showISBNxiii  #1{\unskip}     \fi
\ifx \showISSN     \undefined \def \showISSN      #1{\unskip}     \fi
\ifx \showLCCN     \undefined \def \showLCCN      #1{\unskip}     \fi
\ifx \shownote     \undefined \def \shownote      #1{#1}          \fi
\ifx \showarticletitle \undefined \def \showarticletitle #1{#1}   \fi
\ifx \showURL      \undefined \def \showURL       #1{#1}          \fi

\bibitem[\protect\citeauthoryear{Ahn, Bagrow, and Lehmann}{Ahn
  et~al\mbox{.}}{2010}]%
        {AhnY2010}
{Yong-Yeol Ahn}, {James~P. Bagrow}, {and} {Sune Lehmann}. 2010.
\newblock \showarticletitle{Link communities reveal multiscale complexity in
  networks}.
\newblock {\em Nature\/}  {466} (August 2010), 761--764.
\newblock


\bibitem[\protect\citeauthoryear{Arenas, Fern{\'a}ndez, and G{\'o}mez}{Arenas
  et~al\mbox{.}}{2008}]%
        {Arenas}
{A Arenas}, {A Fern{\'a}ndez}, {and} {S G{\'o}mez}. 2008.
\newblock \showarticletitle{Analysis of the structure of complex networks at
  different resolution levels}.
\newblock {\em New Journal of Physics\/} {10}, 5 (2008), 053039.
\newblock


\bibitem[\protect\citeauthoryear{Bader, Meyerhenke, Sanders, and Wagner}{Bader
  et~al\mbox{.}}{2013}]%
        {sriram}
{David~A. Bader}, {Henning Meyerhenke}, {Peter Sanders}, {and} {Dorothea
  Wagner} (Eds.). 2013.
\newblock {\em Graph Partitioning and Graph Clustering - 10th DIMACS
  Implementation Challenge Workshop, Georgia Institute of Technology, Atlanta,
  GA, USA, February 13-14, 2012. Proceedings}. Contemporary Mathematics, Vol.
  588. American Mathematical Society.
\newblock


\bibitem[\protect\citeauthoryear{Baumes, Goldberg, and Magdon-Ismail}{Baumes
  et~al\mbox{.}}{2005}]%
        {Baumes:2005}
{Jeffrey Baumes}, {Mark Goldberg}, {and} {Malik Magdon-Ismail}. 2005.
\newblock \showarticletitle{Efficient identification of overlapping
  communities}. In {\em Proceedings of the 2005 IEEE international conference
  on Intelligence and Security Informatics} {\em (ISI'05)}. Springer-Verlag,
  Berlin, Heidelberg, 27--36.
\newblock


\bibitem[\protect\citeauthoryear{Berry, Hendrickson, LaViolette, and
  Phillips}{Berry et~al\mbox{.}}{2011}]%
        {Berry_PRE2011}
{Jonathan~W. Berry}, {Bruce Hendrickson}, {Randall~A. LaViolette}, {and}
  {Cynthia~A. Phillips}. 2011.
\newblock \showarticletitle{{Tolerating the community detection resolution
  limit with edge weighting}}.
\newblock {\em Physical Review E\/} {83}, 5 (May 2011), 056119.
\newblock


\bibitem[\protect\citeauthoryear{Blondel, Guillaume, Lambiotte, and
  Lefebvre}{Blondel et~al\mbox{.}}{2008}]%
        {blondel2008}
{Vincent~D Blondel}, {Jean-Loup Guillaume}, {Renaud Lambiotte}, {and} {Etienne
  Lefebvre}. 2008.
\newblock \showarticletitle{{Fast unfolding of communities in large networks}}.
\newblock {\em J. Stat. Mech\/} (2008), P10008.
\newblock


\bibitem[\protect\citeauthoryear{Chakrabort, Sikdar, Tammana, Ganguly, and
  Mukherjee}{Chakrabort et~al\mbox{.}}{2013}]%
        {asonam}
{Tanmoy Chakrabort}, {Sandipan Sikdar}, {Vihar Tammana}, {Niloy Ganguly}, {and}
  {Animesh Mukherjee}. 2013.
\newblock \showarticletitle{Computer Science Fields As Ground-truth
  Communities: Their Impact, Rise and Fall}. In {\em Proceedings of the 2013
  IEEE/ACM International Conference on Advances in Social Networks Analysis and
  Mining} {\em (ASONAM '13)}. ACM, New York, NY, USA, 426--433.
\newblock


\bibitem[\protect\citeauthoryear{Chakraborty}{Chakraborty}{2015}]%
        {1742}
{Tanmoy Chakraborty}. 2015.
\newblock \showarticletitle{Leveraging disjoint communities for detecting
  overlapping community structure}.
\newblock {\em Journal of Statistical Mechanics: Theory and Experiment\/}
  {2015}, 5 (2015), P05017.
\newblock
\showURL{%
\url{http://stacks.iop.org/1742-5468/2015/i=5/a=P05017}}


\bibitem[\protect\citeauthoryear{Chakraborty, Dalmia, Mukherjee, and
  Ganguly}{Chakraborty et~al\mbox{.}}{2016a}]%
        {0002DMG16}
{Tanmoy Chakraborty}, {Ayushi Dalmia}, {Animesh Mukherjee}, {and} {Niloy
  Ganguly}. 2016a.
\newblock \showarticletitle{Metrics for Community Analysis: {A} Survey}.
\newblock {\em CoRR\/}  {abs/1604.03512} (2016).
\newblock


\bibitem[\protect\citeauthoryear{Chakraborty, Kumar, Ganguly, Mukherjee, and
  Bhowmick}{Chakraborty et~al\mbox{.}}{2016b}]%
        {0002KGMB16}
{Tanmoy Chakraborty}, {Suhansanu Kumar}, {Niloy Ganguly}, {Animesh Mukherjee},
  {and} {Sanjukta Bhowmick}. 2016b.
\newblock \showarticletitle{GenPerm: {A} Unified Method for Detecting
  Non-overlapping and Overlapping Communities}.
\newblock {\em CoRR\/}  {abs/1604.03454} (2016).
\newblock


\bibitem[\protect\citeauthoryear{Chakraborty, Srinivasan, Ganguly, Bhowmick,
  and Mukherjee}{Chakraborty et~al\mbox{.}}{2013}]%
        {chakraborty}
{Tanmoy Chakraborty}, {Sriram Srinivasan}, {Niloy Ganguly}, {Sanjukta
  Bhowmick}, {and} {Animesh Mukherjee}. 2013.
\newblock \showarticletitle{{Constant Communities in Complex Networks}}.
\newblock {\em Scientific Reports\/}  {3} (May 2013).
\newblock


\bibitem[\protect\citeauthoryear{Chakraborty, Srinivasan, Ganguly, Mukherjee,
  and Bhowmick}{Chakraborty et~al\mbox{.}}{2014}]%
        {Chakraborty_kdd}
{Tanmoy Chakraborty}, {Sriram Srinivasan}, {Niloy Ganguly}, {Animesh
  Mukherjee}, {and} {Sanjukta Bhowmick}. 2014.
\newblock \showarticletitle{On the Permanence of Vertices in Network
  Communities}. In {\em Proceedings of the 20th ACM SIGKDD International
  Conference on Knowledge Discovery and Data Mining} {\em (KDD '14)}. ACM, New
  York, NY, USA, 1396--1405.
\newblock
\showISBNx{978-1-4503-2956-9}
\showDOI{%
\url{http://dx.doi.org/10.1145/2623330.2623707}}


\bibitem[\protect\citeauthoryear{Chen, Nguyen, and Szymanski}{Chen
  et~al\mbox{.}}{2013}]%
        {Chen_2013}
{Mingming Chen}, {Tommy Nguyen}, {and} {Boleslaw Szymanski}. 2013.
\newblock \showarticletitle{A New Metric for Quality of Network Community
  Structure}.
\newblock {\em ASE Human Journal\/} {1}, 4 (2013), 226--240.
\newblock


\bibitem[\protect\citeauthoryear{Chierichetti, Lattanzi, and
  Panconesi}{Chierichetti et~al\mbox{.}}{2010}]%
        {ChierichettiLP10}
{Flavio Chierichetti}, {Silvio Lattanzi}, {and} {Alessandro Panconesi}. 2010.
\newblock \showarticletitle{Rumour Spreading and Graph Conductance}. In {\em
  SODA}. SIAM, 1657--1663.
\newblock
\showURL{%
\url{http://www.siam.org/proceedings/soda/2010/SODA10_135_chierichettif.pdf}}


\bibitem[\protect\citeauthoryear{Clauset, Newman, , and Moore}{Clauset
  et~al\mbox{.}}{2004}]%
        {Clauset2004}
{Aaron Clauset}, {M.~E.~J. Newman}, {}, {and} {Cristopher Moore}. 2004.
\newblock \showarticletitle{{Finding community structure in very large
  networks}}.
\newblock {\em Phys. Rev. E\/} {70}, 6 (2004), 066111.
\newblock


\bibitem[\protect\citeauthoryear{Danon, Diaz-Guilera, Duch, and Arenas}{Danon
  et~al\mbox{.}}{2005}]%
        {danon2005ccs}
{L. Danon}, {A. Diaz-Guilera}, {J. Duch}, {and} {A. Arenas}. 2005.
\newblock \showarticletitle{{Comparing community structure identification}}.
\newblock {\em Journal of Statistical Mechanics: Theory and Experiment\/}  {9}
  (2005), P09008.
\newblock


\bibitem[\protect\citeauthoryear{De~Meo, Ferrara, Fiumara, and Provetti}{De~Meo
  et~al\mbox{.}}{2013a}]%
        {DeMeo:2013}
{Pasquale De~Meo}, {Emilio Ferrara}, {Giacomo Fiumara}, {and} {Alessandro
  Provetti}. 2013a.
\newblock \showarticletitle{Enhancing community detection using a network
  weighting strategy}.
\newblock {\em Journal of Information Science\/}  {222} (Feb. 2013), 648--668.
\newblock


\bibitem[\protect\citeauthoryear{De~Meo, Ferrara, Fiumara, and Provetti}{De~Meo
  et~al\mbox{.}}{2013b}]%
        {lln2010}
{Pasquale De~Meo}, {Emilio Ferrara}, {Giacomo Fiumara}, {and} {Alessandro
  Provetti}. 2013b.
\newblock \showarticletitle{Enhancing Community Detection Using a Network
  Weighting Strategy}.
\newblock {\em Inf. Sci.\/}  {222} (Feb. 2013), 648--668.
\newblock
\showISSN{0020-0255}
\showDOI{%
\url{http://dx.doi.org/10.1016/j.ins.2012.08.001}}


\bibitem[\protect\citeauthoryear{Delvenne, Yaliraki, and Barahona}{Delvenne
  et~al\mbox{.}}{2010}]%
        {dyb2010}
{J.-C. Delvenne}, {S.~N. Yaliraki}, {and} {M. Barahona}. 2010.
\newblock \showarticletitle{Stability of graph communities across time scales}.
\newblock {\em Proceedings of the National Academy of Sciences\/} {107}, 29
  (2010), 12755--12760.
\newblock


\bibitem[\protect\citeauthoryear{Demers, Greene, Hauser, Irish, Larson,
  Shenker, Sturgis, Swinehart, and Terry}{Demers et~al\mbox{.}}{1987}]%
        {Demers:1987}
{Alan Demers}, {Dan Greene}, {Carl Hauser}, {Wes Irish}, {John Larson}, {Scott
  Shenker}, {Howard Sturgis}, {Dan Swinehart}, {and} {Doug Terry}. 1987.
\newblock \showarticletitle{Epidemic Algorithms for Replicated Database
  Maintenance}. In {\em PODC}. New York, USA, 1--12.
\newblock


\bibitem[\protect\citeauthoryear{Evans and Lambiotte}{Evans and
  Lambiotte}{2009}]%
        {evans:2009}
{T.~S. Evans} {and} {R. Lambiotte}. 2009.
\newblock \showarticletitle{{Line graphs, link partitions, and overlapping
  communities}}.
\newblock {\em Phys. Rev. E\/} {80}, 1 (July 2009), 016105.
\newblock


\bibitem[\protect\citeauthoryear{Farkas, {\'A}bel, Palla, and Vicsek}{Farkas
  et~al\mbox{.}}{2007}]%
        {Vicsek}
{Ill{\'e}s Farkas}, {D{\'a}niel {\'A}bel}, {Gergely Palla}, {and} {Tam{\'a}s
  Vicsek}. 2007.
\newblock \showarticletitle{Weighted network modules}.
\newblock {\em New Journal of Physics\/} {9}, 6 (2007), 180.
\newblock


\bibitem[\protect\citeauthoryear{Fortunato}{Fortunato}{2010}]%
        {Fortunato201075}
{Santo Fortunato}. 2010.
\newblock \showarticletitle{Community detection in graphs}.
\newblock {\em Physics Reports\/} {486}, 3-5 (2010), 75 -- 174.
\newblock
\showURL{%
\url{http://www.sciencedirect.com/science/article/B6TVP-4XPYXF1-1/2/99061fac64%
35db4343b2374d26e64ac1}}


\bibitem[\protect\citeauthoryear{Fortunato and Barthelemy}{Fortunato and
  Barthelemy}{2007}]%
        {Barthelemy}
{Santo Fortunato} {and} {M Barthelemy}. 2007.
\newblock \showarticletitle{Resolution limit in community detection}.
\newblock {\em PNAS\/} (Jan. 2007).
\newblock


\bibitem[\protect\citeauthoryear{Ghosh, Banerjee, Sharma, Agarwal, and
  Ganguly}{Ghosh et~al\mbox{.}}{2011}]%
        {Ghosh}
{Saptarshi Ghosh}, {Avishek Banerjee}, {Naveen Sharma}, {Sanket Agarwal}, {and}
  {Niloy Ganguly}. 2011.
\newblock \showarticletitle{Statistical Analysis of The Indian Railway Network:
  a Complex Network Approach}.
\newblock {\em Acta Physica Polonica B Proceedings Supplement\/}  {4} (March
  2011), 123--137.
\newblock


\bibitem[\protect\citeauthoryear{Girvan and Newman}{Girvan and Newman}{2002}]%
        {GN}
{M. Girvan} {and} {M.~E. Newman}. 2002.
\newblock \showarticletitle{Community structure in social and biological
  networks.}
\newblock {\em PNAS\/} {99}, 12 (June 2002), 7821--7826.
\newblock


\bibitem[\protect\citeauthoryear{Good, Montjoye, and Clauset}{Good
  et~al\mbox{.}}{2010}]%
        {good2010}
{B.H. Good}, {Y.A.~De Montjoye}, {and} {A. Clauset}. 2010.
\newblock \showarticletitle{Performance of modularity maximization in practical
  contexts}.
\newblock {\em Phys. Rev. E\/} {81}, 4 (2010), 046106.
\newblock


\bibitem[\protect\citeauthoryear{Guimera and Amaral}{Guimera and
  Amaral}{2005}]%
        {Guimera}
{Roger Guimera} {and} {Luis A.~Nunes Amaral}. 2005.
\newblock \showarticletitle{Functional cartography of complex metabolic
  networks}.
\newblock {\em Nature\/} {433}, 7028 (Feb 2005), 895--900.
\newblock


\bibitem[\protect\citeauthoryear{He, Liu, Zhang, Jin, and Yang}{He
  et~al\mbox{.}}{2013}]%
        {Dongxiao}
{Dongxiao He}, {Dayou Liu}, {Weixiong Zhang}, {Di Jin}, {and} {Bo Yang}. 2013.
\newblock \showarticletitle{Discovering link communities in complex networks by
  exploiting link dynamics}.
\newblock {\em CoRR\/}  {abs/1303.4699} (2013).
\newblock


\bibitem[\protect\citeauthoryear{Holland and Leinhardt}{Holland and
  Leinhardt}{1971}]%
        {Holland1971}
{Paul~W Holland} {and} {Samuel Leinhardt}. 1971.
\newblock \showarticletitle{{Transitivity in Structural Models of Small
  Groups}}.
\newblock {\em Small Group Research\/} {2}, 2 (1971), 107--124.
\newblock


\bibitem[\protect\citeauthoryear{Hubert and Arabie}{Hubert and Arabie}{1985}]%
        {hubert1985}
{L. Hubert} {and} {P. Arabie}. 1985.
\newblock \showarticletitle{{Comparing partitions}}.
\newblock {\em Journal of classification\/} {2}, 1 (1985), 193--218.
\newblock


\bibitem[\protect\citeauthoryear{Kannan, Vempala, and Veta}{Kannan
  et~al\mbox{.}}{2000}]%
        {Kannan:2000}
{R. Kannan}, {S. Vempala}, {and} {A. Veta}. 2000.
\newblock \showarticletitle{On Clusterings-good, Bad and Spectral}. In {\em
  Proceedings of the 41st Annual Symposium on Foundations of Computer Science}
  {\em (FOCS '00)}. IEEE Computer Society, Washington, DC, USA, 367--.
\newblock
\showISBNx{0-7695-0850-2}
\showURL{%
\url{http://dl.acm.org/citation.cfm?id=795666.796585}}


\bibitem[\protect\citeauthoryear{Kempe, Kleinberg, and Tardos}{Kempe
  et~al\mbox{.}}{2003}]%
        {Kempe:2003}
{David Kempe}, {Jon Kleinberg}, {and} {\'{E}va Tardos}. 2003.
\newblock \showarticletitle{Maximizing the Spread of Influence Through a Social
  Network}. In {\em Proceedings of the Ninth ACM SIGKDD International
  Conference on Knowledge Discovery and Data Mining} {\em (KDD '03)}. ACM, New
  York, NY, USA, 137--146.
\newblock
\showISBNx{1-58113-737-0}
\showDOI{%
\url{http://dx.doi.org/10.1145/956750.956769}}


\bibitem[\protect\citeauthoryear{Lambiotte}{Lambiotte}{2010}]%
        {Renaud}
{Renaud Lambiotte}. 2010.
\newblock \showarticletitle{Multi-scale modularity in complex networks.}. In
  {\em WiOpt}. IEEE, 546--553.
\newblock


\bibitem[\protect\citeauthoryear{Lancichinetti and Fortunato}{Lancichinetti and
  Fortunato}{2009}]%
        {Lancichinetti}
{Andrea Lancichinetti} {and} {Santo Fortunato}. 2009.
\newblock \showarticletitle{Benchmarks for testing community detection
  algorithms on directed and weighted graphs with overlapping communities}.
\newblock {\em Phys. Rev. E\/} {80}, 1 (July 2009), 016118.
\newblock


\bibitem[\protect\citeauthoryear{Lancichinetti and Fortunato}{Lancichinetti and
  Fortunato}{2011}]%
        {santo_11}
{Andrea Lancichinetti} {and} {Santo Fortunato}. 2011.
\newblock \showarticletitle{Limits of modularity maximization in community
  detection}.
\newblock {\em Phys. Rev. E\/}  {84} (2011), 066122.
\newblock


\bibitem[\protect\citeauthoryear{Lancichinetti and Fortunato}{Lancichinetti and
  Fortunato}{2012a}]%
        {Santo}
{Andrea Lancichinetti} {and} {Santo Fortunato}. 2012a.
\newblock \showarticletitle{Consensus clustering in complex networks}.
\newblock {\em Scientific Reports\/}  {2} (2012).
\newblock


\bibitem[\protect\citeauthoryear{Lancichinetti and Fortunato}{Lancichinetti and
  Fortunato}{2012b}]%
        {lf2012}
{Andrea Lancichinetti} {and} {Santo Fortunato}. 2012b.
\newblock \showarticletitle{Consensus clustering in complex networks}.
\newblock {\em CoRR\/}  {abs/1203.6093} (2012).
\newblock
\showURL{%
\url{http://dblp.uni-trier.de/db/journals/corr/corr1203.html#abs-1203-6093}}


\bibitem[\protect\citeauthoryear{Lancichinetti, Fortunato, and
  Kert{\'e}sz}{Lancichinetti et~al\mbox{.}}{2009}]%
        {Andrea}
{Andrea Lancichinetti}, {Santo Fortunato}, {and} {J{\'a}nos Kert{\'e}sz}. 2009.
\newblock \showarticletitle{Detecting the overlapping and hierarchical
  community structure in complex networks}.
\newblock {\em New Journal of Physics\/} {11}, 3 (2009), 033015.
\newblock


\bibitem[\protect\citeauthoryear{Lancichinetti, Radicchi, Ramasco, and
  Fortunato}{Lancichinetti et~al\mbox{.}}{2010}]%
        {DBLP:journals}
{Andrea Lancichinetti}, {Filippo Radicchi}, {Jose~J. Ramasco}, {and} {Santo
  Fortunato}. 2010.
\newblock \showarticletitle{Finding statistically significant communities in
  networks}.
\newblock {\em CoRR\/}  {abs/1012.2363} (2010).
\newblock


\bibitem[\protect\citeauthoryear{Leicht and Newman}{Leicht and Newman}{2008}]%
        {PhysRevLett.}
{E.~A. Leicht} {and} {M.~E.~J. Newman}. 2008.
\newblock \showarticletitle{Community Structure in Directed Networks}.
\newblock {\em Phys. Rev. Lett.\/} {100}, 11 (March 2008), 118703.
\newblock


\bibitem[\protect\citeauthoryear{Leskovec, Lang, Dasgupta, and
  Mahoney}{Leskovec et~al\mbox{.}}{2009}]%
        {cond_09}
{Jure Leskovec}, {Kevin~J. Lang}, {Anirban Dasgupta}, {and} {Michael~W.
  Mahoney}. 2009.
\newblock \showarticletitle{Community Structure in Large Networks: Natural
  Cluster Sizes and the Absence of Large Well-Defined Clusters}.
\newblock {\em Internet Mathematics\/} {6}, 1 (2009), 29--123.
\newblock


\bibitem[\protect\citeauthoryear{Leskovec, Lang, and Mahoney}{Leskovec
  et~al\mbox{.}}{2010}]%
        {Leskovec:2010}
{Jure Leskovec}, {Kevin~J. Lang}, {and} {Michael Mahoney}. 2010.
\newblock \showarticletitle{Empirical comparison of algorithms for network
  community detection}. In {\em Proceedings of the 19th international
  conference on World wide web} {\em (WWW '10)}. ACM, New York, NY, USA,
  631--640.
\newblock


\bibitem[\protect\citeauthoryear{Manning, Raghavan, and Sch\"{u}tze}{Manning
  et~al\mbox{.}}{2008}]%
        {Manning}
{Christopher~D. Manning}, {Prabhakar Raghavan}, {and} {Hinrich Sch\"{u}tze}.
  2008.
\newblock {\em Introduction to Information Retrieval}.
\newblock Cambridge University Press, New York, NY, USA.
\newblock


\bibitem[\protect\citeauthoryear{Newman}{Newman}{2006}]%
        {Newman:2006}
{M~E Newman}. 2006.
\newblock \showarticletitle{Modularity and community structure in networks}.
\newblock {\em PNAS\/} {103}, 23 (June 2006), 8577--8582.
\newblock


\bibitem[\protect\citeauthoryear{Newman}{Newman}{2003}]%
        {Newman-assort-2003}
{M.~E.~J. Newman}. 2003.
\newblock \showarticletitle{Mixing patterns in networks}.
\newblock {\em Phys. Rev. E\/} {67}, 2 (Feb. 2003), 026126.
\newblock
\showDOI{%
\url{http://dx.doi.org/10.1103/PhysRevE.67.026126}}


\bibitem[\protect\citeauthoryear{Newman}{Newman}{2004a}]%
        {PhysRevE.70}
{M.~E.~J. Newman}. 2004a.
\newblock \showarticletitle{Analysis of weighted networks}.
\newblock {\em Phys. Rev. E\/} {70}, 5 (Nov. 2004), 056131.
\newblock


\bibitem[\protect\citeauthoryear{Newman}{Newman}{2004b}]%
        {newman03fast}
{M.~E.~J. Newman}. 2004b.
\newblock \showarticletitle{{Fast algorithm for detecting community structure
  in networks}}.
\newblock {\em Phys. Rev. E\/} {69}, 6 (June 2004), 066133.
\newblock


\bibitem[\protect\citeauthoryear{Newman}{Newman}{2013}]%
        {Newman_13}
{M.~E.~J. Newman}. 2013.
\newblock \showarticletitle{Community detection and graph partitioning}.
\newblock {\em CoRR\/}  {abs/1305.4974} (2013).
\newblock


\bibitem[\protect\citeauthoryear{Newman and Girvan}{Newman and Girvan}{2004}]%
        {NewGir04}
{M.~E.~J. Newman} {and} {M. Girvan}. 2004.
\newblock \showarticletitle{Finding and evaluating community structure in
  networks}.
\newblock {\em Phys. Rev. E\/} {69}, 026113 (2004).
\newblock


\bibitem[\protect\citeauthoryear{Orman, Labatut, and Cherifi}{Orman
  et~al\mbox{.}}{2012}]%
        {Labatut}
{G{\"u}nce~Keziban Orman}, {Vincent Labatut}, {and} {Hocine Cherifi}. 2012.
\newblock \showarticletitle{Comparative Evaluation of Community Detection
  Algorithms: A Topological Approach}.
\newblock {\em CoRR\/}  {abs/1206.4987} (2012).
\newblock


\bibitem[\protect\citeauthoryear{Palla, Derényi, Farkas, and Vicsek}{Palla
  et~al\mbox{.}}{2005}]%
        {PalEtAl05}
{Gergely Palla}, {Imre Derényi}, {Illés Farkas}, {and} {Tamás Vicsek}. 2005.
\newblock \showarticletitle{Uncovering the overlapping community structure of
  complex networks in nature and society}.
\newblock {\em Nature\/} {435}, 7043 (June 2005), 814--818.
\newblock


\bibitem[\protect\citeauthoryear{{Pons} and {Latapy}}{{Pons} and
  {Latapy}}{2006}]%
        {JGAA-124}
{{Pascal} {Pons}} {and} {{Matthieu} {Latapy}}. 2006.
\newblock \showarticletitle{Computing Communities in Large Networks Using
  Random Walks}.
\newblock {\em Journal of Graph Algortihms and Applications\/} {10}, 2 (2006),
  191--218.
\newblock


\bibitem[\protect\citeauthoryear{Psorakis, Roberts, Ebden, and
  Sheldon}{Psorakis et~al\mbox{.}}{2011}]%
        {Psorakis}
{Ioannis Psorakis}, {Stephen Roberts}, {Mark Ebden}, {and} {Ben Sheldon}. 2011.
\newblock \showarticletitle{{Overlapping community detection using Bayesian
  non-negative matrix factorization}}.
\newblock {\em Phys. Rev. E\/} {83}, 6 (June 2011), 066114.
\newblock


\bibitem[\protect\citeauthoryear{Raghavan, Albert, and Kumara}{Raghavan
  et~al\mbox{.}}{2007a}]%
        {Raghavan:1057930}
{Usha~Nandini Raghavan}, {Reka Albert}, {and} {Soundar Kumara}. 2007a.
\newblock \showarticletitle{Near linear time algorithm to detect community
  structures in large-scale networks}.
\newblock {\em Phys. Rev. E\/}  {76} (Sep 2007), 036106.
\newblock


\bibitem[\protect\citeauthoryear{Raghavan, Albert, and Kumara}{Raghavan
  et~al\mbox{.}}{2007b}]%
        {Raghavan-2007}
{Usha~N. Raghavan}, {R{\'e}ka Albert}, {and} {Soundar Kumara}. 2007b.
\newblock \showarticletitle{{Near linear time algorithm to detect community
  structures in large-scale networks}}.
\newblock {\em Phys. Rev. E\/} {76}, 3 (Sept. 2007), 036106.
\newblock


\bibitem[\protect\citeauthoryear{Reichardt and Bornholdt}{Reichardt and
  Bornholdt}{2006}]%
        {reichardt2006smc}
{J. Reichardt} {and} {S. Bornholdt}. 2006.
\newblock \showarticletitle{{Statistical mechanics of community detection}}.
\newblock {\em Arxiv preprint cond-mat/0603718\/} (2006).
\newblock


\bibitem[\protect\citeauthoryear{Richardson, Mucha, and Porter}{Richardson
  et~al\mbox{.}}{2009}]%
        {Thomas}
{Thomas Richardson}, {Peter~J. Mucha}, {and} {Mason~A. Porter}. 2009.
\newblock \showarticletitle{Spectral tripartitioning of networks}.
\newblock {\em Phys. Rev. E\/}  {40} (2009), 027104.
\newblock


\bibitem[\protect\citeauthoryear{Riedy, Bader, Jiang, Pande, and Sharma}{Riedy
  et~al\mbox{.}}{2011}]%
        {seed-set-tr}
{Jason Riedy}, {David~A. Bader}, {Karl Jiang}, {Pushkar Pande}, {and} {Richa
  Sharma}. 2011.
\newblock {\em Detecting Communities from Given Seeds in Social Networks}.
\newblock {T}echnical {R}eport GT-CSE-11-01. Georgia Institute of Technology.
\newblock


\bibitem[\protect\citeauthoryear{Rosvall and Bergstrom}{Rosvall and
  Bergstrom}{2007}]%
        {rosvall2007}
{M. Rosvall} {and} {C.T. Bergstrom}. 2007.
\newblock \showarticletitle{{An information-theoretic framework for resolving
  community structure in complex networks}}.
\newblock {\em PNAS\/} {104}, 18 (2007), 7327.
\newblock


\bibitem[\protect\citeauthoryear{Rosvall and Bergstrom}{Rosvall and
  Bergstrom}{2008}]%
        {Rosvall29012008}
{Martin Rosvall} {and} {Carl~T. Bergstrom}. 2008.
\newblock \showarticletitle{Maps of random walks on complex networks reveal
  community structure}.
\newblock {\em PNAS\/} {105}, 4 (2008), 1118--1123.
\newblock


\bibitem[\protect\citeauthoryear{Seifi, Junier, Rouquier, Iskrov, and
  Guillaume}{Seifi et~al\mbox{.}}{2013}]%
        {sjri2012}
{Massoud Seifi}, {Ivan Junier}, {Jean-Baptiste Rouquier}, {Svilen Iskrov},
  {and} {Jean-Loup Guillaume}. 2013.
\newblock \showarticletitle{Stable Community Cores in Complex Networks}.
\newblock In {\em Complex Networks}, {Ronaldo Menezes}, {Alexandre Evsukoff},
  {and} {Marta~C. González} (Eds.). Studies in Computational Intelligence,
  Vol. 424. Springer Berlin Heidelberg, 87--98.
\newblock
\showISBNx{978-3-642-30286-2}
\showDOI{%
\url{http://dx.doi.org/10.1007/978-3-642-30287-9_10}}


\bibitem[\protect\citeauthoryear{Shi and Malik}{Shi and Malik}{2000}]%
        {Shi:2000}
{Jianbo Shi} {and} {Jitendra Malik}. 2000.
\newblock \showarticletitle{Normalized Cuts and Image Segmentation}.
\newblock {\em IEEE Trans. Pattern Anal. Mach. Intell.\/} {22}, 8 (Aug. 2000),
  888--905.
\newblock
\showISSN{0162-8828}
\showDOI{%
\url{http://dx.doi.org/10.1109/34.868688}}


\bibitem[\protect\citeauthoryear{Sun, Gao, and Han}{Sun et~al\mbox{.}}{2011}]%
        {Sun_11}
{Peng-Gang Sun}, {Lin Gao}, {and} {Shan~Shan Han}. 2011.
\newblock \showarticletitle{Identification of overlapping and non-overlapping
  community structure by fuzzy clustering in complex networks}.
\newblock {\em Inf. Sci.\/} {181}, 6 (2011), 1060--1071.
\newblock


\bibitem[\protect\citeauthoryear{Xie and Szymanski}{Xie and Szymanski}{2011}]%
        {Xie_11}
{Jierui Xie} {and} {Boleslaw~K. Szymanski}. 2011.
\newblock \showarticletitle{Community Detection Using A Neighborhood Strength
  Driven Label Propagation Algorithm}.
\newblock {\em CoRR\/}  {abs/1105.3264} (2011).
\newblock


\bibitem[\protect\citeauthoryear{Xie and Szymanski}{Xie and Szymanski}{2012}]%
        {Xie_12}
{Jierui Xie} {and} {Boleslaw~K. Szymanski}. 2012.
\newblock \showarticletitle{Towards Linear Time Overlapping Community Detection
  in Social Networks}.
\newblock {\em CoRR\/}  {abs/1202.2465} (2012).
\newblock


\bibitem[\protect\citeauthoryear{Yang and Leskovec}{Yang and Leskovec}{2012}]%
        {Yang:2012}
{Jaewon Yang} {and} {Jure Leskovec}. 2012.
\newblock \showarticletitle{Defining and evaluating network communities based
  on ground-truth}. In {\em Proceedings of the ACM SIGKDD Workshop on Mining
  Data Semantics} {\em (MDS '12)}. ACM, New York, NY, USA, 3:1--3:8.
\newblock


\bibitem[\protect\citeauthoryear{Yang and Leskovec}{Yang and Leskovec}{2014}]%
        {Yang14}
{Jaewon Yang} {and} {Jure Leskovec}. 2014.
\newblock \showarticletitle{Overlapping Communities Explain Core-Periphery
  Organization of Networks}.
\newblock {\em Proceedings of IEEE\/}  {102} (2014), 1892 -- 1902.
\newblock


\end{thebibliography}

\end{document}